\documentclass[10pt,journal]{IEEEtran}

\usepackage[T1]{fontenc}
\usepackage{amssymb}
\usepackage{amsmath}
\usepackage{graphicx}
\usepackage{algorithm, algcompatible}
\usepackage{hyperref}
\usepackage{bm}
\usepackage{colortbl}
\usepackage{xcolor}
\usepackage{multirow}
\usepackage{booktabs}
\usepackage{adjustbox}
\usepackage{multicol}
\usepackage{mathrsfs}
\usepackage{mathtools}
\usepackage{hhline}
\usepackage{subfigure}
\usepackage{framed}
\usepackage[percent]{overpic}
\usepackage{threeparttable}
\usepackage{eurosym}
\usepackage{orcidlink}
\usepackage[mathlines,switch]{lineno}
\usepackage[noadjust]{cite}
\usepackage{soul}
\usepackage{eso-pic}

\date{August 3, 2023}

\begin{document}
\AddToShipoutPictureBG{%
  \AtPageUpperLeft{%
    \setlength\unitlength{1in}%
    \hspace*{\dimexpr0.5\paperwidth\relax}
    \makebox(0,-0.75){\parbox{0.9\textwidth}{\small This article has been accepted for publication in IEEE Journal of Selected Topics in Applied Earth Observations and Remote Sensing. This is the author's version which has not been fully edited and content may change prior to final publication. Citation information: DOI 10.1109/JSTARS.2024.3362688}}%
}}
\AddToShipoutPictureBG{%
  \AtPageLowerLeft{%
    \setlength\unitlength{1in}%
    \hspace*{\dimexpr0.5\paperwidth\relax}
    \makebox(0,0.75)[c]{\parbox{0.9\textwidth}{\small This work is licensed under a Creative Commons Attribution-NonCommercial-NoDerivatives 4.0 License. For more information, see https://creativecommons.org/licenses/by-nc-nd/4.0}}%
}}

\title{Monitoring of Urban Changes with multi-modal Sentinel 1 and 2 Data in Mariupol, Ukraine, in 2022/23}

\author{\IEEEauthorblockN{1\textsuperscript{st} Georg Zitzlsberger \orcidlink{0000-0001-7467-8218}}
\IEEEauthorblockA{\textit{IT4Innovations}\\
\textit{VSB -- Technical University}\\
17.listopadu 15\\
70833 Ostrava, Czech Republic}
\and
\IEEEauthorblockN{2\textsuperscript{nd} Michal Podhoranyi \orcidlink{0000-0001-8405-6658}}
\IEEEauthorblockA{\textit{IT4Innovations}\\
\textit{VSB -- Technical University}\\
17.listopadu 15\\
70833 Ostrava, Czech Republic}
}

\author{\IEEEauthorblockN{Georg Zitzlsberger\IEEEauthorrefmark{1} \orcidlink{0000-0001-7467-8218} and
Michal Podhoranyi\IEEEauthorrefmark{2} \orcidlink{0000-0001-8405-6658}}\\
\IEEEauthorblockA{IT4Innovations,
VSB -- Technical University\\
17.listopadu 15\\
70833 Ostrava, Czech Republic}}

\maketitle

\begin{abstract}
The ability to constantly monitor urban changes is of significant socio-economic interest, like detecting trends in urban expansion or tracking the vitality of urban areas. Especially in present conflict zones or disaster areas, such insights provide valuable information to keep track of the current situation. However, they are often subject to limited data availability in space and time.

We built on our previous work, which used a transferred Deep Neural Network (DNN) operating on multi-modal Sentinel 1 and 2 data. In the current study, we have demonstrated and discussed its applicability in monitoring the present conflict zone of Mariupol, Ukraine, with high-temporal resolution Sentinel time series for the years 2022/23. A transfer to that conflict zone was challenging due to the limited availability of recent Very High Resolution (VHR) data.

The current work had two objectives. First, transfer learning with older and publicly available VHR data was shown to be sufficient. That guaranteed the availability of more and less expensive data as time constraints were relaxed. Second, in an ablation study, we analyzed the effects of loss of observations to demonstrate the resiliency of our method. That was of particular interest due to the malfunctioning of Sentinel 1B shortly before the selected conflict.

Our study demonstrated that urban change monitoring is possible for present conflict zones after transferring with older VHR data. It also indicated that, despite the multi-modal input, our method was more dependent on optical multispectral than Synthetic Aperture Radar (SAR) observations but resilient to loss of observations.

\end{abstract}


\section{Introduction}
The detection of changes with the use of satellite based remote sensing data has a history of almost six decades, with the first mentioning of a \textit{change detector device} by \cite{Shepard1964}. Since then many methods have been developed to detect changes~\cite{doi:10.1080/01431168908903939, rs13152869}. While many methods have been proposed to detect changes, in the last decade the focus moved more towards using neural networks~\cite{rs12101688, rs12152460, 9136674, doi:10.1080/10095020.2022.2085633, rs14040871, rs14071552, rs15082092} with the advent of DNNs in that time frame. The types of changes across those works vary drastically. Some works detect changes in general, including vegetation and water bodies, others consider only buildings but ignore other infrastructure. Only a subset considers urban structure types (UST) as defined by \cite{rs11020173}. In addition, the majority of works operate on observation pairs that are required to be of sufficient quality. The result was a large occurrence of so-called Siamese network architectures which replicate the network on the input side for each image as a pair. Overall, they are limited for broad use due to requiring high-quality VHR data, and they reduce the temporal resolution to detect and monitor urban changes.

\begin{figure*}[ht]
\begin{center}
\newcommand{\citelink}[2]{\hyperlink{cite.#1}{#2}}
\begin{overpic}[width=0.9\textwidth]{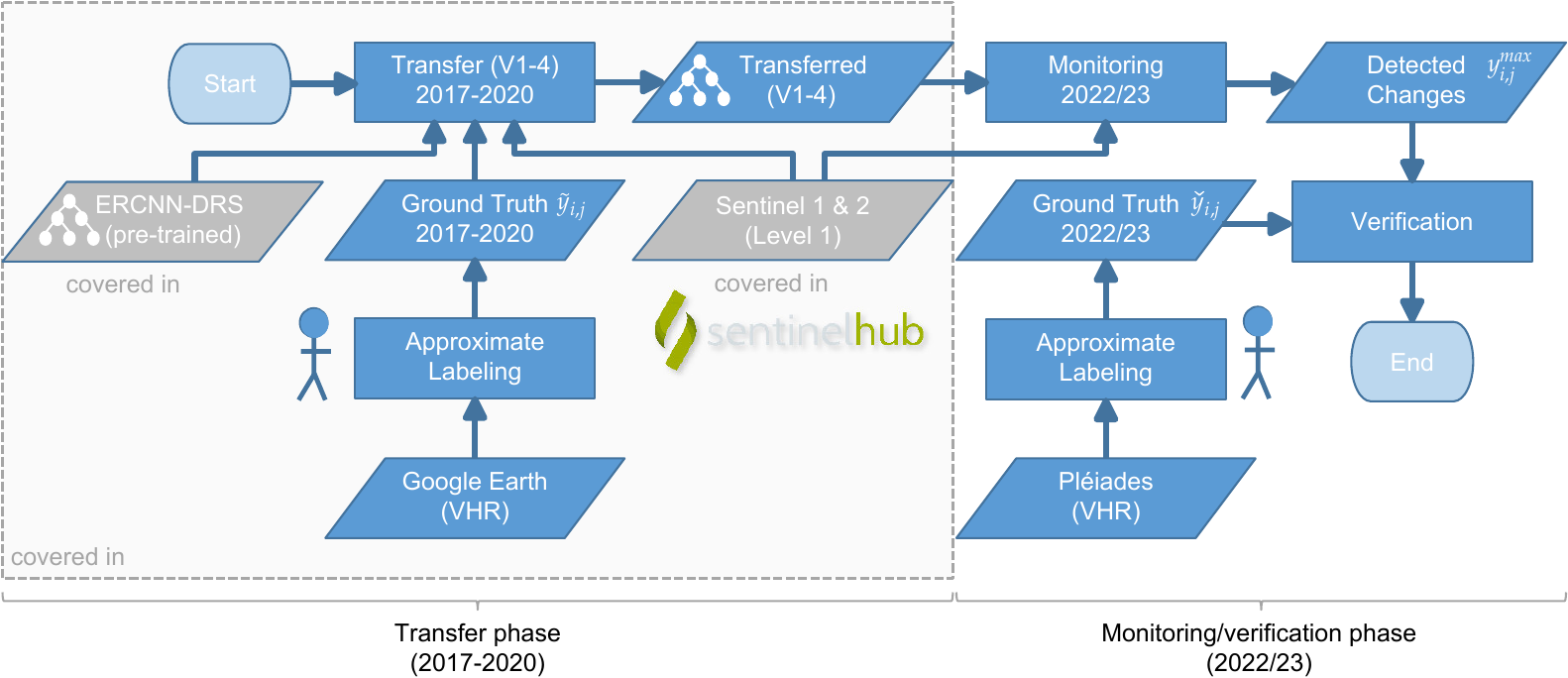}
 \put (12,24.7) {\textcolor{gray}{\tiny\cite{rs13153000}}}
 \put (8.5, 7.2) {\textcolor{gray}{\tiny\cite{doi:10.1080/01431161.2023.2243021}}}
 \put (53.5,24.7) {\textcolor{gray}{\tiny\cite{ZITZLSBERGER2023101369}}}
\end{overpic}
\end{center}
\caption{Flowchart of the transfer learning and monitoring process for the AoI of Mariupol. In blue are data and processing steps of our current work; gray denotes previous work.}
\label{flowchart_ukraine}
\end{figure*}

In our previous works, summarized in Figure~\ref{flowchart_ukraine}, we have addressed these problems by introducing a new approach. First, we have introduced a method that leveraged an ensemble of neural networks for level 1 Sentinel 1 and 2 multi-modal remote sensing data (i.e., SAR and  optical multispectral). Its design was tailored for the objective to continuously monitor urban changes~\cite{rs13153000}. This method operated on time series observations, partitioned into half-year windows, to provide enough context for allowing low quality level 1 data and to localize changes over time. It pre-trained a model, called \textit{Ensemble of Recurrent Convolutional Neural Networks for Deep Remote Sensing} (ERCNN-DRS), with synthetic but noisy labels to avoid manual labor. In a follow-up study~\cite{doi:10.1080/01431161.2023.2243021}, a further optimization with transfer learning was demonstrated to fine-tune the pre-trained network towards a specific Area of Interest (AoI) to increase the detection quality and allow more control of the UST changes to detect. For practical feasibility, the transfer learning used a set of windows simultaneously in order to simplify the manual ground truth generation guided by public VHR data, i.e., Google Earth\textsuperscript{\texttrademark} historic imagery, and spanned a larger time frame of multiple years. Both works were trained and verified only for a fixed time frame, i.e., 2017-2020. It therefore raised the question whether a model, transferred to one time frame, would also be applicable for a different time frame.

In this work, we reused the pre-trained ERCNN-DRS model and applied the same transfer learning method but to the AoI of Mariupol for the time frame 2017-2020. We subsequently evaluated the performance of the transferred model for the years 2022/23. This was done starting three months before the Russian invasion on February 24, 2022, and until mid-2023. Due to being an active war zone, VHR remote sensing data was limited, even if commercially accessible\footnote{Higher resolution Maxar WorldView data (0.15-0.3 m/pixel) for Ukraine was under embargo at the time of writing.}. However, medium resolution data like Sentinel 1 and Sentinel 2 were still available for that region without limitations and allowed the monitoring of ongoing urban changes. We analyzed the applicability of transferring to the AoI of Mariupol for the time frame 2017-2020 and its use for 2022/23. Because of the lack of public VHR data for that time frame, commercial Airbus Pl\'eiades observations were used for validation purposes. These would be expensive for transfer learning due to the required amount but we used them only for evaluation which does not require a larger area and kept costs low\footnote{An ESA NoR sponsorship worth 600 \euro\ for the Airbus Pl\'eiades data enabled the verification of our work.}. As we demonstrate, the transfer with an earlier time frame using public data can be sufficient which helps to keep costs low.

Furthermore, due to the outage of Sentinel 1B on December 23, 2021, we also addressed the question of the impact of loss of observations to the overall solution. As we will show in an ablation study, the chosen method is resilient and does not instantly break down if final observation patterns diverge from the ones seen during training.

This work was subject to two objectives. First, we analyzed whether a transferred model for a new AoI for a specific time frame can be used for a later time frame, even though observation patterns change. Second, the resiliency of our method to a decaying number of observations for the different observation modes, SAR and optical multispectral, were studied. We addressed both objectives with the case study to monitor urban changes of the, at the time of writing, besieged and occupied city of Mariupol in Ukraine where limited data is available despite the increased need to monitor changes.

In this work, section~\ref{method} describes our approach from the transfer of an existing pre-trained ERCNN-DRS to verification. Section~\ref{study_area} provides details on the selected study area, observation data, and data processing. The training procedure is explained in section~\ref{training}. Training and verification results are discussed in section~\ref{results} with both quantitative and qualitative analysis. This section also contains the ablation study to understand the resiliency of our method. In section~\ref{limitations} we summarize the shortcomings of our approach that require further work. Section~\mbox{\ref{discussion}} contains a discussion on the peculiarities and trade-offs of our methods to give guidance to adapting it to other scenarios. The key results and areas of improvement are summarized in section~\mbox{\ref{conclusion}}.

\section{Methodology}\label{method}
We built on top of three existing works to enable urban change monitoring and combined these in the current work for the different AoI of Mariupol. Figure~\ref{flowchart_ukraine} summarizes our two-step approach from transfer to validation. Reused data from other works is in gray, and items covered in this work are in blue. Most of the steps were automatized with only the labeling procedures carried out manually. Our applied methodology is described in the following, separated by reuse of existing methods and their extensions needed by this work.

\subsection{Existing Methods}
The pre-trained baseline ERCNN-DRS model stems from \cite{rs13153000}. The model architecture is shown in Figure~\ref{ercnn-drs_arch} in the Appendix for completeness. That work laid the foundation of utilizing large observation counts, so-called deep-temporal windows $\bm{w}^{t}_{i,j}$ of a fixed duration, with multi-modal remote sensing data for identifying urban changes. These windows start at time $t$ and are tiled with $i$ and $j$ being tile coordinates. Furthermore, these windows were applied in a sliding window approach, allowing for a longer observation period than the fixed window duration.

To create the windowed time series the \textit{rsdtlib} library~\cite{ZITZLSBERGER2023101369} was utilized. It retrieves Sentinel 1 and Sentinel 2 remote sensing level 1 observations from \textit{Sentinel Hub} and pre-processes them. The pre-processing involves temporal stacking, assembling, tiling and windowing to retrieve the final multi-modal windows $\bm{w}^{t}_{i,j}$. Important parameters of the data pipeline stages are shown in Table~\ref{parameters}, characterizing the data, tiles, and windows. In the following we briefly describe these steps.

\begin{table}[tb]
\begin{center}
{
\begin{tabular}{clll}
\toprule
&\textbf{Parameter} & \textbf{Mnemonic} & \textbf{Value}\\
\midrule
\multirow{2}{*}{\rotatebox[origin=c]{90}{\parbox[c]{0.4cm}{\centering data}}}
&$b^{\textrm{[}asc \mid dsc\textrm{]}}_{SAR}$ & SAR bands       & 2 (VV+VH) \\
&$b_{OPT}$                                        & optical bands   & 13\\
\midrule
\multirow{2}{*}{\rotatebox[origin=c]{90}{\parbox[c]{0.6cm}{\centering tiles}}}
&$x$ & x-dimension           & 32 pixel\\
&$y$ & y-dimension           & 32 pixel\\
\midrule
\multirow{3}{*}{\rotatebox[origin=c]{90}{\parbox[c]{1.8cm}{\centering windows}}}
&$\rho$   & stride                                                  & 1 (\# obs.)\\
&$\delta$ & step $\Big(\frac{\cdot}{\textnormal{observation}}\Big)$ & 2 days\\
&$\Delta$ & window period                                           & 6 months\\
&$\omega$ & min. window size                                        & 35 (\# obs.)\\
&$\Omega$ & max. window size                                        & 92 (\# obs.)\\
\bottomrule
\end{tabular}}
\end{center}
\caption{Important remote sensing data, tile, and window parameters. These were unchanged from the pre-trained network stage and used identically in our current work.}
\label{parameters}
\end{table}

The multi-modal data was defined by the number of bands, which were the two polarizations Vertical-Vertical (VV) and Vertical-Horizontal (VH) for SAR, and spectral channels for optical observations (13 bands in the range of ca. 440-2200 nm). For SAR, we considered ascending and descending orbit directions as individual observation modes. These have different observation directions and hence cannot be directly compared where larger elevation differences are present. Each observation mode was temporally stacked to only update pixels in the observations that were not masked due to clouds or out-of-swath. If masked, the value of the pixel in the previous observation was carried forward.

Due to memory constraints, the entire AoI was not considered at once. Instead, it was tiled into non-overlapping $32 \times 32$ pixels patches. This was a configuration used for the pre-training, and so was used in this work as well\footnote{Since ERCNN-DRS is fully convolutional, changes of tile sizes are possible for transfer or inference.}.

\begin{figure}[t]
\begin{center}
\includegraphics[width=0.45\textwidth]{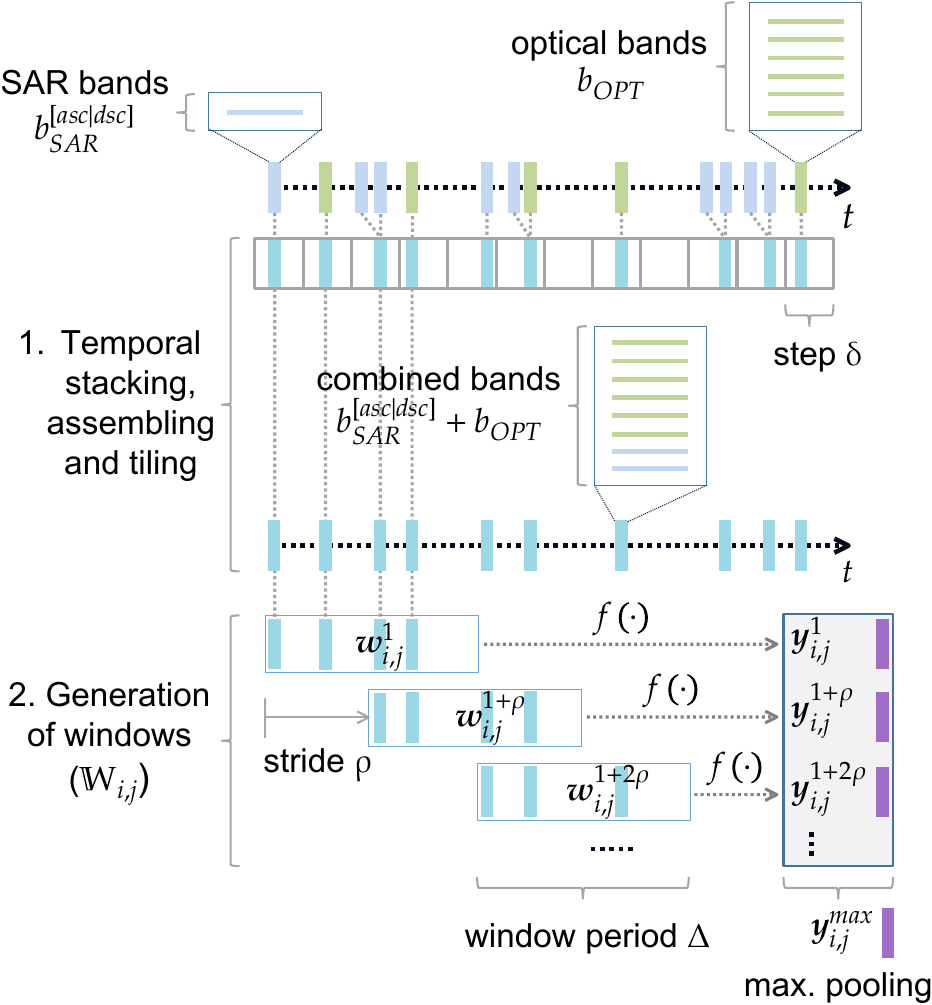}
\end{center}
\caption{The two steps of generating the set of windows $\mathbb{W}_{i, j}$ for each tile with coordinates $i$ and $j$. The window predictions were used in combination with \textit{Maximum Pooling over Time} to retrieve a combined prediction $\bm{y}^{max}_{i,j}$ during the transfer phase.}
\label{window_maxpool}
\end{figure}

As shown in Figure~\ref{window_maxpool}, windows were constructed from these multi-mode observations by assembling them into observations with a sampling step of $\delta$. Using two days showed a good compromise of avoiding redundant observations (e.g., due to swath overlaps) and retaining high temporal resolution for the purpose of urban change monitoring. In the following, we use $\delta = 2 \textrm{ days}$ unless otherwise noted. Windows were of a fixed period $\Delta$ of half a year and windows with fewer than $\omega = 35$ observations were discarded to ensure enough data points were available. Due to the window period and step size, there is a natural upper bound of observations per window $\Omega = 92 \approx \frac{\Delta}{\delta}$. A unit-stride ($\rho = 1$) was used for the sliding windows, which derived windows starting at every next sampled observation. Hence every $\delta$-sampled observation defined the start of a new window, i.e., every two days or longer, depending on the amount of observations over time. As a result, a set of windows $\bm{w}^{t}_{i,j} \in \mathbb{W}_{i, j}$ with the same parameters was retrieved that can be used for transfer learning or inference.

In the previous work \cite{doi:10.1080/01431161.2023.2243021}, the transfer and optimization of the pre-trained ERCNN-DRS model was already demonstrated. It used a different AoI (Li\`ege in Belgium) compared to the pre-training AoIs for the time frame 2017-2020. The transfer was realized with a small amount of manually labeled tiles. It was demonstrated that, even with an already low effort ground truth guided by Google Earth historic imagery, the transfer showed an improved performance. Instead of labeling each window, a set of windows spanning a larger time frame of four years, was considered for labeling. To be able to train with a set of windows at each step, a maximum pooling method was applied. The maximum pooling followed the principle of \textit{Maximum Pooling over Time}~\cite{10.5555/1953048.2078186}. Figure~\ref{window_maxpool} shows the maximum pooling of the individual window predictions $\bm{y}^{t}_{i,j}$ to retrieve $\bm{y}^{max}_{i,j}$.

To leverage the limited dataset size, bagging (bootstrap aggregating)~\cite{breiman96} was utilized, using a non-exhaustive cross-validation. Three model variants were trained with this cross-validation approach that used disjunct validation data from the overall dataset for every transfer variant. The bagging resulted in an ensemble of weak learners from the variants which provided a better performance and precision/recall balance than each individual variant. However, this was only executed and validated for the same time frame as the pre-training.

\subsection{Extension and Modification of Methods}
In this work, we applied transfer learning to the pre-trained ERCNN-DRS and fine-tuned it for the AoI of Mariupol. ERCNN-DRS is transferred four times, receiving the trained model variants V1-4. Similar as to the previous work~\cite{doi:10.1080/01431161.2023.2243021}, the transfer is done with a non-exhaustive cross-validation approach where disjunct validation data are used. All four model variants were used for monitoring the urban changes in 2022/23 for the AoI of Mariupol. In this work, the performance of each variant, as well as the bagging ensemble of all four variants, is analyzed.

Since no public VHR data were available during the time frame 2022/23\footnote{Google Earth historic imagery ended early 2021 at the time of writing.}, we applied the transfer learning to the period of the beginning of 2017 until the end of 2020. The practical feasibility due to easy labeling with Google Earth historic VHR imagery has already been demonstrated in the aforementioned work. However, in this work, we analyzed the quality of predictions when the transferred model was applied to years outside the transfer period, in order to be able to monitor the urban changes during the recent years when public data was not yet available.

For the verification of changes in that time frame, we used commercial Airbus Pl\'eiades VHR observations from the beginning of 2022 to early 2023. Opposed to Maxar WorldView, which was under embargo for Ukraine at the time of writing, Airbus Pl\'eiades data were still commercially available. However, this restricted the best resolution for verification to 0.5 m/pixel (panchromatic). These observations were used identically to the earlier ground truth generation ($\tilde{\bm{y}}_{i,j}$) for the transfer but with labeling the changes in the monitoring time frame ($\check{\bm{y}}_{i,j}$). A final verification step compared the model predictions against the manually identified changes. This comparison is covered in section~\ref{results}, which gives a quantitative and qualitative analysis.

The monitoring past the transfer time frame imposed additional challenges due to changing observation patterns over time. This was especially impacted by the failure of Sentinel 1B on December 23, 2021\footnote{\url{https://www.esa.int/Applications/Observing_the_Earth/Copernicus/Sentinel-1/Mission_ends_for_Copernicus_Sentinel-1B_satellite}; accessed 16-10-2023\nopagebreak}. The unavailability of Sentinel 1B led to a significant reduction of SAR observations with only Sentinel 1A left in operation. Both Sentinel 1 satellites were placed on the same orbit plane with a difference of $180^{\circ}$ in orbit phase. As a result, the cycle time for Europe increased from six days to twelve days after the malfunction. Since the transferred models were trained with both Sentinel 1A and 1B satellite observations, this change could have had a significant impact on the operation of the transferred models. As we will show, our method did not break down by the change of observation patterns due to the reduction of SAR observations as a result of the loss of Sentinel 1B. We studied the resiliency and scalability of changes in observation frequency for each mode and the combination of modes with a simulated SAR and optical data loss in an ablation study (see section~\ref{results}).

Overall, three different observation data sources were used: i) \textit{Sentinel Hub} for Sentinel 1 \& 2 data, ii) Google Earth historic VHR imagery, and iii) \textit{Sentinel Hub} for Airbus Pl\'eiades VHR observations. The Sentinel 1 \& 2 data were the primary data used for the transfer of the four model variants and their inference (monitoring). Their processing is described in detail in section~\ref{primary_data}. The other two data sources were used for generating ground truth maps for transfer and verification, and are covered in sections~\ref{ground_truth_transfer} and \ref{ground_truth_verification}, respectively.

\section{Study Area and Remote Sensing Data}\label{study_area}
We applied our methods to the area of Mariupol (Ukraine) to monitor urban-related changes and activities with the Russian invasion that began on February 24, 2022. During the first months of the siege, approximately up to 95\% of the city and infrastructure were damaged or destroyed\footnote{\url{https://web.archive.org/web/20221031202808/https://www.abc.net.au/news/2022-05-26/damage-data-reveals-extent-of-vicious-russian-tactics/101070918}; accessed 16-10-2023\nopagebreak}. Up to the writing of this work, Mariupol was still under Russian occupation, and heavy reconstruction of buildings and infrastructure was observed. We monitored not only the city of Mariupol but also the surrounding area with over 536 $\textnormal{km}^2$, covering suburbs, rural areas, the sea, mines, and farming regions. The area is also subject to frequent overcast due to its location by the Black Sea and winters with snow and ice.

\begin{figure}[t]
\begin{center}
\includegraphics[width=1.0\columnwidth]{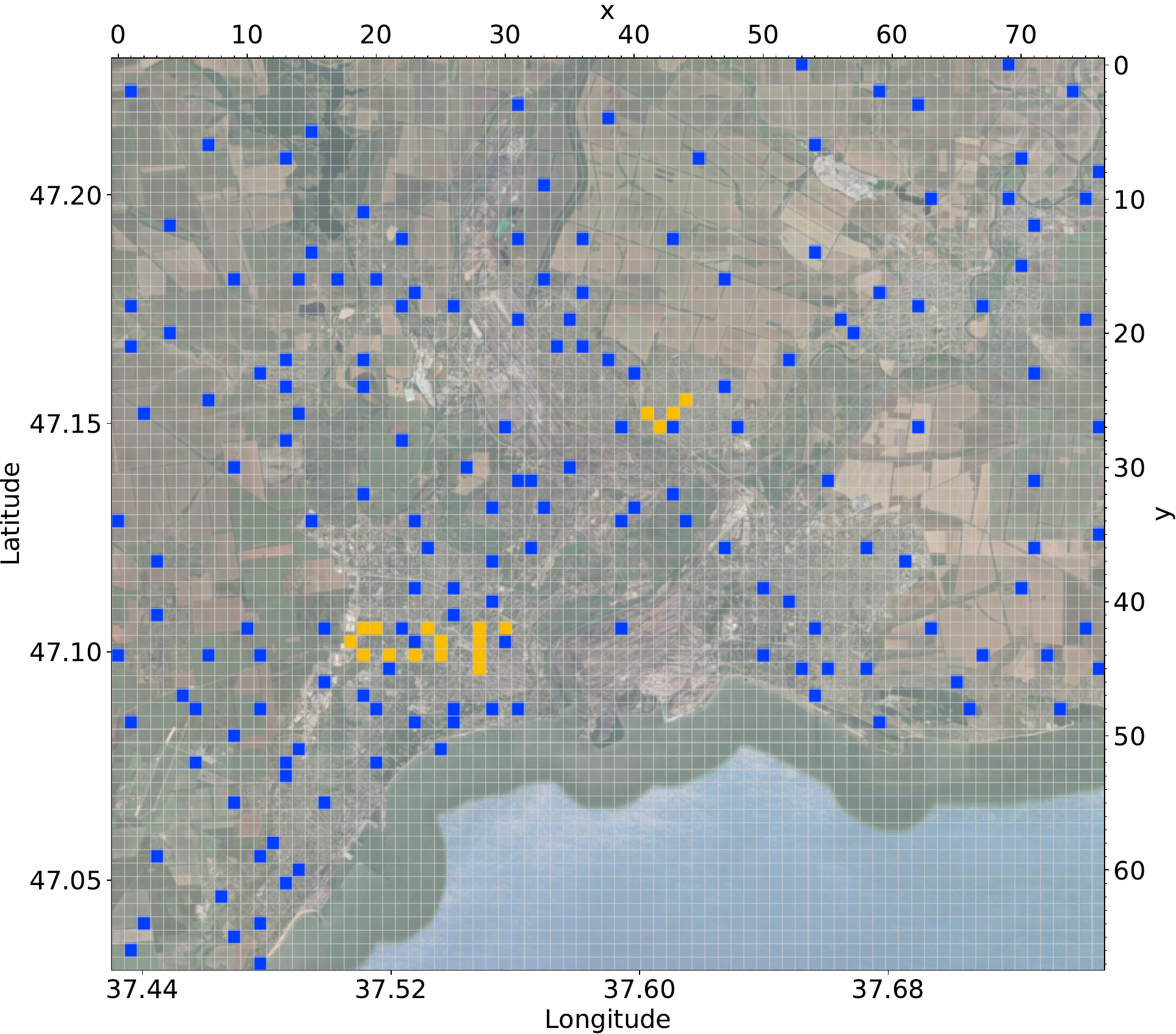}
\end{center}
\caption{Tiles for the AoI of Mariupol. The blue tiles covering 2017-2020 were used for training and validation, referred to as \textit{trainval} dataset (164 in total). The 18 tiles in orange for 2022/23 were used for verification purposes, referred to as \textit{testing} dataset. Geographic coordinates are in EPSG:4326 and tile coordinates are in $(y, x)$ dimensions. Background image \copyright 2019/20 Google Earth, for reference only.}
\label{grid_mariupol}
\end{figure}

Figure~\ref{grid_mariupol} shows the tiled AoI, with training tiles for the transfer and verification tiles for the monitoring phase. To avoid spatial bias we used only disjunct sets of transfer and verification tiles, even though both covered different time frames. The processing of the three different data sources and types used in this work is described in the following.

\subsection{Primary Data}\label{primary_data}
\begin{table*}[htb]
\begin{center}
\begin{tabular}{lllllllll}
\toprule
&& \textbf{Site}     & \multicolumn{2}{l}{\textbf{SAR observations}} & \multicolumn{2}{l}{\textbf{Optical multispectral}} & \textbf{Area} & \textbf{Stage}\\
&&                   & \multicolumn{2}{l}{\textbf{(ascending \& descending)}} & \multicolumn{2}{l}{\textbf{observations}} & \textbf{($\textnormal{km}^2$)}\\
\midrule
\multirow{4}{*}{\rotatebox[origin=c]{90}{\parbox[c]{3.0cm}{\centering Sentinel 1 \& 2}}} & \multirow{3}{*}{\rotatebox[origin=c]{90}{\parbox[c]{2cm}{\centering 2017-2020}}}
& \cellcolor{gray!20}Rotterdam       & \multicolumn{2}{l}{\cellcolor{gray!20}1,603 (-4)} & \multicolumn{2}{l}{\cellcolor{gray!20}278 (-10)}                    & \cellcolor{gray!20}523.6 & \multirow{2}{*}{${\setlength\arraycolsep{0pt}\left.\begin{array}{l}\rule{0ex}{0ex}\\\rule{0ex}{0ex}\end{array}\right\rbrace\textrm{pre-training}}$}\\
&& \cellcolor{gray!20}Limassol        & \multicolumn{2}{l}{\cellcolor{gray!20}468 (-0)}   & \multicolumn{2}{l}{\cellcolor{gray!20}407 (-35)}                    & \cellcolor{gray!20}576.2\vspace{4mm}\\
&& Mariupol                           & \multicolumn{2}{l}{648 (-0)} & \multicolumn{2}{l}{431 (-131)}                    & 536.2 & \multirow{1}{*}{${\setlength\arraycolsep{0pt}\left.\begin{array}{l}\rule{0ex}{0ex}\end{array}\right\rbrace\textrm{transfer}}$}\vspace{4mm}\\
\cmidrule{2-8}\rule{0pt}{6mm}
&\multirow{1}{*}{\rotatebox[origin=c]{90}{\parbox[c]{.6cm}{\centering 2022/23}}} & Mariupol                           & \multicolumn{2}{l}{155 (-0)} & \multicolumn{2}{l}{232 (-84)} & 536.2 & \multirow{1}{*}{${\setlength\arraycolsep{0pt}\left.\begin{array}{l}\rule{0ex}{0ex}\end{array}\right\rbrace\textrm{monitoring}}$}\vspace{4mm}\\
\midrule
\multicolumn{2}{l}{Sources}         &  SAR:     & \multicolumn{4}{l}{10 m/pixel, Sentinel 1, \textit{SENTINEL1\_IW\_[ASC|DSC]}} & \multicolumn{2}{r}{\multirow{2}{*}{$\vcenter{\hbox{\includegraphics[height=0.6cm]{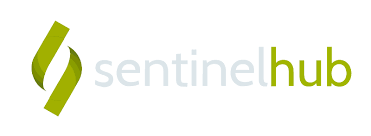}}}$}}\\
&                &  Optical: & \multicolumn{4}{l}{10 m/pixel, Sentinel 2, \textit{L1C}}                      &\\
\bottomrule
\end{tabular}
\end{center}
\caption{Used AoIs, the covered areas, and the number of their available observations, with removed ones in parentheses. Only level 1 products were used. AoIs in grey were used for pre-training only, which is outside the scope of this work.}
\label{sources}
\end{table*}

The core data of our method comprised multi-modal SAR and optical multispectral observations from Sentinel 1 and Sentinel 2, respectively. They were used as input to the selected DNN and span the time frames of 2017-2020 for transfer, as well as the time frame November 24 2021, up to mid 2023 for the actual monitoring phase. All data were retrieved from \textit{Sentinel Hub} as \textit{level 1} products. Table~\ref{sources} summarizes the primary data products used.

The \textit{level 1} products provided by \textit{Sentinel Hub} were already orthorectified and co-registered. In our method, all available bands and polarizations were used, i.e., no dimensionality reduction was applied; that is both polarizations (VV+VH) for Sentinel 1 observations and 13 spectral bands for Sentinel 2 observations were available. The pre-processing followed the same way as used for the pre-training of ERCNN-DRS and is executed with the \textit{rsdtlib} library as described earlier.

\begin{figure*}[htb]
\begin{center}
\includegraphics[width=0.9\textwidth]{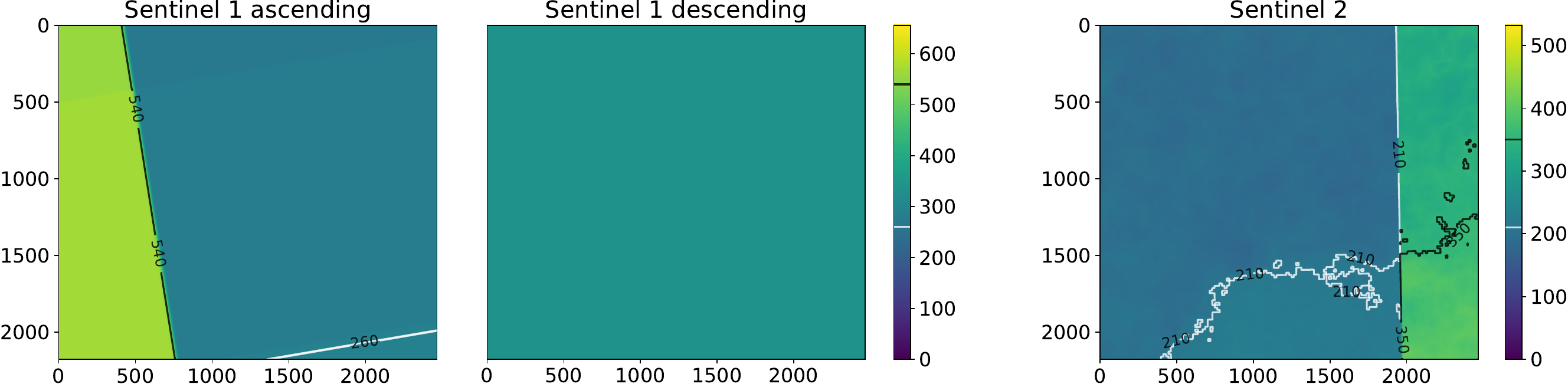}
\end{center}
\caption{Number of observations for each pixel within the AoI of Mariupol, separate for Sentinel 1 in ascending and descending orbit direction (left), and Sentinel 2 (right). Sentinel 1 has a range of [220, 546] and [324, 324] (no variation) for ascending and descending orbit directions, respectively. Sentinel 2 observations are within [166, 383]. Contours are shown for selected observation numbers.}
\label{num_obs_loc}
\end{figure*}

The available Sentinel observations for the period from 2017 until mid-2023 vary over time and by location within the selected AoI. Figure~\ref{num_obs_loc} shows the available observations for each pixel within the AoI for $\delta = 1 \textrm{ second}$. For Sentinel 1 in ascending orbit direction, different and overlapping swaths were visible. These result in more observations where swaths overlapped and less where the surface was scanned less frequently or irregularly. Sentinel 1 in descending orbit direction did not show any variance due to full coverage of the AoI by the swaths. For Sentinel 2, in addition to swath patterns, cloud masking (as provided by \textit{Sentinel Hub}) added to the irregularity of the observations. Also the coast of the Black Sea became visible, which was a result of the applied cloud detection method to overestimate clouds over land surfaces.

\begin{figure*}[htb]
\begin{center}
\includegraphics[width=0.8\textwidth]{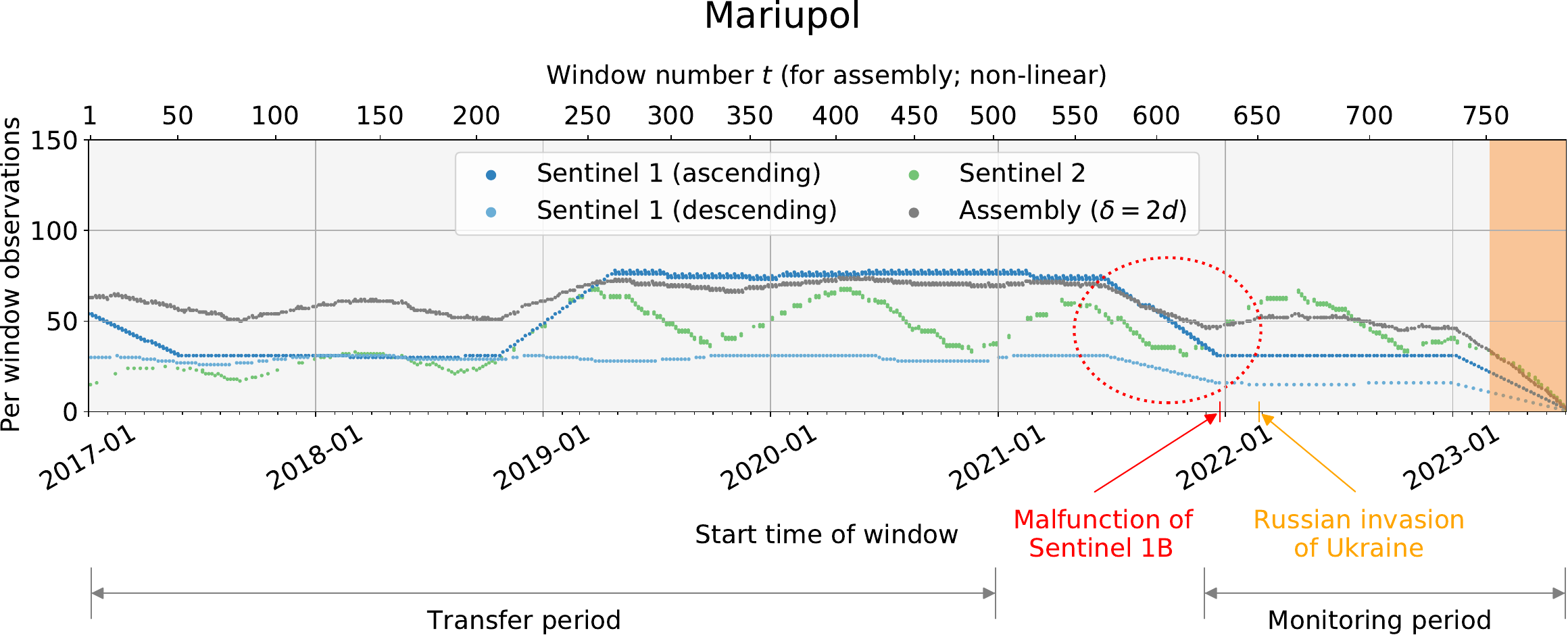}
\end{center}
\caption{Windows and their observations. In our work, we combined Sentinel 1 (blue) and 2 (green) observations in a two-day ($\delta = 2 \textrm{ days}$) interval (grey). Highlighted in orange and discarded were windows with less than 35 ($\omega$) observations. A malfunction of Sentinel 1B occured on 23-12-2021 and the Russian invasion on 24-02-2022, as indicated on the timeline.}
\label{num_obs}
\end{figure*}

In Figure~\ref{num_obs}, the available observations per each six-month window ($\Delta$) are plotted, using $\delta = 2 \textrm{ days}$. The loss of Sentinel 1B and the decrease of available SAR observations is clearly visible. It should be noted that due to the windowed time series used, the sudden drop of Sentinel 1B observations led to a gradual reduction of the observations over a half-year period prior to the day of malfunction.

\begin{figure}[h]
\begin{center}
\begin{adjustbox}{width=0.5\textwidth,center}
\setlength{\fboxsep}{1ex}
\newcommand{\STAB}[1]{\begin{tabular}{@{}c@{}}#1\end{tabular}}
\begin{tabular}{cccc}
& Visual A & Visual B & Approximated\\
&          &          & urban changes\\
\midrule
\multirow{1}{*}[6.25ex]{\rotatebox[origin=c]{90}{\fbox{\parbox[c]{0.232\columnwidth}{\centering Tile: 37:3}}}}
& \makebox[0.255\columnwidth]{$\vcenter{\hbox{\frame{\includegraphics[width=0.255\columnwidth]{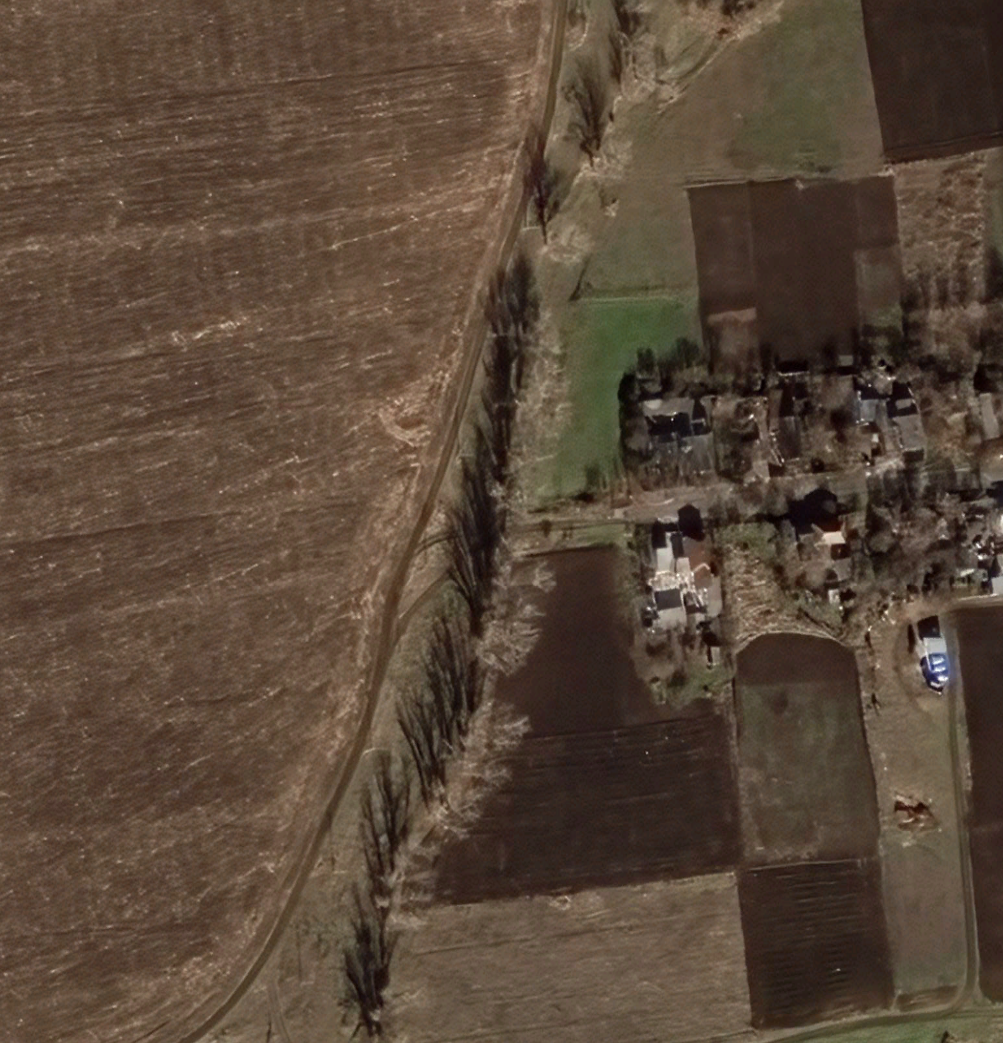}}}}$\hfill} & \makebox[0.255\columnwidth]{$\vcenter{\hbox{\frame{\includegraphics[width=0.255\columnwidth]{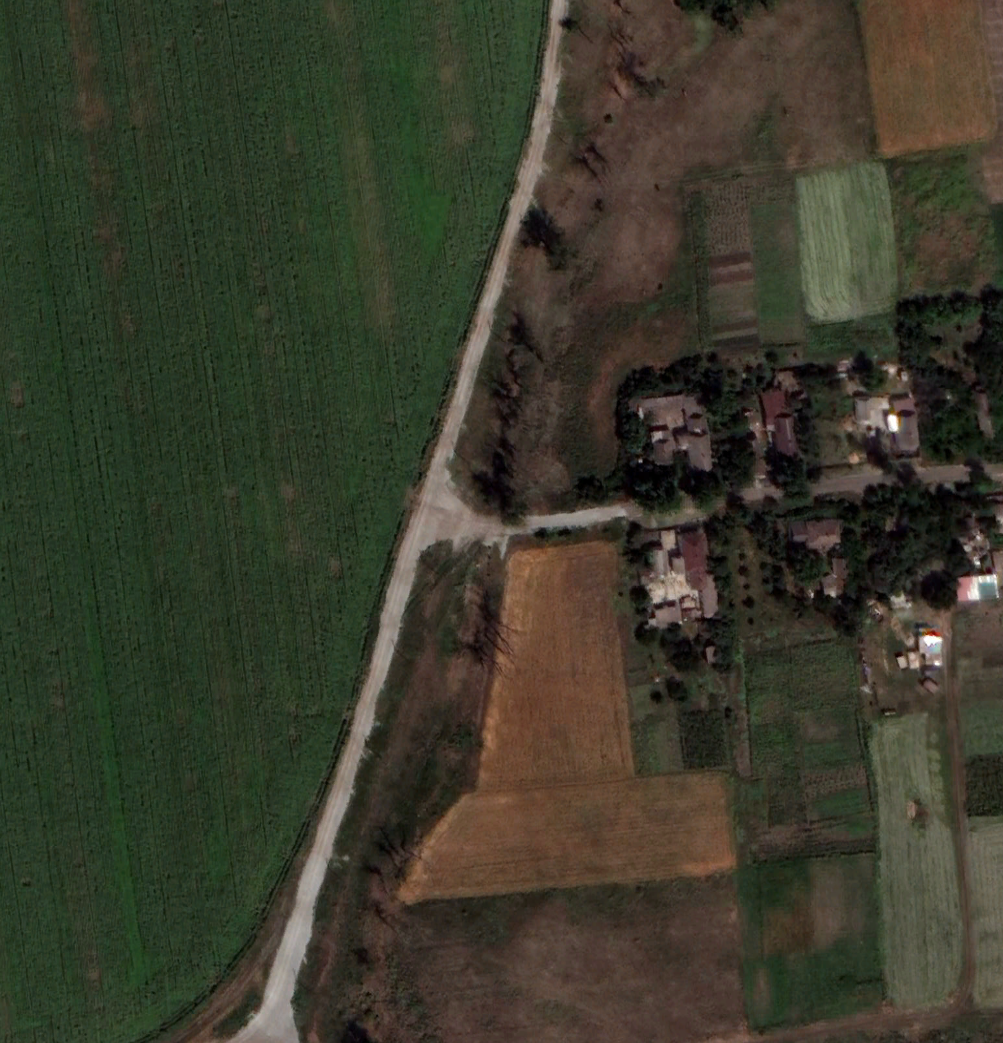}}}}$\hfill} & \makebox[0.255\columnwidth]{$\vcenter{\hbox{\frame{\includegraphics[width=0.255\columnwidth]{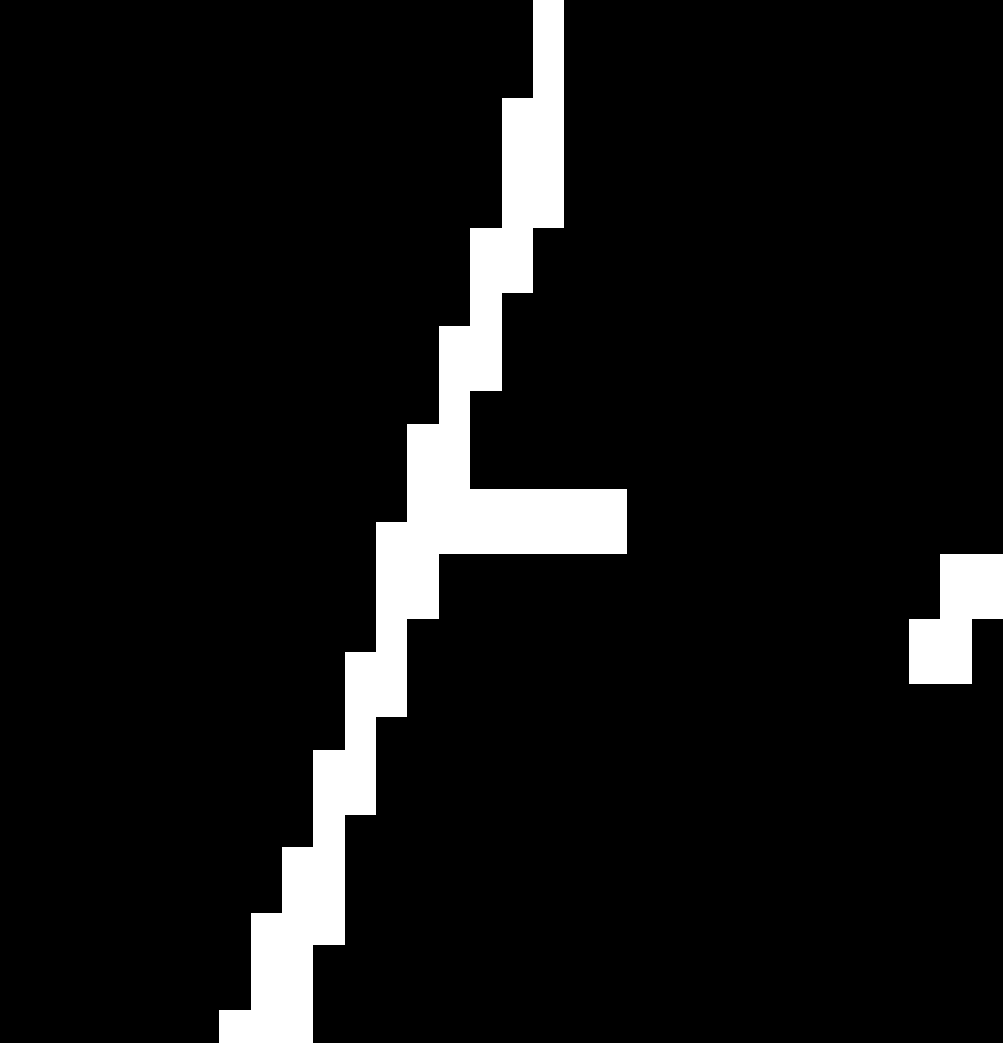}}}}$\hfill}\vspace{1mm}\\
& 2016-11 & 2020-07 & \\
\multirow{1}{*}[6.25ex]{\rotatebox[origin=c]{90}{\fbox{\parbox[c]{0.232\columnwidth}{\centering Tile: 3:31}}}}
& \makebox[0.255\columnwidth]{$\vcenter{\hbox{\frame{\includegraphics[width=0.255\columnwidth]{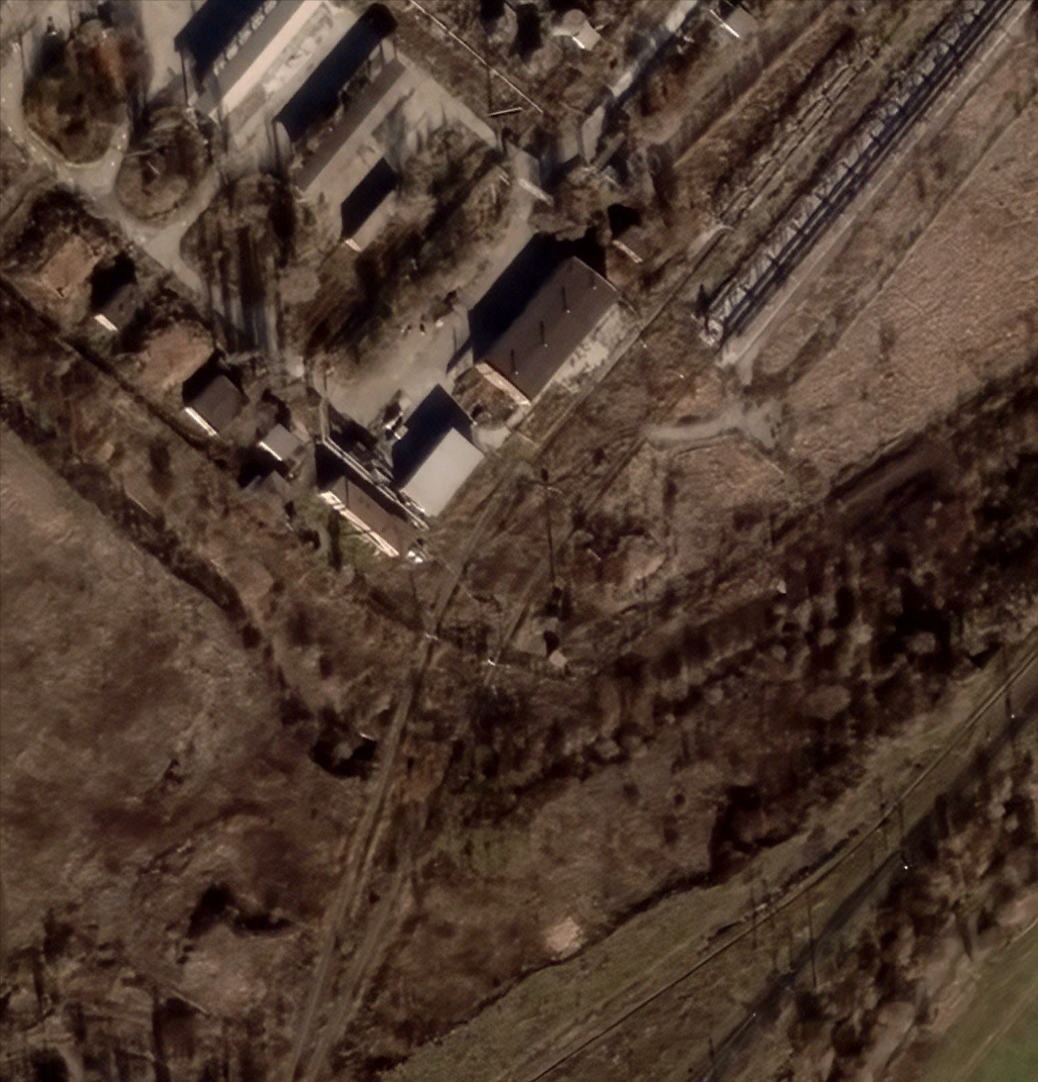}}}}$\hfill} & \makebox[0.255\columnwidth]{$\vcenter{\hbox{\frame{\includegraphics[width=0.255\columnwidth]{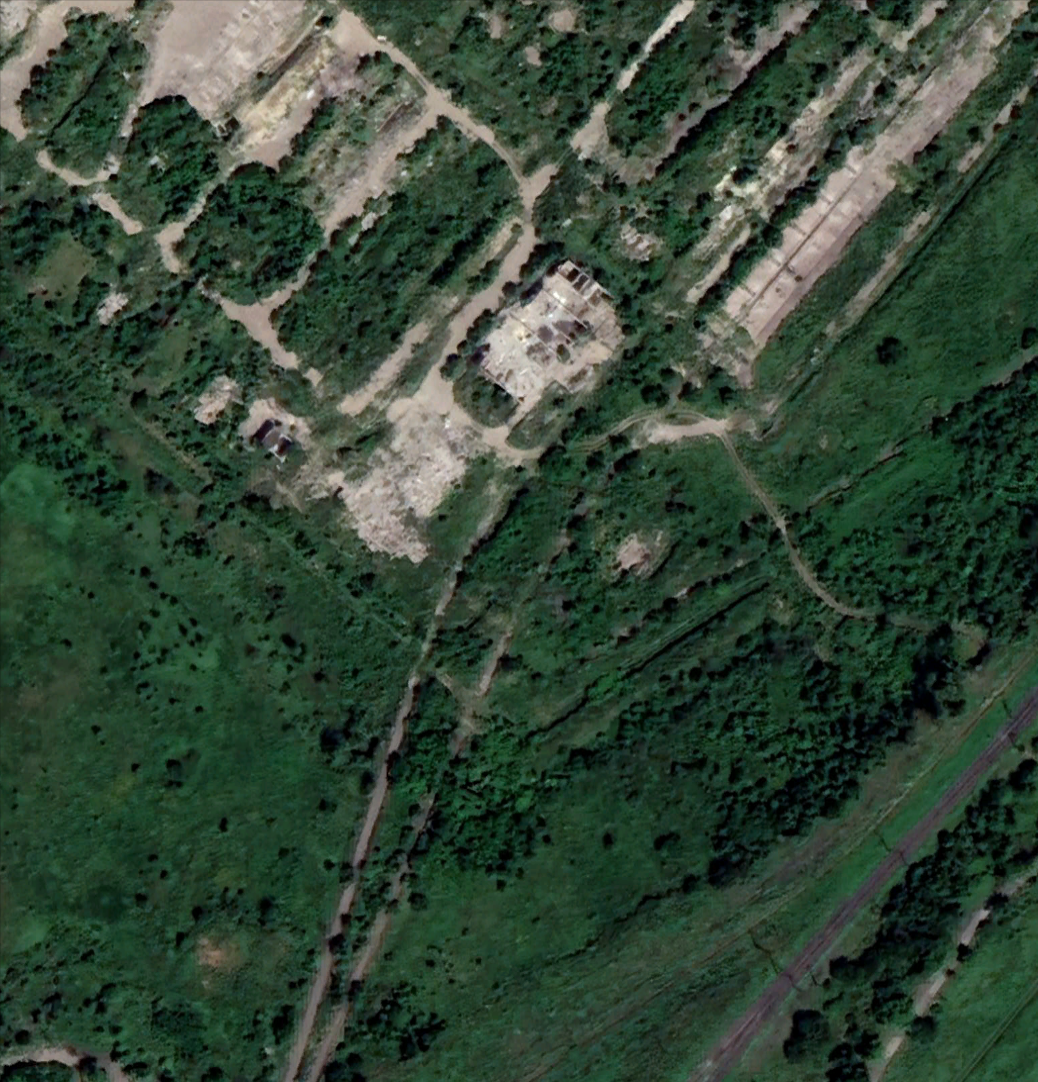}}}}$\hfill} & \makebox[0.255\columnwidth]{$\vcenter{\hbox{\frame{\includegraphics[width=0.255\columnwidth]{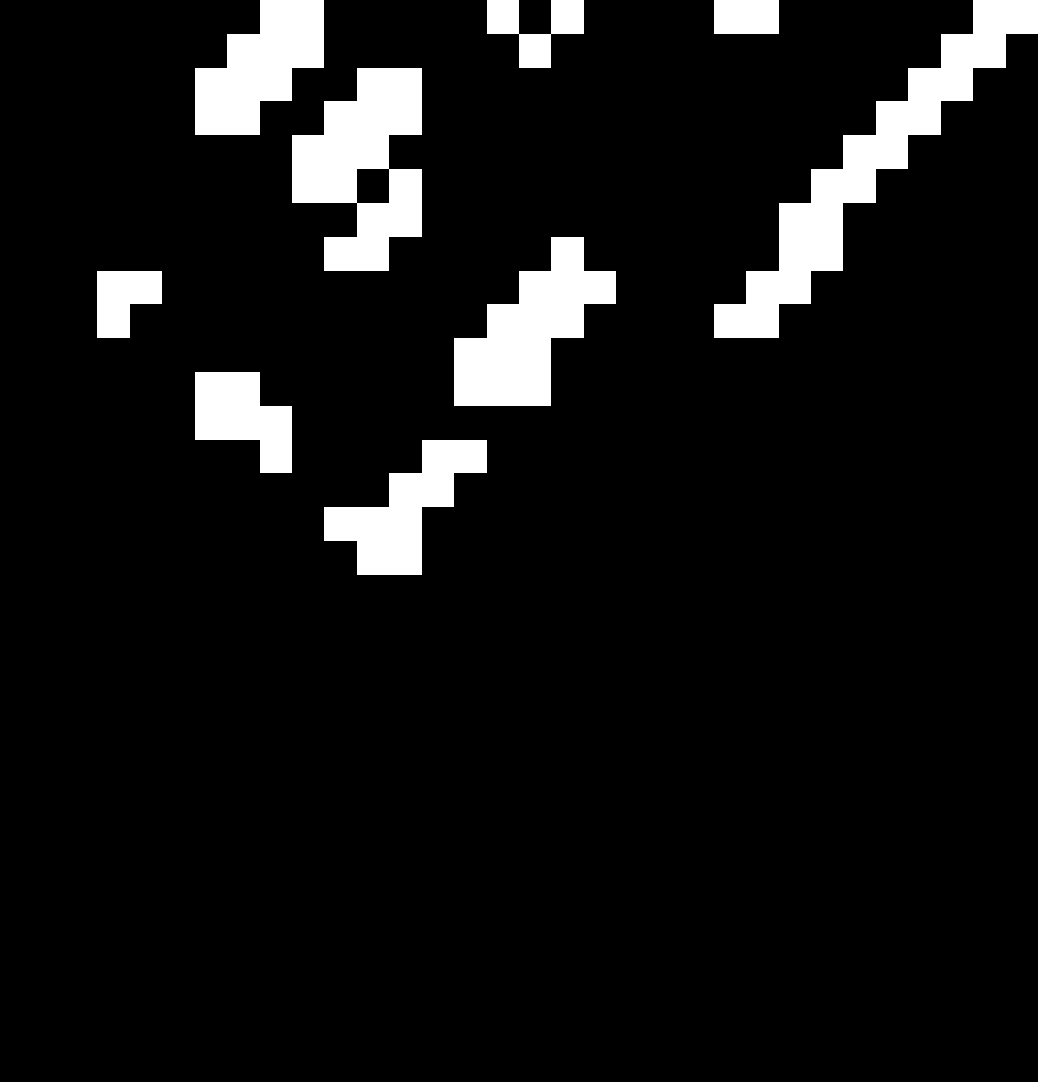}}}}$\hfill}\vspace{1mm}\\
& 2016-11 & 2021-06 & \\
\multirow{1}{*}[6.25ex]{\rotatebox[origin=c]{90}{\fbox{\parbox[c]{0.232\columnwidth}{\centering Tile: 31:31}}}}
& \makebox[0.255\columnwidth]{$\vcenter{\hbox{\frame{\includegraphics[width=0.255\columnwidth]{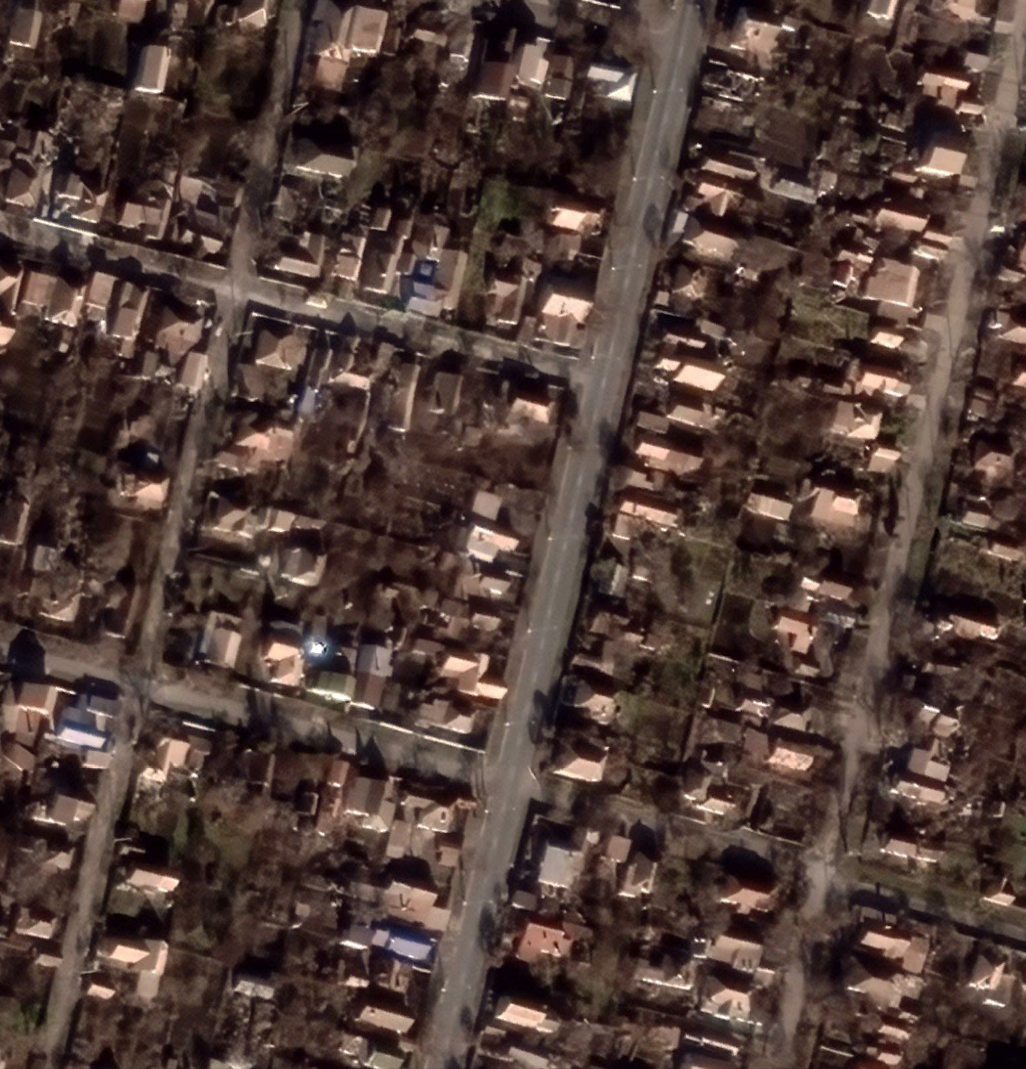}}}}$\hfill} & \makebox[0.255\columnwidth]{$\vcenter{\hbox{\frame{\includegraphics[width=0.255\columnwidth]{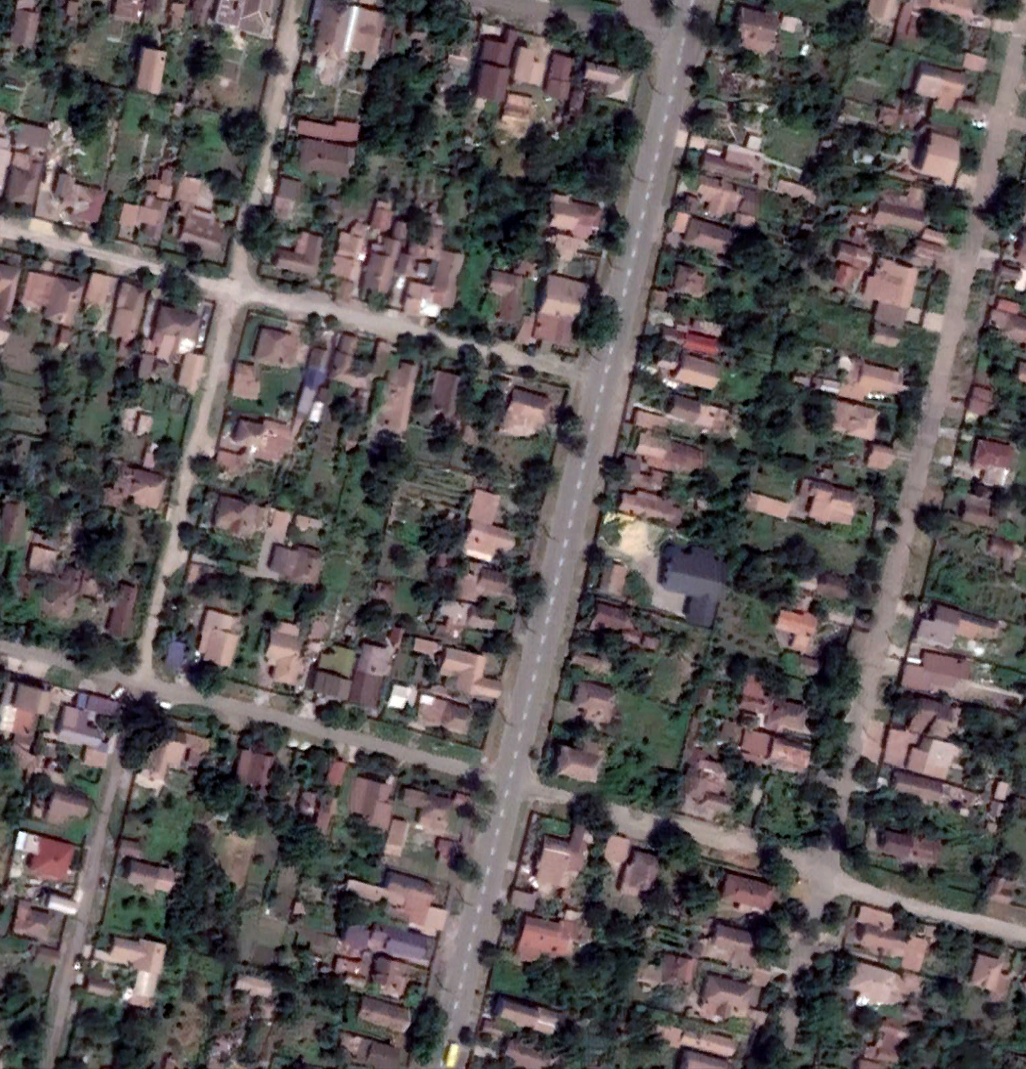}}}}$\hfill} & \makebox[0.255\columnwidth]{$\vcenter{\hbox{\frame{\includegraphics[width=0.255\columnwidth]{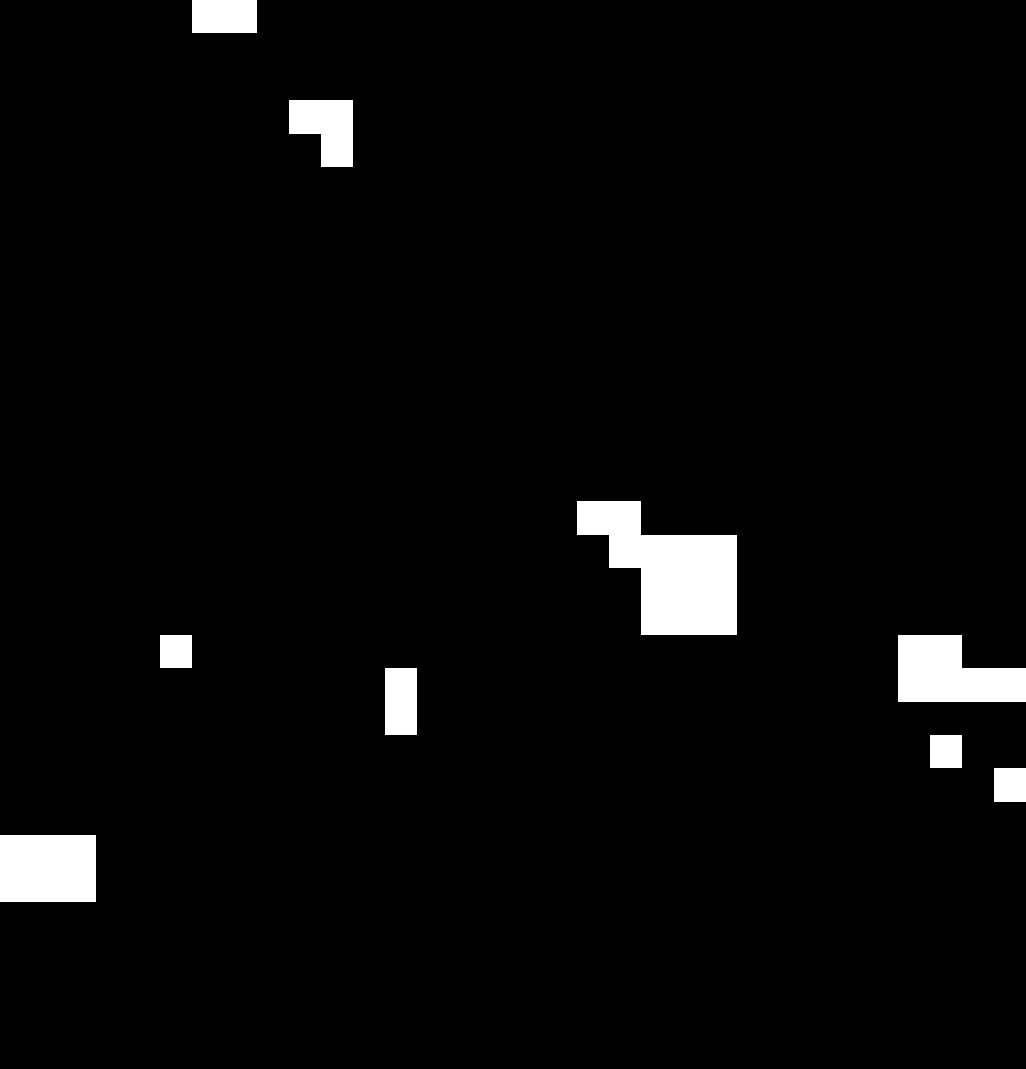}}}}$\hfill}\vspace{1mm}\\
& 2016-11 & 2021-06 &
\end{tabular}
\end{adjustbox}
\end{center}
\caption{At least two historic Google Earth very-high resolution images near the beginning of 2017 (left) and the end of 2020 (middle) were used to approximate the ground truth ($32 \times 32$ pixel) for urban changes $\tilde{\bm{y}}_{i,j}$ (right).}
\label{labeling_samples}
\end{figure}

\subsection{Ground Truth for Transfer Period}\label{ground_truth_transfer}
To generate the labels $\tilde{\bm{y}}_{i,j}$, as used for the transfer learning process, we utilized VHR historic satellite and aerial imagery from Google Earth. These were used for a selection of 164 random tiles (referred to as \textit{trainval} dataset) in the AoI of Mariupol to approximately identify past urban changes in the time frame 2017-2020. Using a longer time frame, like four years, simplifies the labeling process due to increased chances of more historic VHR observations being available. A higher number aids in identifying and localizing changes over time. Depending on the exact location within the selected AoI, between 10-20 historic observations were available.

Tiles that contained changes that were too large and homogeneous (e.g., open pit mines, destruction of large storage buildings, and steel factories), were removed during the random selection process. This avoided an undesired bias towards specific change patterns and instead balanced the transferred network towards a more diverse set of changes. In addition, this simplified the labeling process since areas that have constantly been a subject of change were hard to label. Rapid and spatially constraint changes were easier to notice and mark.

The created ground truth $\tilde{\bm{y}}_{i,j}$ is binary with an assigned value of \texttt{1.0} if a change was related to man-made UST objects, indicated by visual inspection of at least two VHR images. The value of \texttt{0.0} was assigned otherwise. Also, side effects of constructions were treated as changes, like modified soil around construction sites or paved roads. Any change below the sensor resolution ($< 10 \textrm{ m/pixel}$) was ignored. Figure~\ref{labeling_samples} shows three tiles as examples with the closest observations at the beginning of 2017 and end of 2020. The manually created binary ground truth is shown next to the visual samples. Due to the low amount of available VHR observations, it is not possible to label the full extent of the changes. Hence, we refer to the labels as approximate. The three examples show the construction of a road (tile 37:3), destruction of factory buildings (tile 3:31), and (re-)construction of buildings in a suburban area (tile 31:31).

\subsection{Ground Truth for Monitoring Period}\label{ground_truth_verification}
To verify the urban changes during the years 2022/23, recent Airbus Pl\'eiades data were leveraged. Since these data were involved with significant costs, we selected 18 tiles (referred to as \textit{testing} dataset) in two locations: i) the city center (14 tiles) and ii) a suburb to the north (4 tiles). Pl\'eiades data were used in panchromatic mode, resulting in 0.5m/pixel resolution. This was sufficient to identify changes and evaluate the predictions. Depending on the location, six to seven observations at different times from March 2022 until January 2023 were screened.

The generation of the verification reference for the monitoring period $\check{\bm{y}}_{i,j}$ followed the same rule as with the ground truth $\tilde{\bm{y}}_{i,j}$. Both were binary, with values decided upon visual inspection of the available observations. However, $\check{\bm{y}}_{i,j}$ only spans the monitoring period of 2022/23.

\section{Training}\label{training}
The transfer phase was executed on the Karolina GPU cluster\footnote{\url{https://docs.it4i.cz/karolina/hardware-overview/}; accessed 16-10-2023} at IT4Innovations. One compute node with eight NVIDIA A100 GPUs, each with 40 GB of memory, was used for training. The training environment comprised Tensorflow 2.7, including Keras, and Horovod~\cite{sergeev2018horovod} 0.23.0 for leveraging multiple GPUs. We used synchronous SGD~\cite{chen:2016:iclr, NIPS2010_abea47ba} with a momentum of 0.8 and set the learning rate $\alpha$ to 0.008. The loss function $\mathcal{L}$ was identical to the pre-training, that is the \textit{Tanimoto loss with complement}~\cite{tanimoto-compl}. It compared the maximum pooled prediction $\bm{y}^{max}_{i,j}$ against the ground truth $\tilde{\bm{y}}_{i,j}$, expressed as $\mathcal{L}(\tilde{\bm{y}}_{i,j}, \bm{y}^{max}_{i,j})$. For every tile, pixels at the border (dead area) were ignored, and only the center $30 \times 30$ pixels of each tile were considered. This is due to the general problem of tiled data and convolutional networks which increase errors towards the borders~\cite{10.3389/fnins.2020.00065, DBLP:journals/corr/abs-1805-12219, huang2019tiling, Isensee2021}. We did not mitigate this in our current work, but an overlapping of tiles can easily be applied to remove these errors. Empirically, we found that removing only the direct border pixels is sufficient. Removing a larger border would result in reduction of label data. Hence, the use of only the center $30 \times 30$ pixels was a compromise between the avoidance of a higher loss of label data and increased computational and storage needs with the use of overlapping tiles. We, however, applied inference with 8 pixel overlapping tiles of size $93 \times 93$ for the final monitoring to ensure a complete coverage of the entire AoI and more efficient inference due to larger tiles (the result is shown in Figure~\mbox{\ref{mariupol_urban_changes}} in the Appendix).

For the transfer phase, the pre-trained model was used, with no applied layer freezing. The shallow structure of the pre-trained model architecture did not develop patterns found in deeper networks. In deeper networks, more general features are extracted in layers closer to the input and more specialized features towards the output~\cite{10.5555/2969033.2969197}. Freezing layers closer to the input is a common practice for transfer learning so that only more specialized layers are transferred and the general ones are only reused. This results in lower resource needs and faster training. However, for our shallow network architecture such generalization and specialization patterns are less likely to develop. Instead, in our case we benefited from transfer learning by using a more specific definition of urban changes with coarse temporal information. Since the pre-trained model was trained with more temporal information, i.e., one label per window, transfer learning can build on top of this.

Furthermore, a batch size of eight for each GPU was used, totalling an effective batch size of 64 (8 * 8 GPUs). Due to limited memory, for every tile (training sample) only ten partially overlapping windows were randomly selected to span the (almost) entire four years of 2017-2020. This follows the previously proposed approach to avoid bias of the network under training to specific windows and their observation patterns. In this work, the first window started at $t = 21$ and the following nine windows were selected based on uniform random relative offsets within the range of $[40, 49]$. The initial offset was needed; as with the temporal stacking not every pixel had a value at $t \in [1, 20]$. This originated from the cloud masks and out-of-swath where no previous value was available that leaves a zeroed gap. Since we started the training time frame at the beginning of 2017, during winter, the window with $t = 21$ started in March 2017 (see Figure~\ref{num_obs}). This was acceptable as urban changes are less likely during winter. Conversely, the last window does not always end with 2020. With the selected range, the time frame covered reaches into the second half of 2020. Again, the likeliness of changes is lower towards the end of the year in autumn or winter, which we considered a compromise between providing a variability of windows and full coverage. Since the ground truth was approximate and we focused mostly on changes within the four years, rather than changes towards the beginning or end, this was an acceptable trade-off. The overlapping of windows ensured that all observations were visible by the network under training, except for observations at the beginning of 2017 and towards the end of 2020. Ultimately, the variance of selected windows adds augmentation to the training samples which reduces variance and improves generalization of the trained model.

Further augmentation was carried out on each sample with random horizontal flipping (factor two), rotation in 90 degree increments (factor four), and binary application of a \textit{temporal comb filter}~\cite{doi:10.1080/01431161.2023.2243021} (factor two). Altogether, these augmentations increased the dataset size \textit{trainval} by a factor of 16.

We would like to clarify that the chosen method for transferring was practically simple but had challenging memory requirements. The preparation of the ground truth was simplified by aggregating a set of windows and labeling changes visible over a longer time frame like multiple years. However, each prediction of a window needed to be maximum pooled for every training step. In turn, more windows over a longer transfer time frame needed to be concurrently trained with shared weights. In our case of using four years and ten randomly chosen windows, ca. 250 GB of memory was needed. We hence used multiple GPUs in a distributed data parallel training setting to increase the available memory.

\begin{figure*}
\begin{center}
\includegraphics[width=0.99\textwidth]{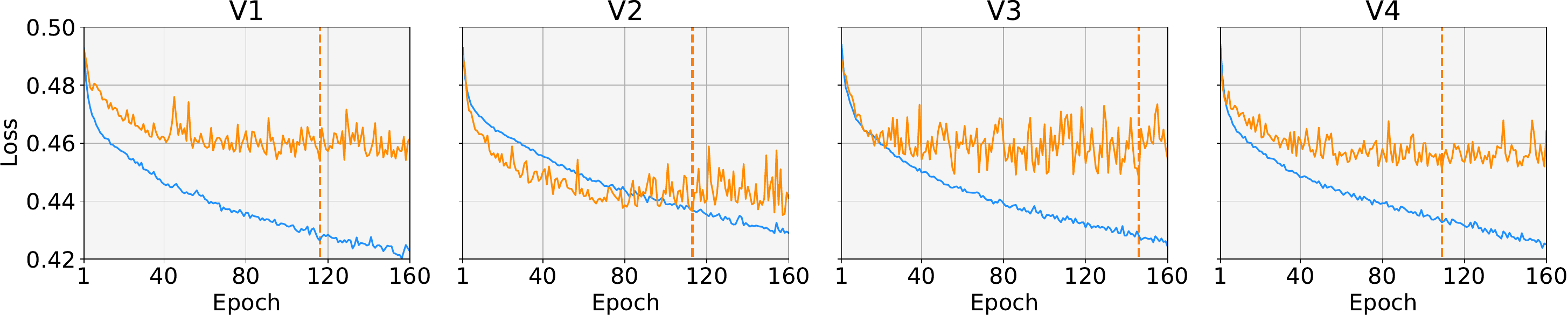}
\end{center}
\caption{Loss values over epochs for all four transferred variants. The transfer training losses are represented by the blue curve, and the validation losses by the orange curve. The orange dashed lines show the best epochs based on validation data.}
\label{losses}
\end{figure*}

Four different transfers were carried out on the same pre-trained model but with different training/validation splits. We used a non-exhaustive cross validation approach with all validation sets being disjunct. The loss curves of the transfers of the variants V1-4 are shown in Figure~\ref{losses}. The best validation loss is highlighted with an orange dashed vertical line. For V1, epoch 116 showed the best validation loss. For V2, V3, and V4, the best epochs were 113, 146, and 109, respectively. We hence define the per-tile predictions of the four individual transfer models used in the monitoring phase as

\begin{displaymath}
\bm{y}_{i,j}^{V1} \coloneqq \max\big\{f_{V1}(\bm{w}^{t}_{i,j}) : \bm{w}^{t}_{i,j} \in \mathbb{W}_{i, j}\big\}
\end{displaymath}

with the element-wise maximum operation over the two-dimensional predictions of each window. The trained parameters of V1 are used by the forward propagation $f_{V1}(\cdot)$. Similarly, the predictions $\bm{y}_{i,j}^{V2}$, $\bm{y}_{i,j}^{V3}$, and $\bm{y}_{i,j}^{V4}$ are defined for V2-4.

\begin{table*}[h]
\begin{center}
\begin{threeparttable}
{\small
\begin{tabular}{lllll}
\toprule
\textbf{Infrastructure}               & \textbf{Cluster}      & \textbf{Hardware}             & \textbf{Transfer (h)} & \textbf{Environment}\\
\midrule
IT4Innovations                        & 1 Karolina            & 2 AMD EPYC 7H12 (64 cores)    & 29:13h            & TF 2.7,\\
\href{https://docs.it4i.cz}{\nolinkurl{docs.it4i.cz}} & GPU node              & 8 NVIDIA A100 (40 GB)         &                   & Horovod 0.23.0\\
\midrule
LUMI                                  & 1 node of             & 1 AMD EPYC 7A53 (64 cores)    & 38:30h            & TF 2.10,\\
    \href{https://lumi-supercomputer.eu}{\nolinkurl{lumi-supercomputer.eu}} & LUMI-G                & 4 AMD MI250X\tnote{*} (128 GB)    &                   & Horovod 0.26.1\\
\midrule
CESNET MetaCentrum                    & DGX H100              & 2 Intel Xeon 8480C (56 cores) & 29:26h            & TF 2.12,\\
\href{https://metavo.metacentrum.cz/en}{\nolinkurl{metavo.metacentrum.cz}}        &                       & 8 NVIDIA H100 (80 GB)    &                   & Horovod 0.27.0\\
\bottomrule
\end{tabular}
\begin{tablenotes}\footnotesize
\item[*] Multi-chip module (MCM) with two GPUs each -- effectively 8 GPUs
\end{tablenotes}
}
\end{threeparttable}
\end{center}
\caption{Computing systems used to verify the transfer phase. The transfer time was for one variant up to epoch 160.}
\label{platforms}
\end{table*}

The predictions of a combination of V1-4 are defined as
\begin{displaymath}
\bm{y}_{i,j}^{C} \coloneqq \sqrt[4]{\bm{y}_{i,j}^{V1} \cdot \bm{y}_{i,j}^{V2} \cdot \bm{y}_{i,j}^{V3} \cdot \bm{y}_{i,j}^{V4}}
\end{displaymath}
with an element-wise maximum, the Hadamard multiplication and fourth root. This follows the bagging methodology to create an ensemble of weak learners. The combined predictions $\bm{y}_{i,j}^{C}$ were constructed with all windows starting in the monitoring period (unit-stride $\rho = 1$). This is different to the transfer phase where only ten partially overlapping random windows were considered.

While the main development system was one node of the IT4Innovations' Karolina GPU cluster, other deep learning systems have also been used to confirm the transfer phase. Table~\ref{platforms} shows the verified systems with their hardware and software environment used. Transfer times were similar, except for the LUMI-G node. The reason was the I/O bound workload of our method. While the network only contained ca. 69k parameters, the data samples we worked with were comparatively more complex. Since LUMI-G contained only a single CPU socket, I/O was limited compared to the other systems.

\section{Results}\label{results}
The transferred model variants V1-4, as well as their bagged ensemble, were analyzed quantitatively with commonly used metrics in the machine learning and remote sensing domains. In addition, qualitative analysis was carried out with selected examples that also demonstrated the temporal localization of changes. Finally, an ablation study, which varies the number of observations, gave further insights into the transferred model's performance and resiliency.

\subsection{Quantitative Analysis}\label{qanalysis}
For quantitative analysis, we used three different metrics: i) receiver operating characteristic (ROC) curve, ii) precision recall (PR) curve, and iii) Cohen's Kappa $\kappa$ with varying thresholds.

ROC curves are a very common metric for binary classifiers~\cite{10.5555/2595566.2595576}. Since we received predictions that are of continuous values and not binary, ROC is a useful choice. They do not require a certain threshold value to be defined, but apply different thresholds at once. The area under the curve (AUC) forms a metric that allows for comparison of different models.

ROC curves do not work well for unbalanced classes, which was also true for our case. The amount of no-changes were dominant to the amount of changes since most of the pixels did not change. The choice of PR curves~\cite{10.1145/65943.65945} are more suitable for such skewed datasets as they can provide more insight \cite{10.1145/1143844.1143874}. Similarly to ROC curves, PR curves consider varying thresholds and the area under the curve is a well suited metric for model comparison.

Cohen's Kappa~\cite{cohen1960coefficient} $\kappa$, provides the agreement of two raters. In this work it was used for the two classes of change and no-change. Since it does not apply varying thresholds itself and expects binary raters, it required the selection of a threshold value upfront. We studied $\kappa$ with thresholds in the range of $0.0$ to $1.0$. Thresholds with the highest $\kappa$ values would finally be used since they represent the best agreement of the prediction output and the compared ground truth.

\begin{figure*}[ht]
\begin{center}
\includegraphics[height=0.3\textwidth]{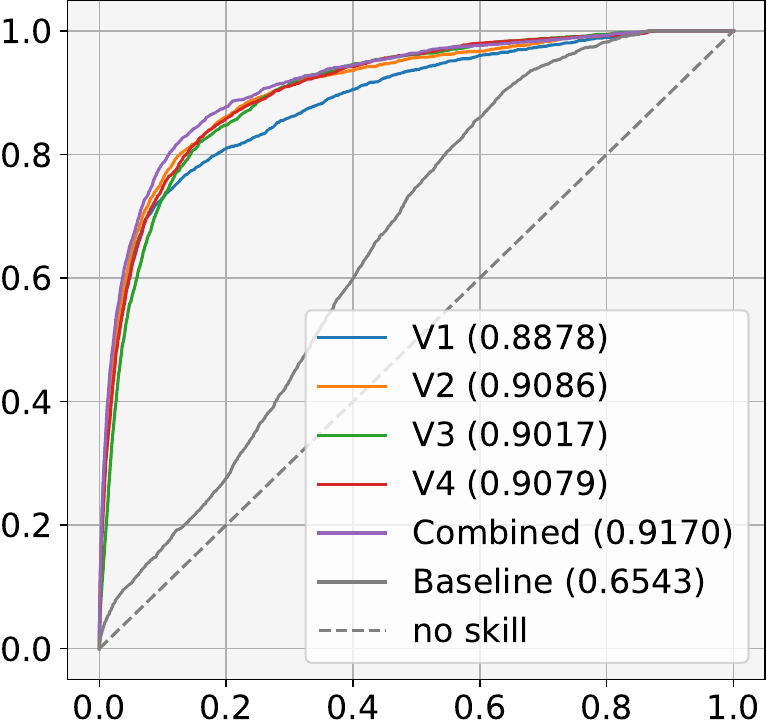}\hspace{2mm}\includegraphics[height=0.3\textwidth]{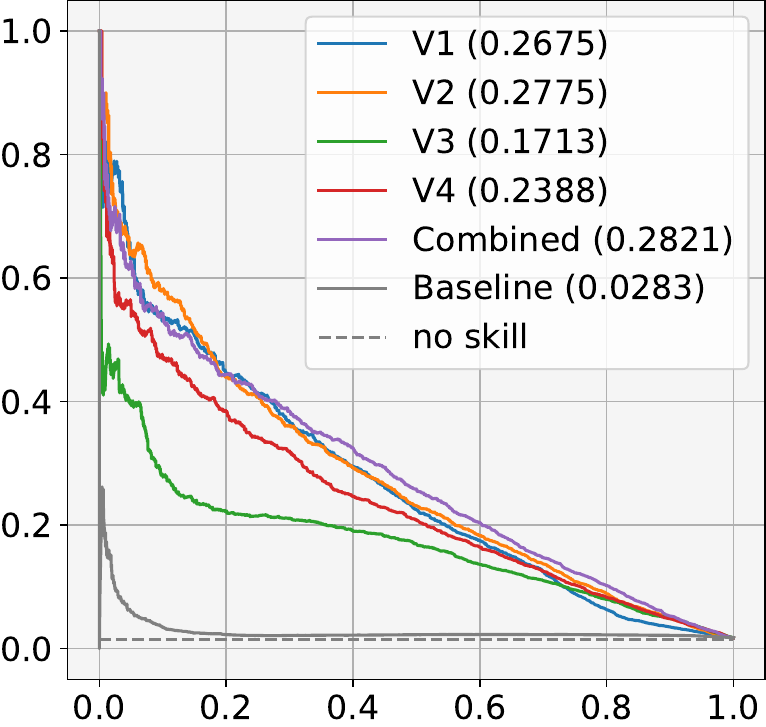}\hspace{2mm}\includegraphics[height=0.3\textwidth]{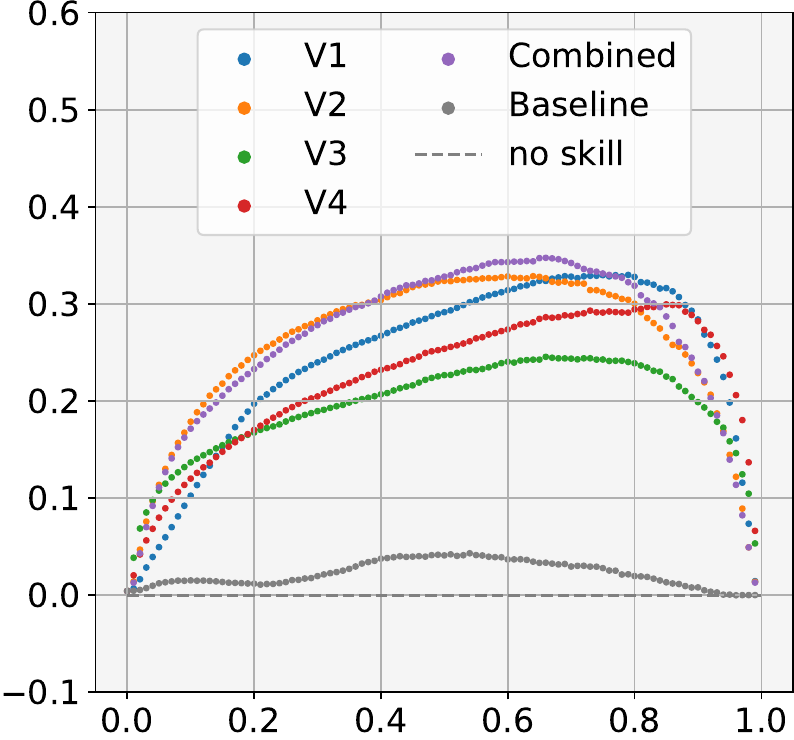}
\end{center} \caption{For the \textit{trainval} set: ROC (left) and PR (middle) curves; Cohen's Kappa is shown for different thresholds (right). Area under the ROC/PR curves are in parenthesis.}
\label{trainval_curves}
\end{figure*}

\begin{figure*}[ht]
\begin{center}
\includegraphics[height=0.3\textwidth]{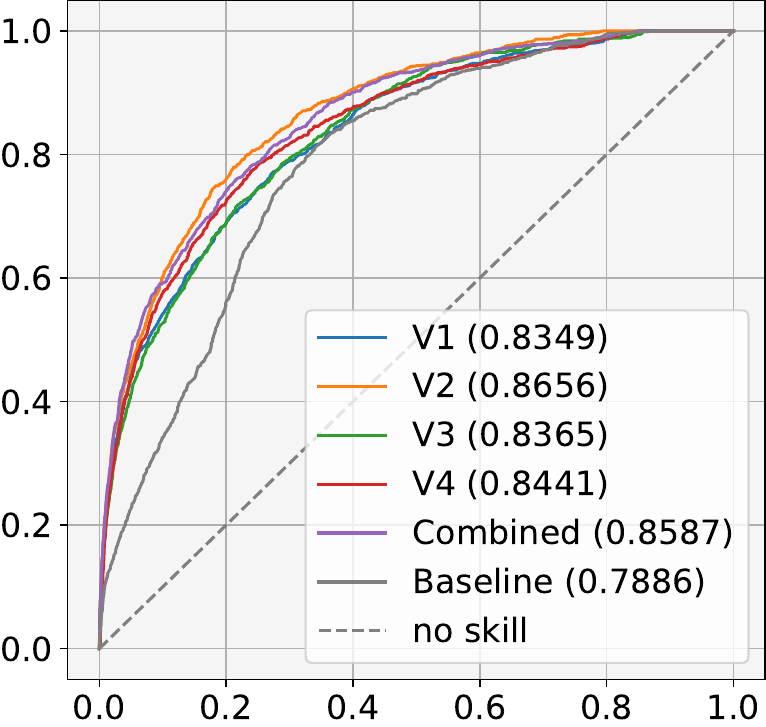}\hspace{2mm}\includegraphics[height=0.3\textwidth]{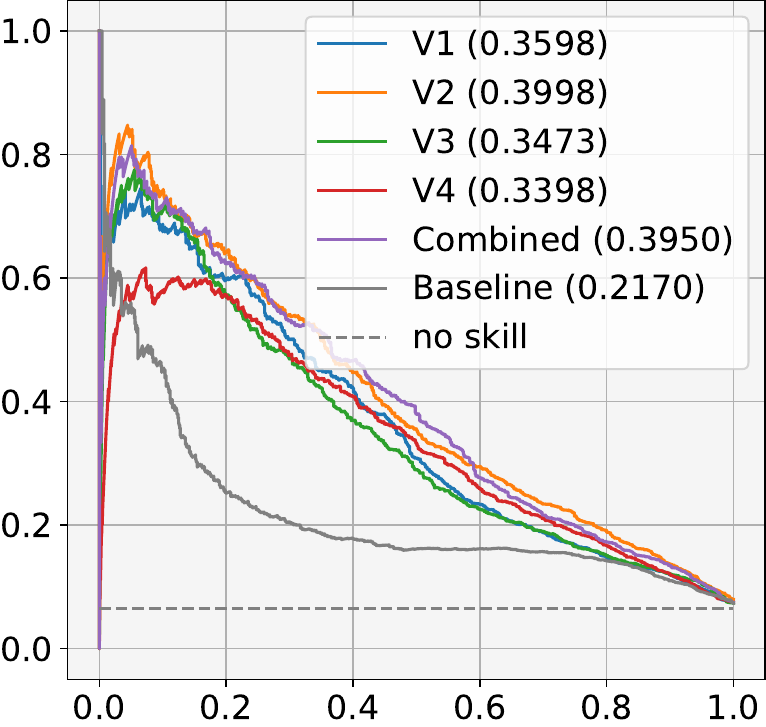}\hspace{2mm}\includegraphics[height=0.3\textwidth]{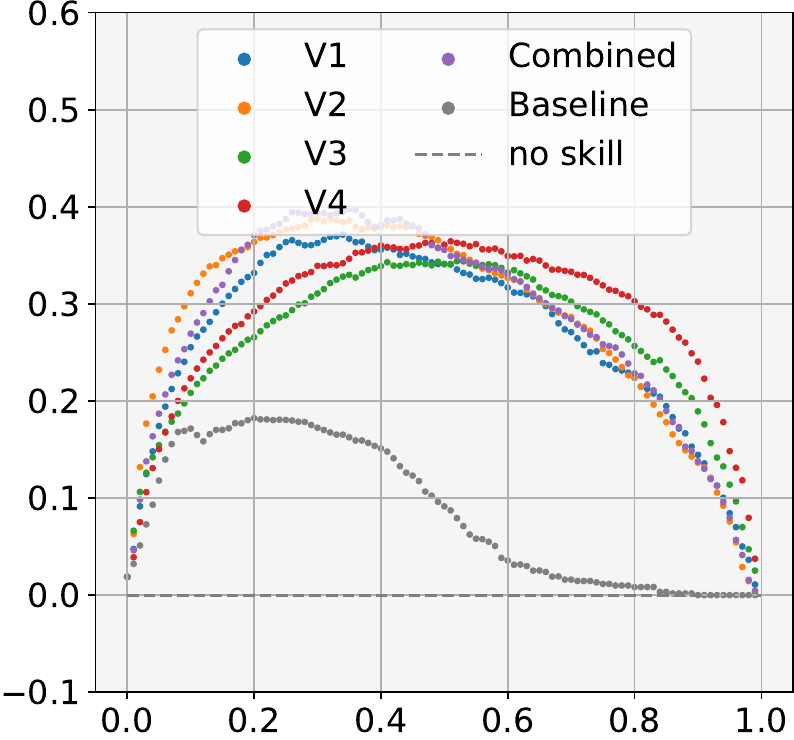}
\end{center} \caption{For the \textit{testing}$^-$ dataset set: ROC (left) and PR (middle) curves; Cohen's Kappa is shown for different thresholds (right). Area under the ROC/PR curves are in parenthesis.}
\label{testing_curves}
\end{figure*}

Figure~\ref{trainval_curves} shows the ROC and PR curves, as well as $\kappa$ for the \textit{trainval} dataset used for the transfer variants. Whilst all variants show a similar performance, variant V3 scores lower for all three metrics. This likely stems from a worse representation of training samples that were randomly drawn for the non-exhaustive cross validation. Nevertheless, all variants are performing significantly better than the pre-trained ERCNN-DRS (baseline). The combination of all methods showed the highest scores in all three metrics. Noticeable is also the higher $\kappa$ for larger thresholds for the variants and their combination.

\begin{figure*}[p]
\begin{center}
\begin{tabular}{c}
\includegraphics[width=0.96\textwidth]{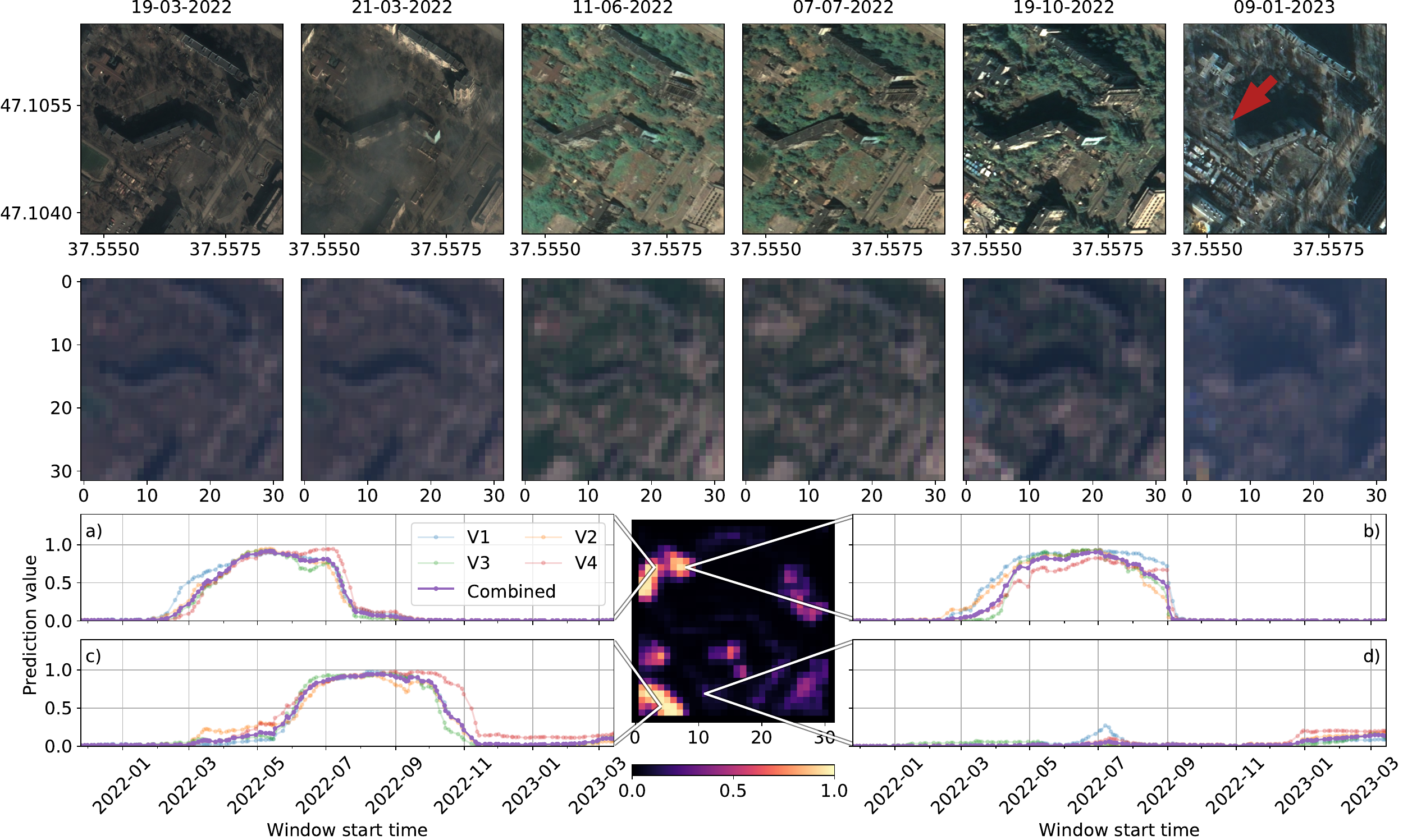}\vspace{-4mm}\\
{\footnotesize \hspace{8mm}\textbf{i) Tile 42:30}}\\\hline\\\vspace{2mm}
\includegraphics[width=0.96\textwidth]{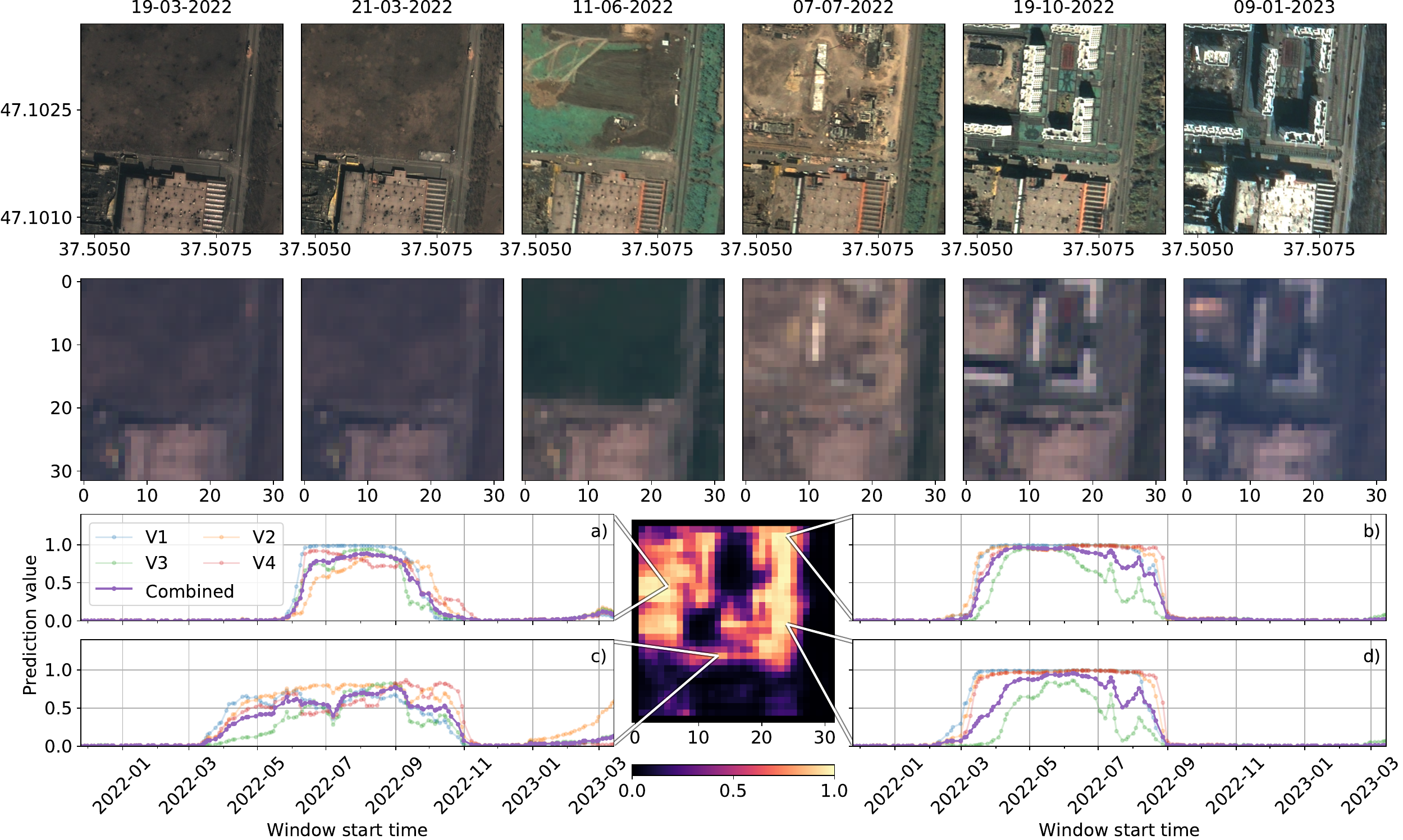}\vspace{-4mm}\\
{\footnotesize \hspace{8mm}\textbf{ii) Tile 43:18}}
\end{tabular}
\end{center}
\caption{Two examples for the verification of changes 2022/23 with a limited number of Airbus Pl\'eiades observations (top rows). Second rows show Sentinel 2 true color data at similar observation times of the Pl\'eiades counterparts ($\pm3 \textnormal{days}$). Bottom rows show the prediction $\bm{y}_{i,j}^{C}$ with prediction value time series of four selected pixels.}
\label{example2}
\end{figure*}

In comparison, Figure~\ref{testing_curves} shows the metrics for the \textit{testing}$^-$ dataset. This dataset is the \textit{testing} dataset with tile 43:18 removed. Since the \textit{testing} dataset was characterized by only 18 different tiles (samples), outliers caused by tiles with larger changes skewed the results. In particular, tile 43:18 (see Figure~\ref{example2}) significantly increased the AUCs due to large and intense construction activities. We therefore decided to remove it for our analysis and analyze smaller and heterogeneous changes. The curves for the full \textit{testing} dataset can be found in Figure~\ref{full_testing_curves} in the Appendix. For the \textit{testing}$^-$ dataset, all variants produce similar results, with V4 showing a slightly different behavior for the PR curves and $\kappa$. The ERCNN-DRS baseline scores better here but still lacks behind its transferred variants. Their combinations do not score highest in the ROC and PR curves, but are close to the best variant (V2). The combined variants, however, score highest for the $\kappa$. What is noticeable here are the different thresholds that result in the highest $\kappa$ between the \textit{trainval} and \textit{testing}$^-$ datasets. The former suggests thresholds of over 0.7, whereas the latter produces higher $\kappa$ scores around 0.3. Nevertheless, the $\kappa$ values are around 0.3 for thresholds of 0.7 in both cases, suggesting that outliers caused higher $\kappa$ values for lower thresholds in the much smaller \textit{testing}$^-$ dataset. After all, the time frame 2022/23 was subject to large and highly frequent urban changes with a deviation from the regular urban development in 2017-2020.

\begin{figure*}[p]
\begin{center}
\begin{tabular}{c}
\includegraphics[width=0.96\textwidth]{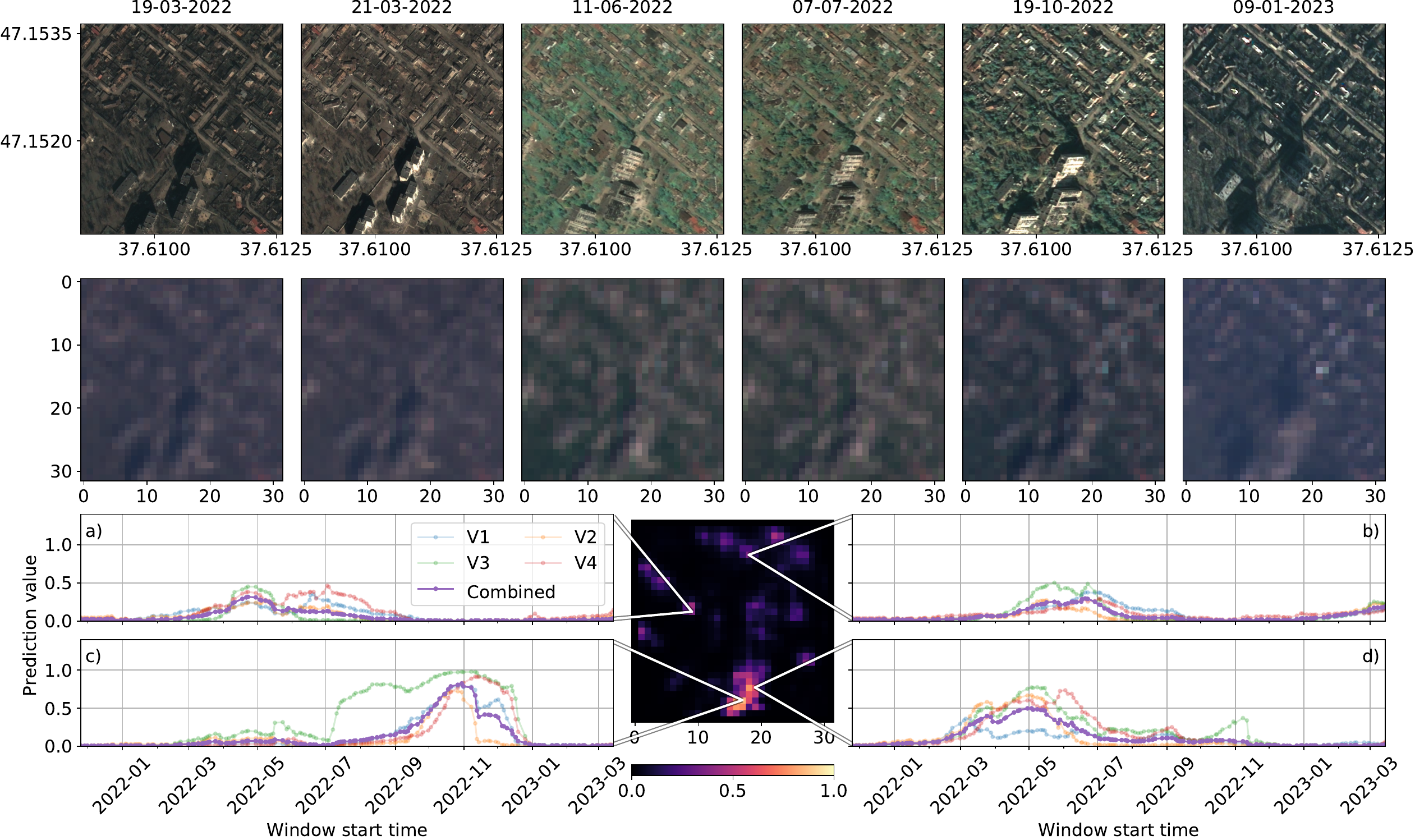}\vspace{-4mm}\\
{\footnotesize \hspace{8mm}\textbf{i) Tile 26:43}}\\\hline\\\vspace{2mm}
\includegraphics[width=0.96\textwidth]{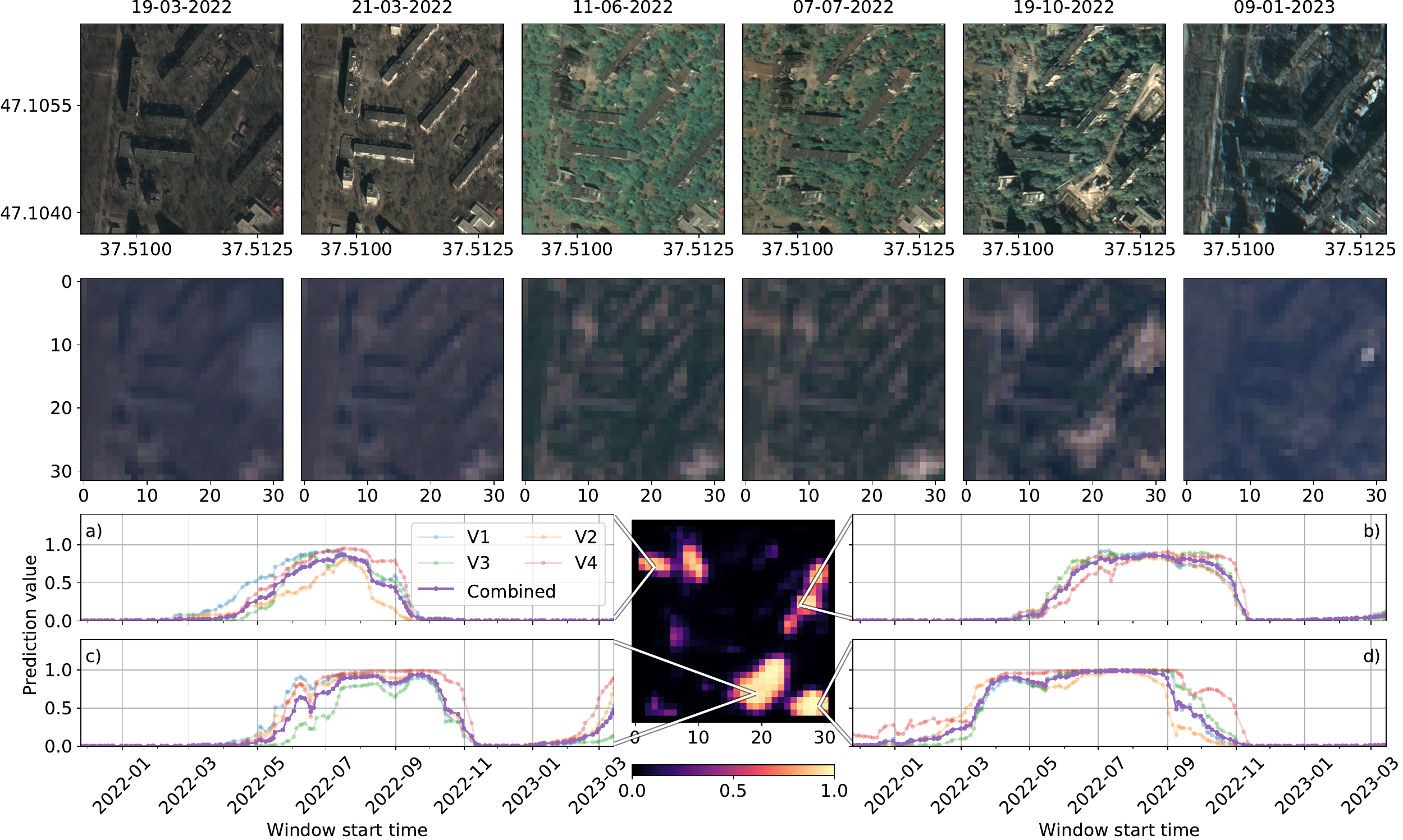}\vspace{-4mm}\\
{\footnotesize \hspace{8mm}\textbf{ii) Tile 42:19}}
\end{tabular}
\end{center}
\caption{Two examples for the verification of changes 2022/23 with a limited number of Airbus Pl\'eiades observations (top rows). Second rows show Sentinel 2 true color data at similar observation times of the Pl\'eiades counterparts ($\pm3 \textnormal{days}$). Bottom rows show the prediction $\bm{y}_{i,j}^{C}$ with prediction value time series of four selected pixels.}
\label{example1}
\end{figure*}

\subsection{Qualitative Analysis}
We selected four different tiles from the monitoring period that showed representative and diverse change patterns. The quality of the predictions were analyzed. Figures~\ref{example1} and \ref{example2} show two tiles that detected different urban changes and activities. For all four examples, we also depict six VHR Airbus Pl\'eiades and Sentinel 2 (true color) observations over 2022/23. The VHR observations were used for creating the ground truth for verification and only the latter were used for the predictions. The differences in spatial resolution are clearly noticeable. As a result, the only changes that were detectable by our method are larger than the sensor resolution of 10 m/pixel. The time stamp on the top corresponds to the Airbus Pl\'eiades observation with the closest Sentinel 2 counterparts $\pm 3$ days apart. For each tile, one pixel of four different change regimes was selected to show how their prediction values changed over time. We selected the regions to provide a diverse and balanced set of change patterns with clear indications given by the VHR data. Tracing the predictions over time enabled the localization of when changes happened. This also shows how the transferred model variants identified changes differently, based on different learned patterns (i.e., \textit{trainval} splits). Nevertheless, the detected changes were coherent and directly attributable to change events.

Tile 26:43 in Figure~\ref{example1} i) shows smaller changes and a large one (bottom center). While the smaller changes were harder to analyze -- see charts a) and b) -- the larger one shows the destruction of a large building. Depending on the selected pixels, changes took place at different times. This building was severely damaged between March and June 2022 according to the VHR Airbus Pl\'eiades observations. The effects of this damage were also detected by the transferred models with a pixel selected in front of the building as shown in d). The building itself was torn down between the middle of October 2022 and early January 2023. Again, this was detected by the model variants, with V3 being overly sensitive to these changes in charts c).

The tile 42:19 in Figure~\ref{example1} ii) shows many large changes. Construction of building in b) and c) took place in the second half of 2022. The selected pixel d) shows a reconstruction of a large building in the middle of 2022. In a), the nearby building was torn down in the middle of 2022. The selected pixel belonged to an area that was temporarily used for the destruction process.

The example i) in Figure~\ref{example2} addressing the tile 42:30 shows other urban-related changes with different patterns. In a) and b) the same building is monitored where the roof was damaged and reconstructed throughout 2022. A larger building was erected of which a quarter is visible in the lower left that led to changes in the second half of 2022, as shown in c). This scene also contains the destruction of two larger buildings. The one shown in d) only results in low prediction values; the other one is not detected at all (see the red arrow at rightmost Airbus Pl\'eiades observation).

The last example ii) in Figure~\ref{example2} shows the excluded tile 43:18 from the \textit{testing}$^-$ dataset. This contains an exceptionally large set of changes with the construction of a building complex. These constructions happened swiftly after the Russian occupation of Mariupol at end of May 2022. The charts a), b), and d) show concurrent constructions of different buildings of the complex. The chart c) shows a supply corridor which was later converted into a road.

\subsection{Ablation Study}
The unavailability of Sentinel 1B gave rise to the question as to how resilient our methods were to a change in the number of observations in relation to the overall prediction performance. In an ablation study, we selectively reduced the number of observations by increasing the step size $\delta$. This effectively samples fewer observations and simulates scenarios where fewer real observations are available due to atmospheric disturbance or mission outages. We applied this approach separately for each mode to also analyze the impact of the modes to the predictions. The ablation study was carried out with both the \textit{trainval} and the \textit{testing}$^-$ dataset and hence to two different time frames. To avoid an influence by a shrunken number of observations per window (i.e., to fall below $\omega$), we retained the same number of observations but only updated them at a step size interval. For example, moving from the default $\delta = 2 \textrm{ days}$ to $\delta = 120 \textrm{ days}$ resulted in windows with the same number of observations as for $\delta = 2 \textrm{ days}$, but until 120 days had been reached, the observation values remained unchanged.

\begin{figure*}[htb]
\begin{center}
\begin{tabular}{cccc}
& SAR & Optical & SAR \& Optical\\
& ($\delta^{SAR} \coloneqq \delta$, $\delta^{OPT} \coloneqq 2 \textrm{ days}$) & ($\delta^{SAR} \coloneqq 2 \textrm{ days}$, $\delta^{OPT} \coloneqq \delta$) & ($\delta^{SAR} \coloneqq \delta$, $\delta^{OPT} \coloneqq \delta$)\\
\multirow{1}{*}[22ex]{\rotatebox[origin=c]{90}{\parbox[c]{3cm}{\centering $\delta = 120 \textrm{ days}$}}} & \includegraphics[width=4.5cm]{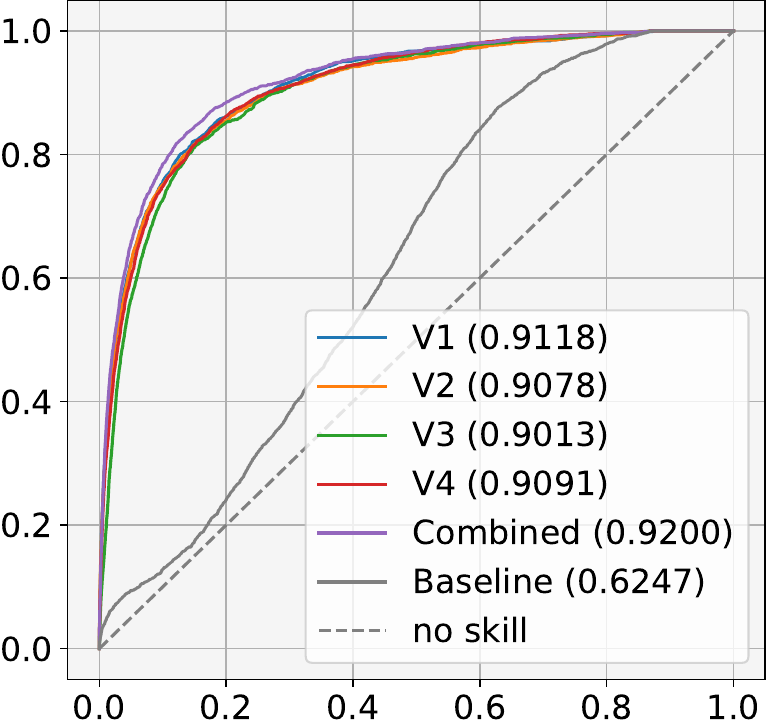} & \includegraphics[width=4.5cm]{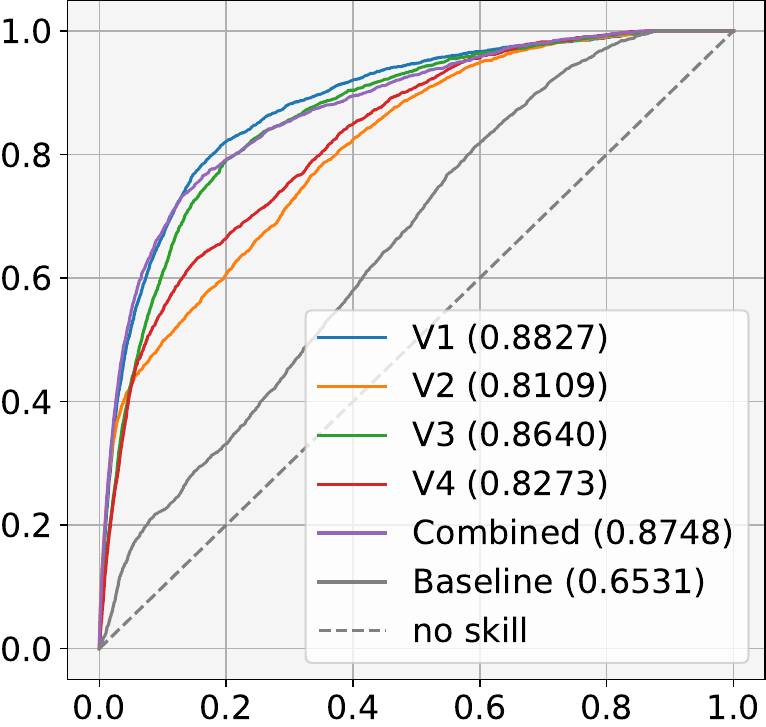} & \includegraphics[width=4.5cm]{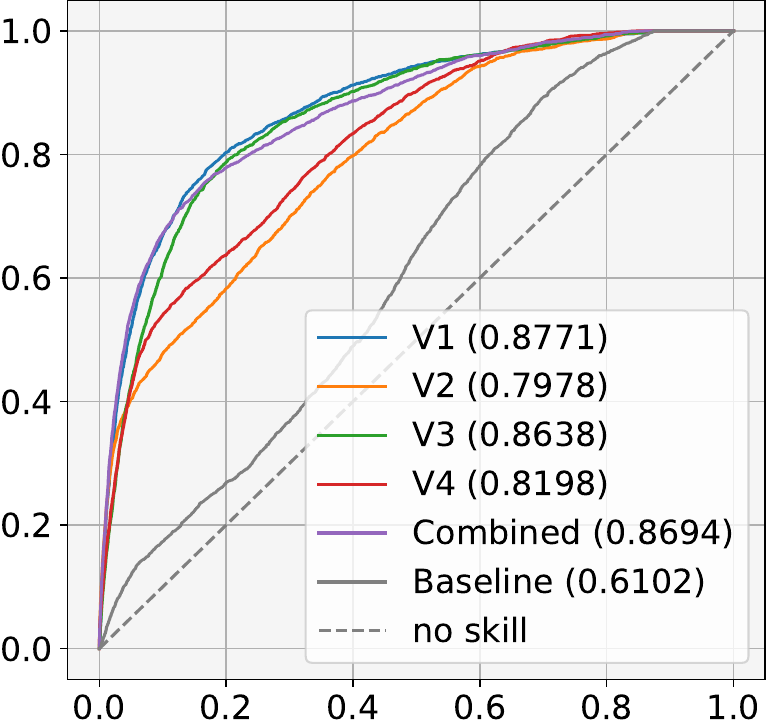}\\
\multirow{1}{*}[22ex]{\rotatebox[origin=c]{90}{\parbox[c]{3cm}{\centering $\delta = 600 \textrm{ days}$}}} & \includegraphics[width=4.5cm]{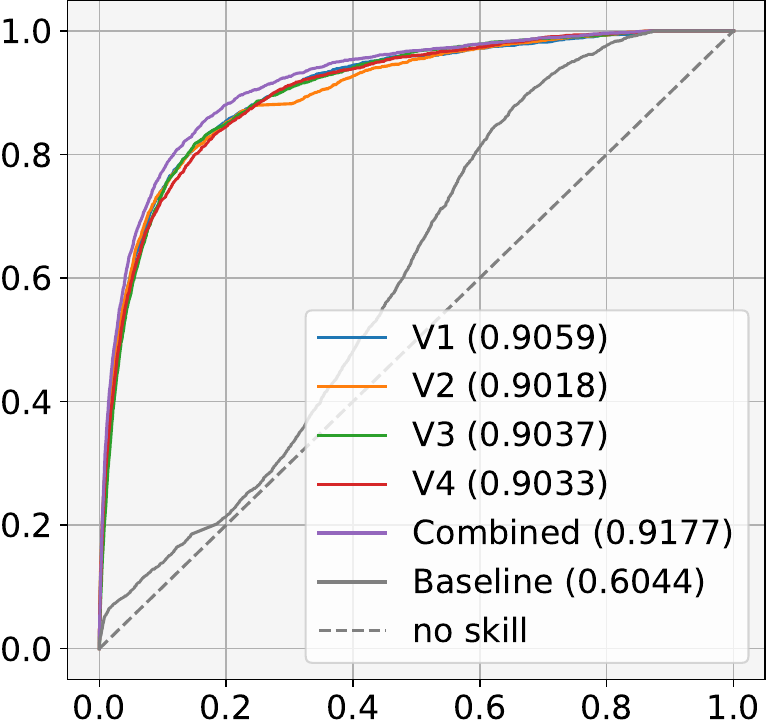} & \includegraphics[width=4.5cm]{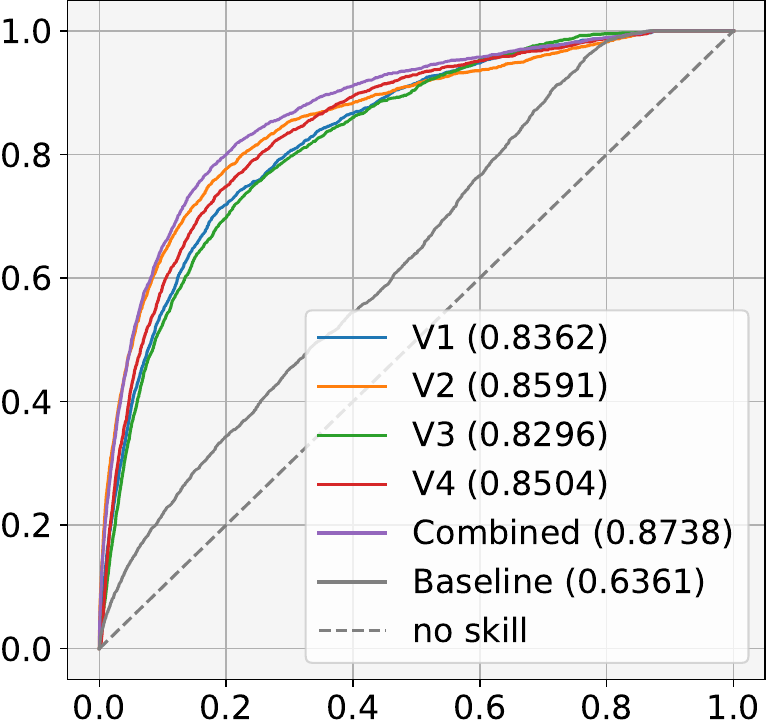} & \includegraphics[width=4.5cm]{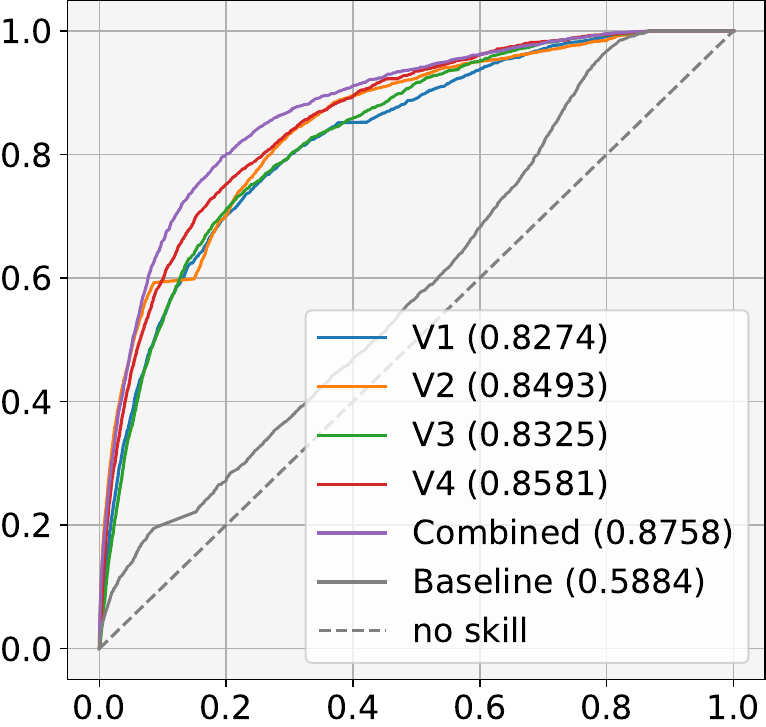}\\
\multirow{1}{*}[22ex]{\rotatebox[origin=c]{90}{\parbox[c]{3cm}{\centering $\delta = \infty \textrm{ days}$}}} & \includegraphics[width=4.5cm]{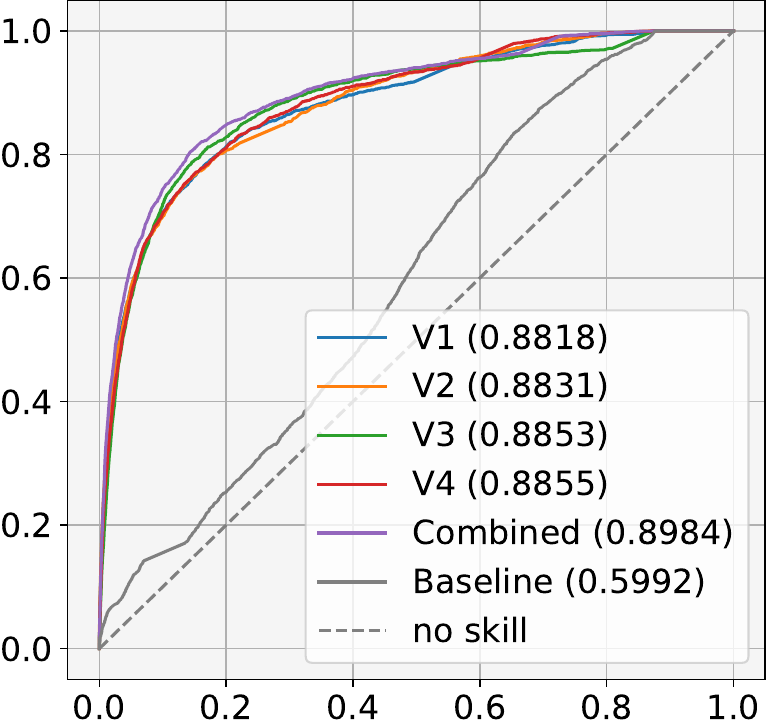} & \includegraphics[width=4.5cm]{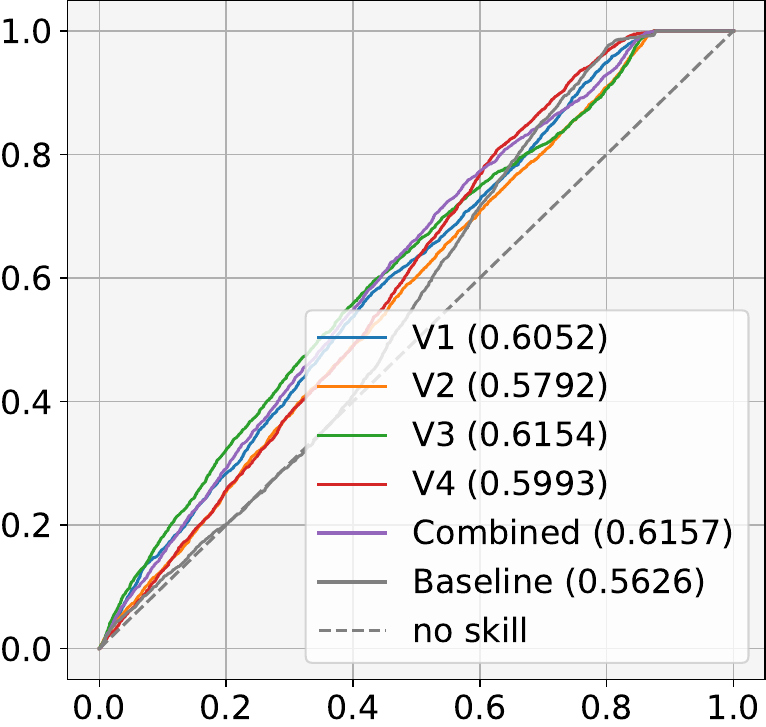} & \includegraphics[width=4.5cm]{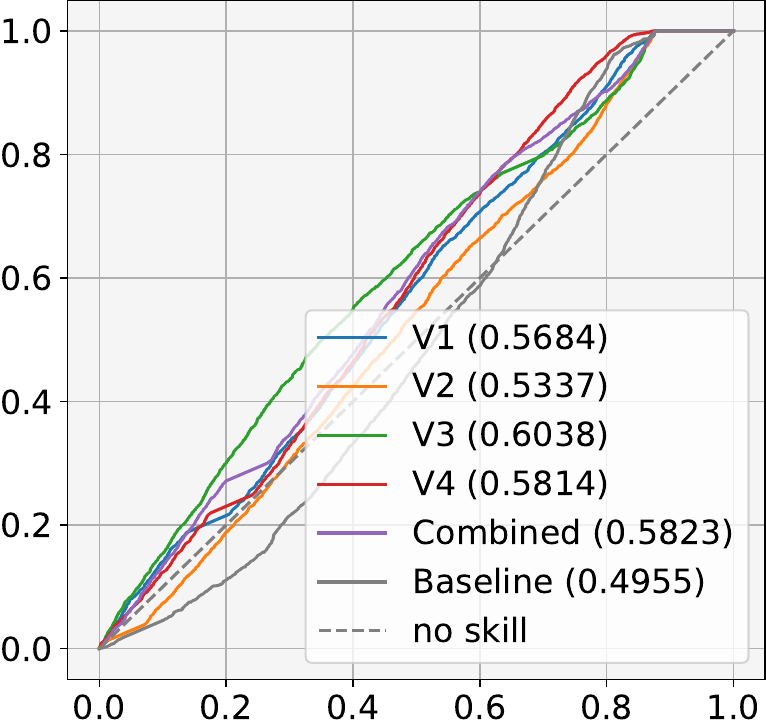}\\
\end{tabular}
\end{center}
\caption{ROC curves of the different models applied on \textit{trainval} with different sampling steps.}\label{ablation_roc_trainval}
\end{figure*}

\begin{figure*}[htb]
\begin{center}
\begin{tabular}{cccc}
& SAR & Optical & SAR \& Optical\\
& ($\delta^{SAR} \coloneqq \delta$, $\delta^{OPT} \coloneqq 2 \textrm{ days}$) & ($\delta^{SAR} \coloneqq 2 \textrm{ days}$, $\delta^{OPT} \coloneqq \delta$) & ($\delta^{SAR} \coloneqq \delta$, $\delta^{OPT} \coloneqq \delta$)\\
\multirow{1}{*}[22ex]{\rotatebox[origin=c]{90}{\parbox[c]{3cm}{\centering $\delta = 120 \textrm{ days}$}}} & \includegraphics[width=4.5cm]{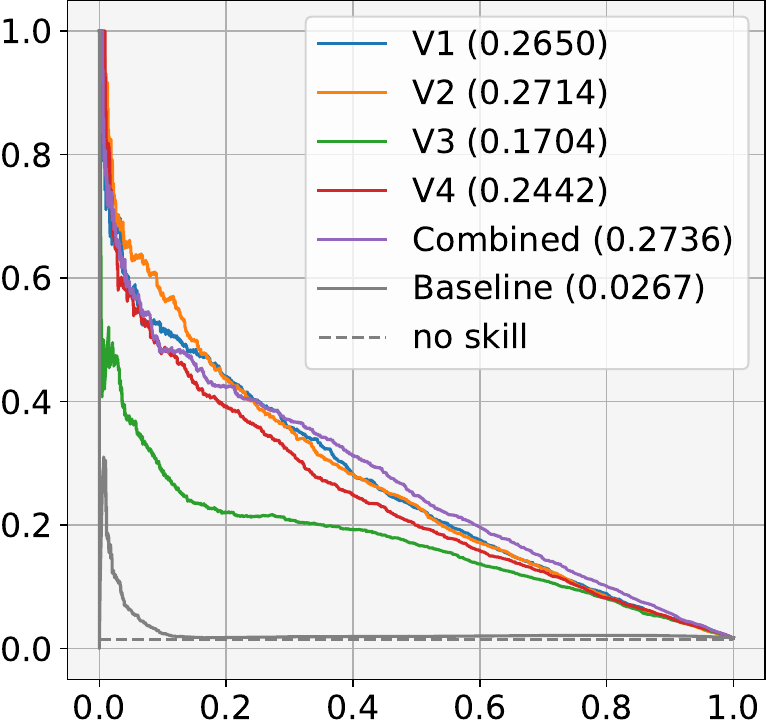} & \includegraphics[width=4.5cm]{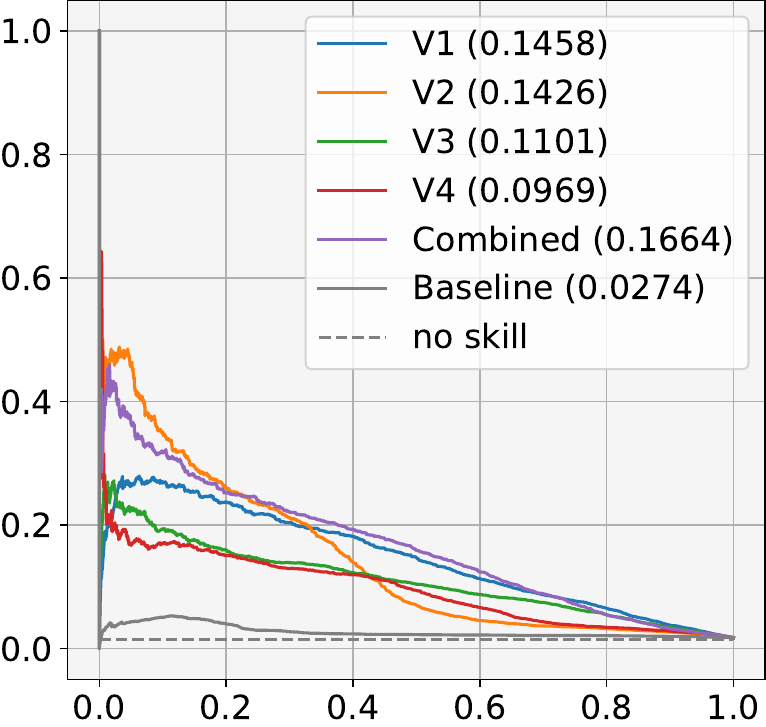} & \includegraphics[width=4.5cm]{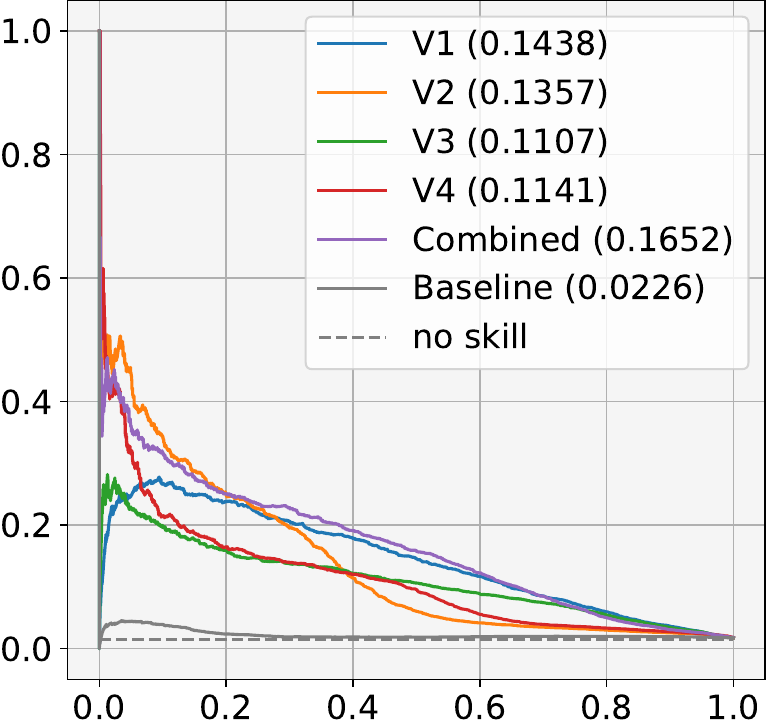}\\
\multirow{1}{*}[22ex]{\rotatebox[origin=c]{90}{\parbox[c]{3cm}{\centering $\delta = 600 \textrm{ days}$}}} & \includegraphics[width=4.5cm]{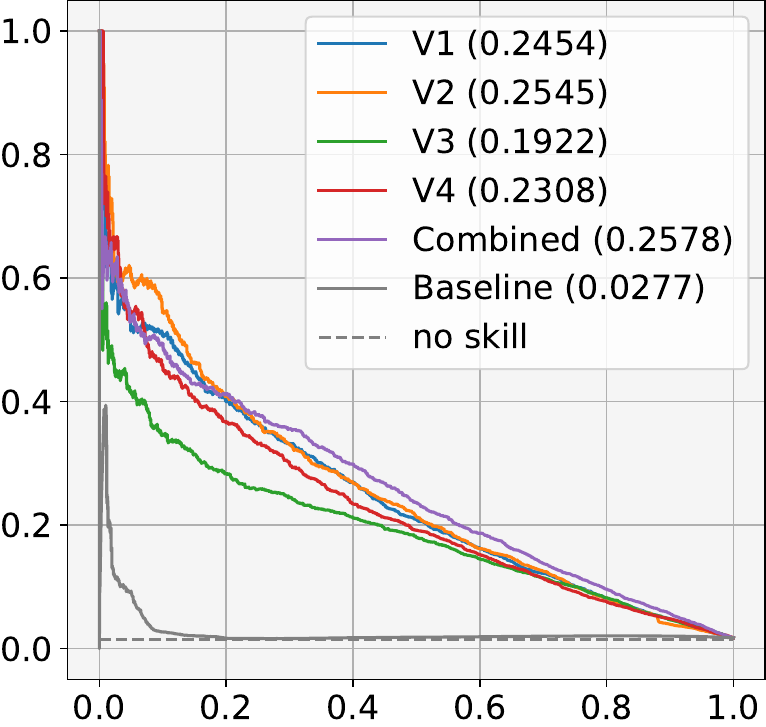} & \includegraphics[width=4.5cm]{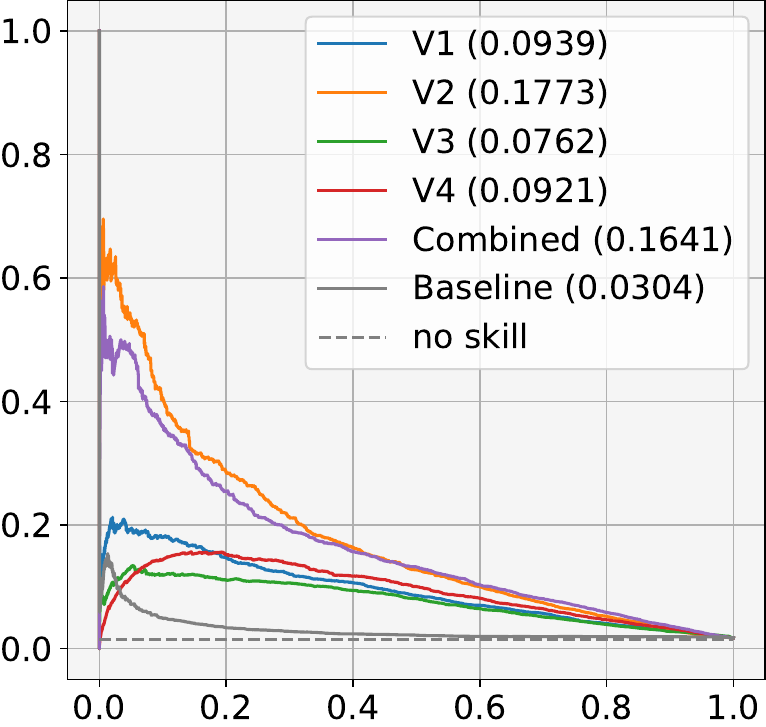} & \includegraphics[width=4.5cm]{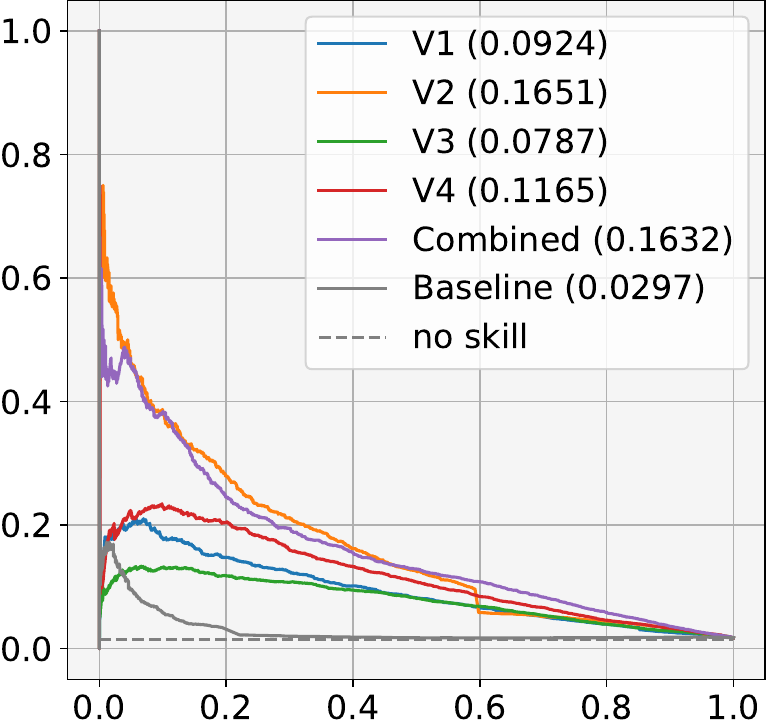}\\
\multirow{1}{*}[22ex]{\rotatebox[origin=c]{90}{\parbox[c]{3cm}{\centering $\delta = \infty \textrm{ days}$}}} & \includegraphics[width=4.5cm]{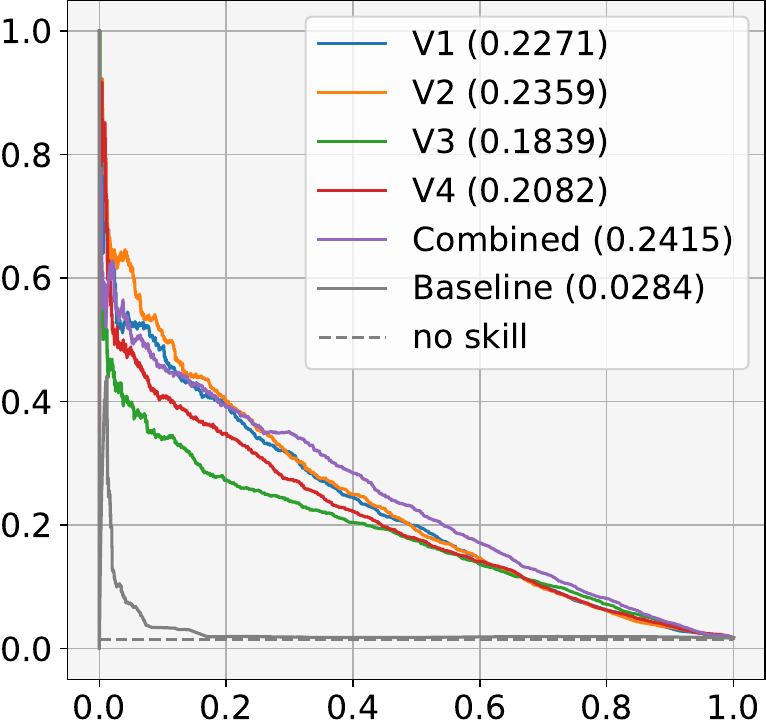} & \includegraphics[width=4.5cm]{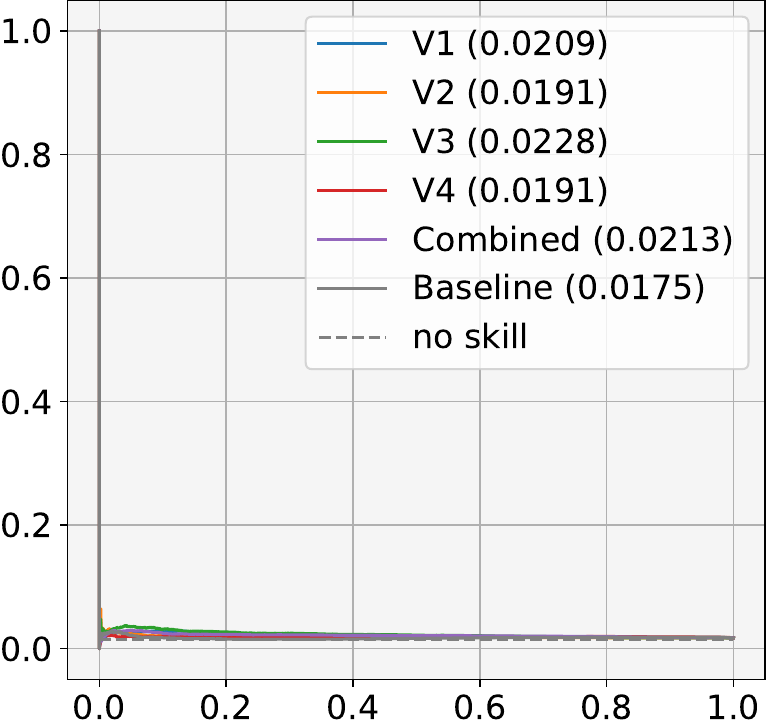} & \includegraphics[width=4.5cm]{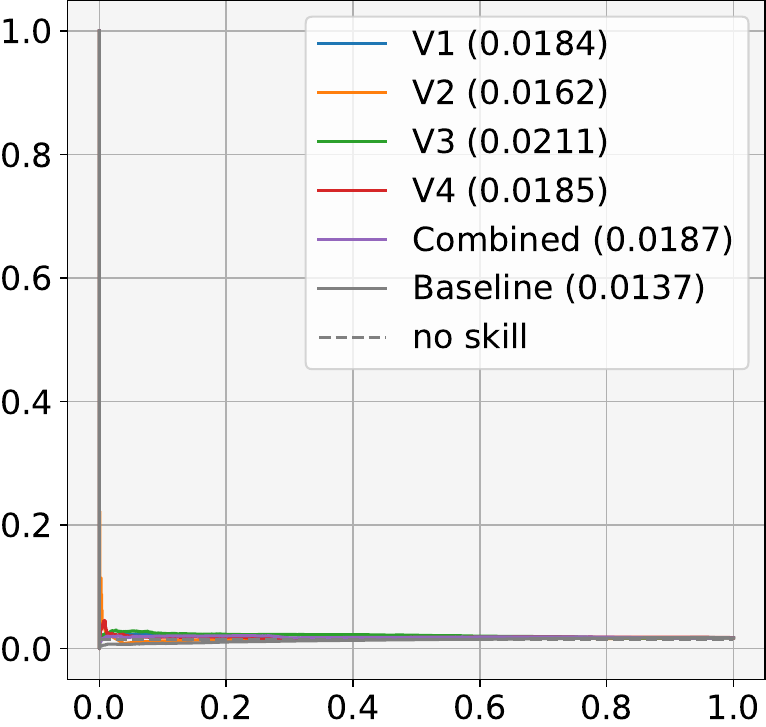}\\
\end{tabular}
\end{center}
\caption{PR curves of the different models applied on \textit{trainval} with different sampling steps.}\label{ablation_pr_trainval}
\end{figure*}

\begin{figure*}[htb]
\begin{center}
\begin{tabular}{cccc}
& SAR & Optical & SAR \& Optical\\
& ($\delta^{SAR} \coloneqq \delta$, $\delta^{OPT} \coloneqq 2 \textrm{ days}$) & ($\delta^{SAR} \coloneqq 2 \textrm{ days}$, $\delta^{OPT} \coloneqq \delta$) & ($\delta^{SAR} \coloneqq \delta$, $\delta^{OPT} \coloneqq \delta$)\\
\multirow{1}{*}[22ex]{\rotatebox[origin=c]{90}{\parbox[c]{3cm}{\centering $\delta = 120 \textrm{ days}$}}} & \includegraphics[width=4.5cm]{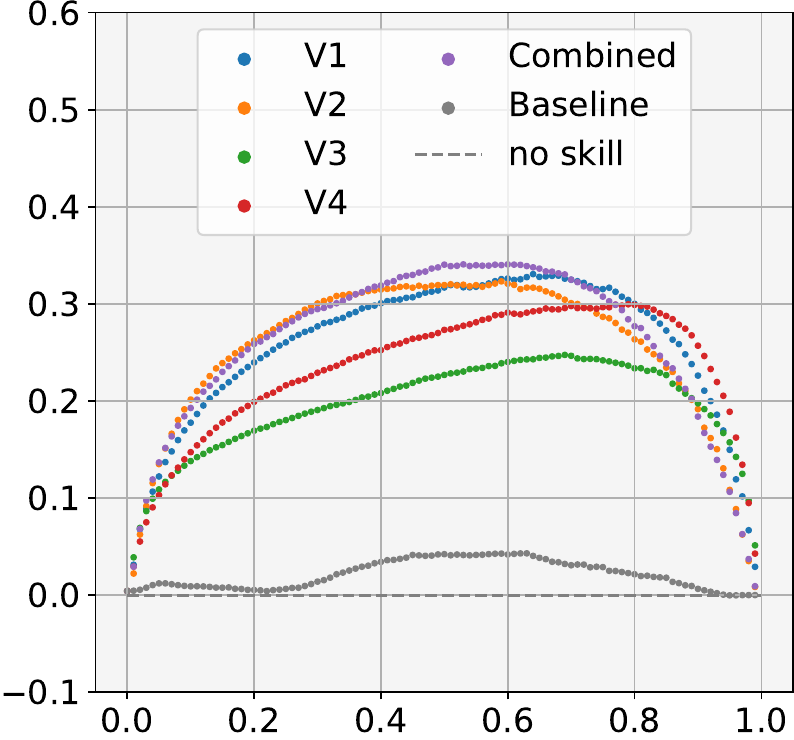} & \includegraphics[width=4.5cm]{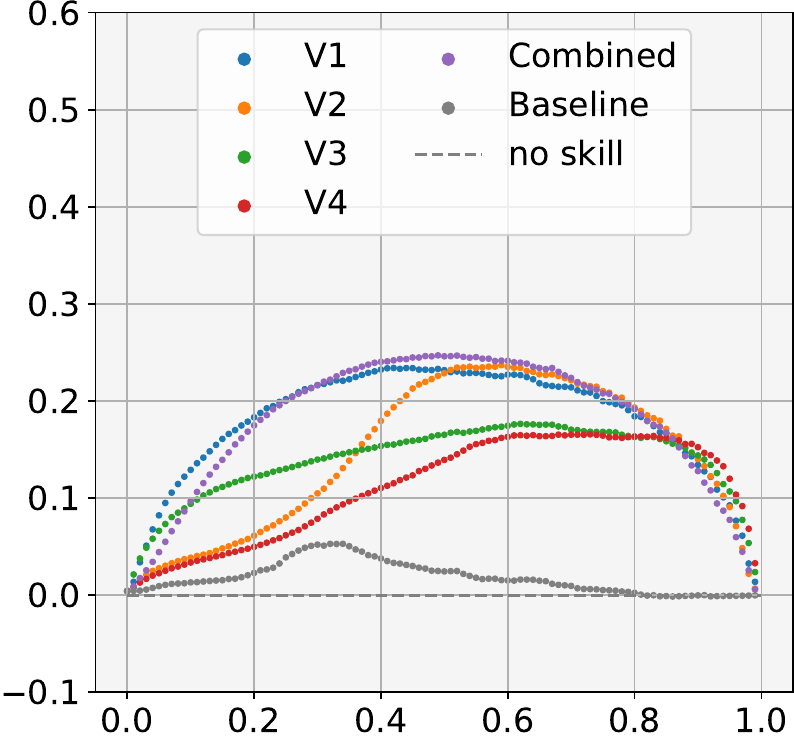} & \includegraphics[width=4.5cm]{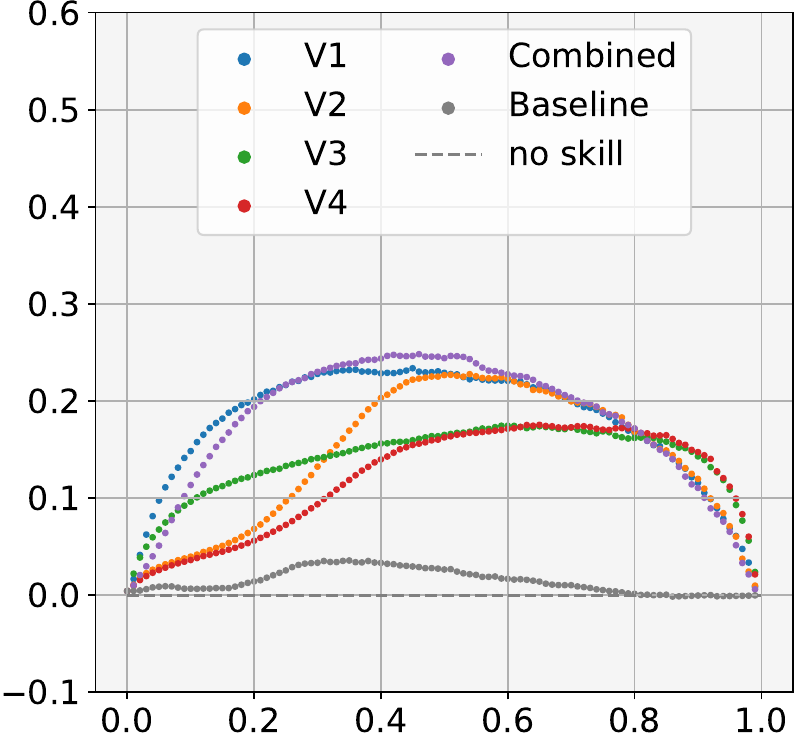}\\
\multirow{1}{*}[22ex]{\rotatebox[origin=c]{90}{\parbox[c]{3cm}{\centering $\delta = 600 \textrm{ days}$}}} & \includegraphics[width=4.5cm]{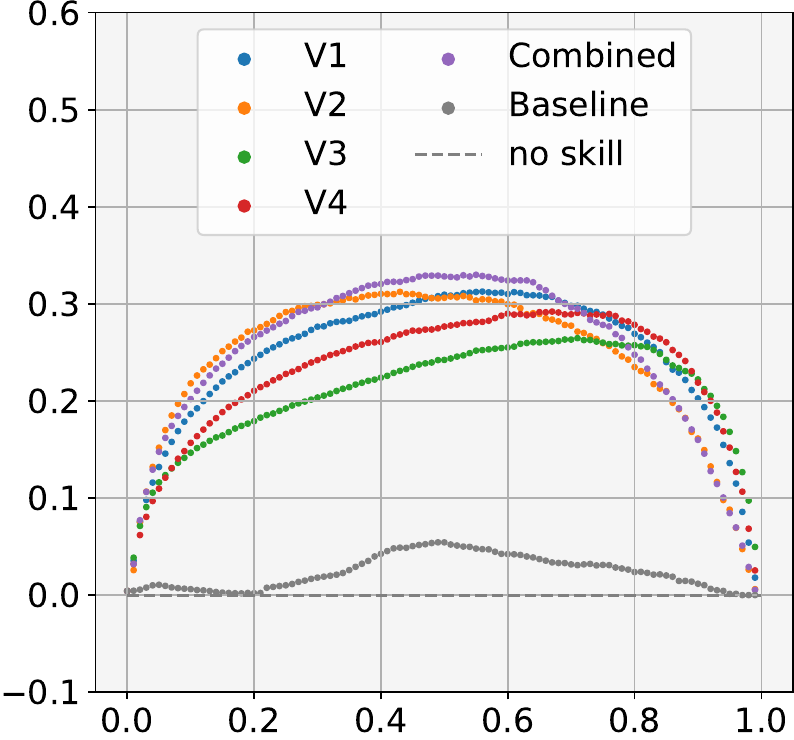} & \includegraphics[width=4.5cm]{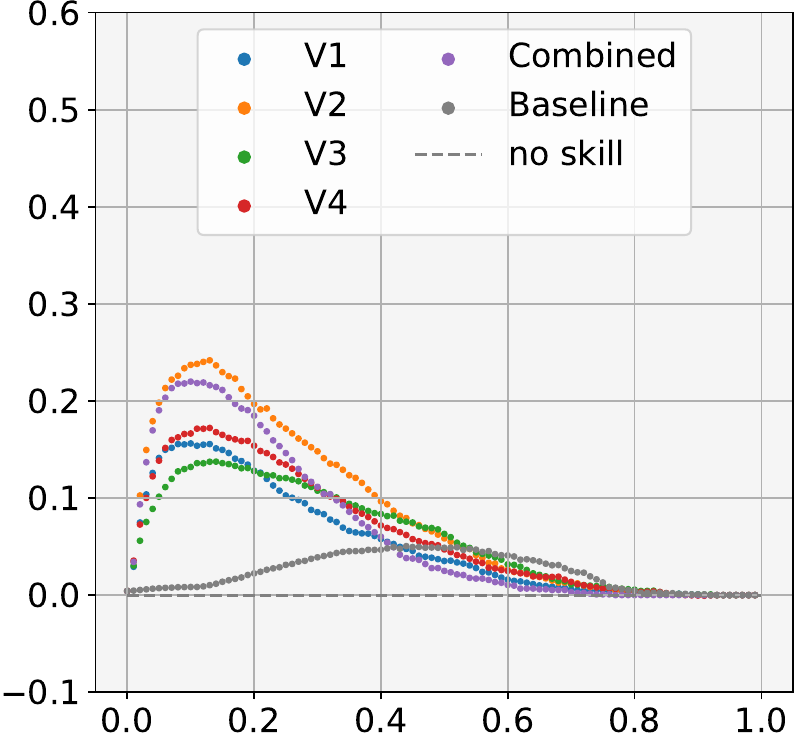} & \includegraphics[width=4.5cm]{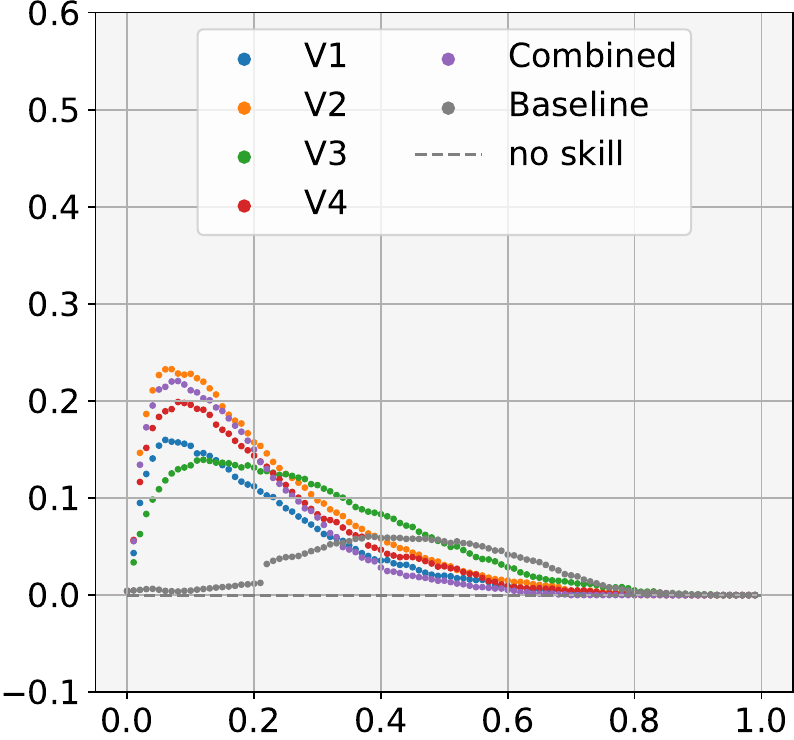}\\
\multirow{1}{*}[22ex]{\rotatebox[origin=c]{90}{\parbox[c]{3cm}{\centering $\delta = \infty \textrm{ days}$}}} & \includegraphics[width=4.5cm]{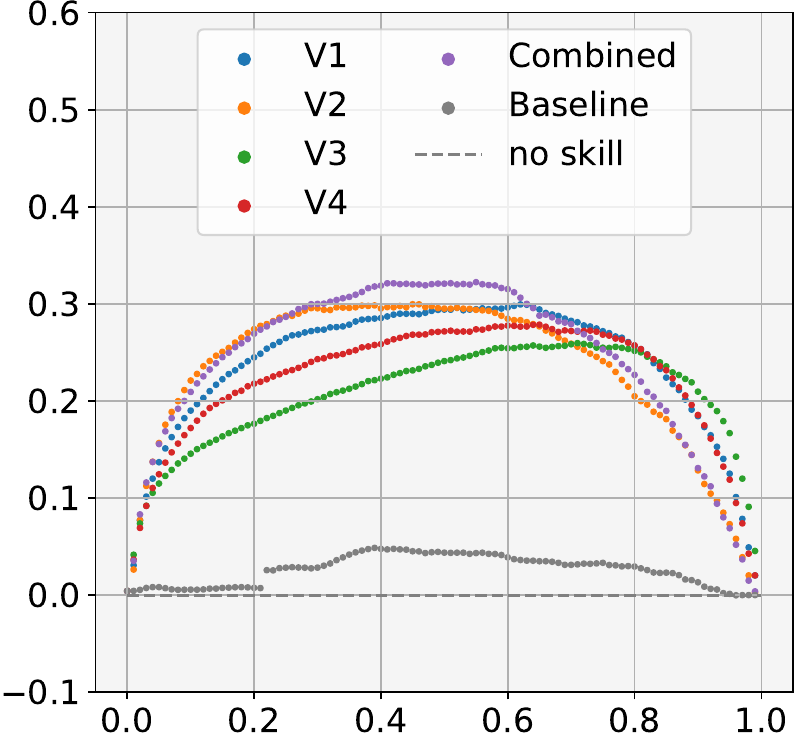} & \includegraphics[width=4.5cm]{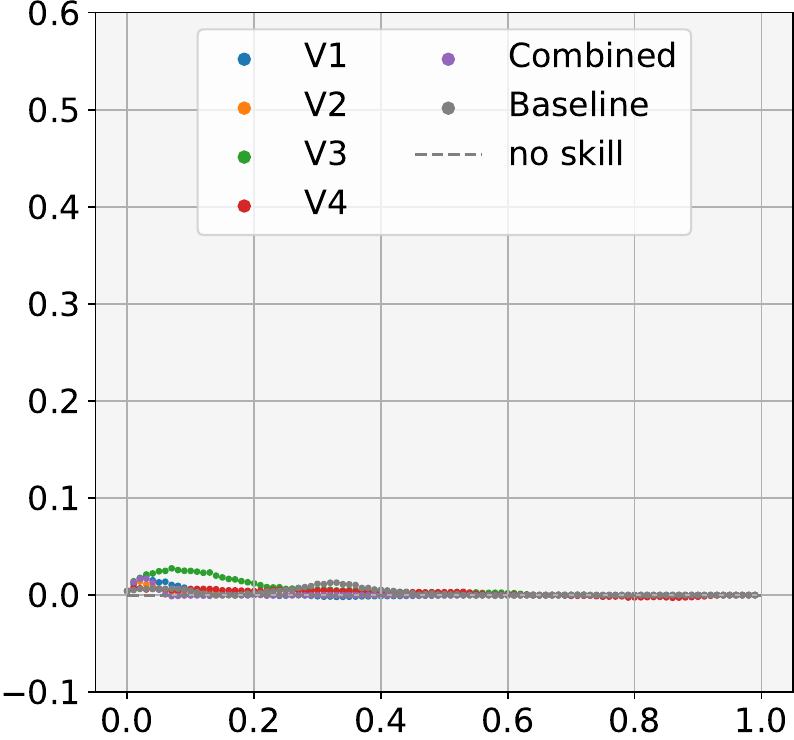} & \includegraphics[width=4.5cm]{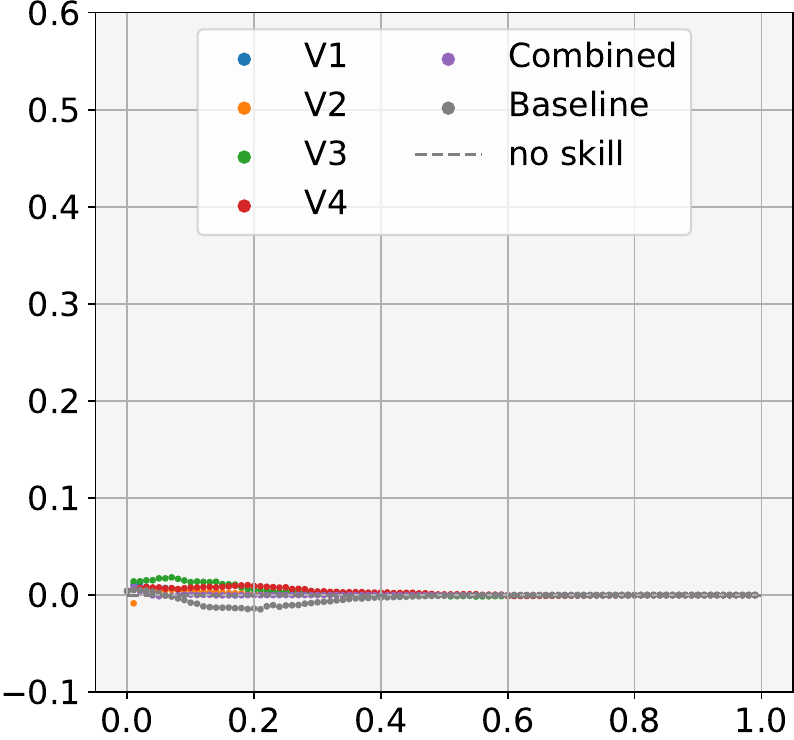}\\
\end{tabular}
\end{center}
\caption{Cohen's Kappa of the different models applied on \textit{trainval} with different sampling steps and a changing threshold.}\label{ablation_kappa_trainval}
\end{figure*}

For the \textit{trainval} dataset, which spans the years starting from early 2017 until the end of 2020, $\delta$ was changed from the default of 2 days to 120, 600 and $\infty$ days. The latter only contained one invariant observational state as no update has been done (infinite sampling step). Figures~\ref{ablation_roc_trainval}, \ref{ablation_pr_trainval}, and \ref{ablation_kappa_trainval} show the resulting ROC and PR curves as well as the $\kappa$ for different thresholds, respectively.

\begin{figure*}[p] 
\begin{center}
\begin{tabular}{cccc}
& SAR & Optical & SAR \& Optical\\
& ($\delta^{SAR} \coloneqq \delta$, $\delta^{OPT} \coloneqq 2 \textrm{ days}$) & ($\delta^{SAR} \coloneqq 2 \textrm{ days}$, $\delta^{OPT} \coloneqq \delta$) & ($\delta^{SAR} \coloneqq \delta$, $\delta^{OPT} \coloneqq \delta$)\\
\multirow{1}{*}[22ex]{\rotatebox[origin=c]{90}{\parbox[c]{3cm}{\centering $\delta = 120 \textrm{ days}$}}} & \includegraphics[width=4.5cm]{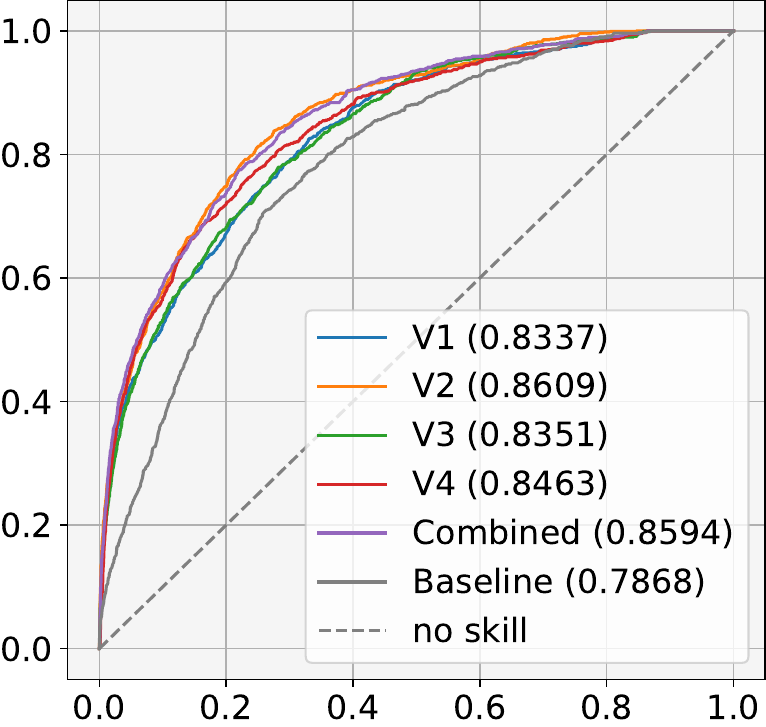} & \includegraphics[width=4.5cm]{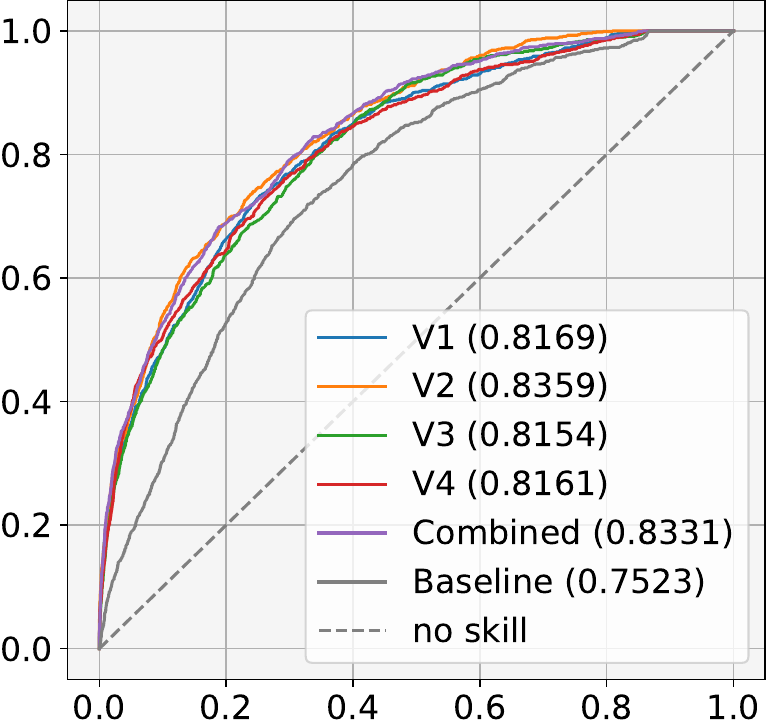} & \includegraphics[width=4.5cm]{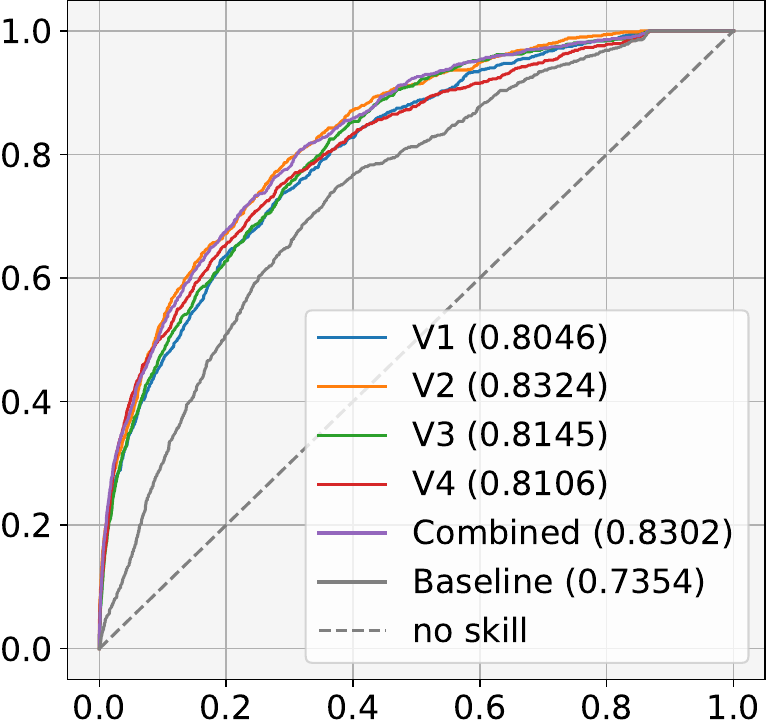}\\
\multirow{1}{*}[22ex]{\rotatebox[origin=c]{90}{\parbox[c]{3cm}{\centering $\delta = \infty \textrm{ days}$}}} & \includegraphics[width=4.5cm]{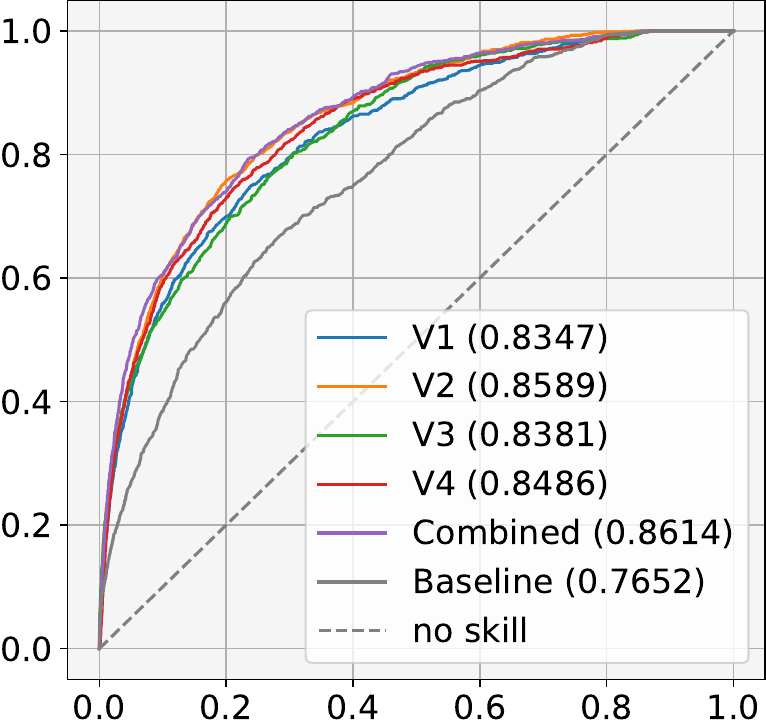} & \includegraphics[width=4.5cm]{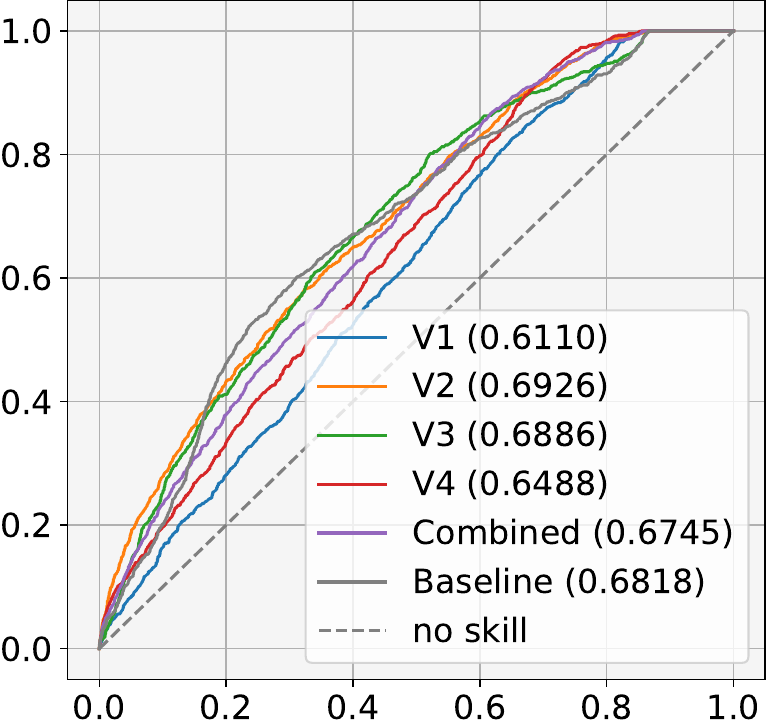} & \includegraphics[width=4.5cm]{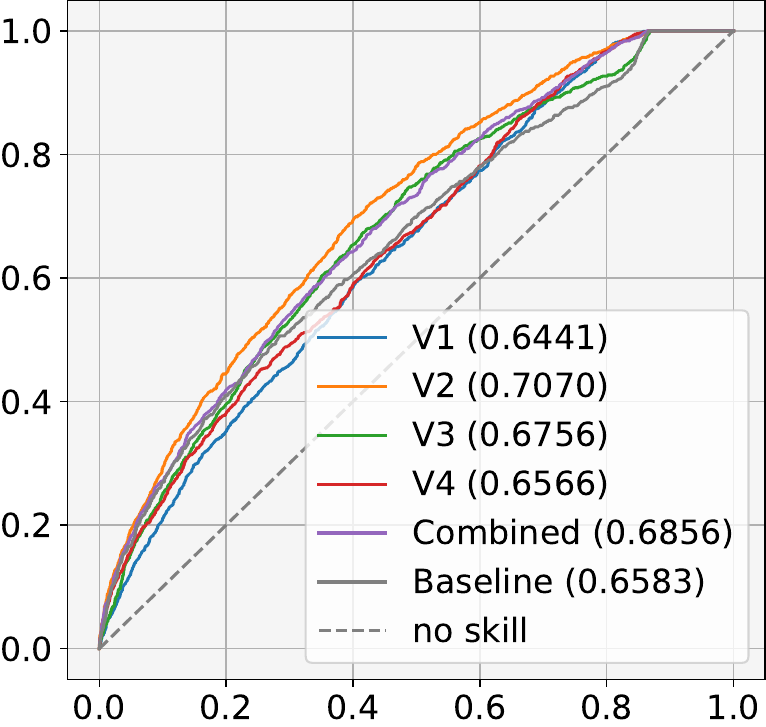}\\
\end{tabular}
\end{center}
\caption{ROC curves of the different models applied on \textit{testing}$^-$ with different sampling steps.}\label{ablation_roc_testing}
\end{figure*}

\begin{figure*}[p] 
\begin{center}
\begin{tabular}{cccc}
& SAR & Optical & SAR \& Optical\\
& ($\delta^{SAR} \coloneqq \delta$, $\delta^{OPT} \coloneqq 2 \textrm{ days}$) & ($\delta^{SAR} \coloneqq 2 \textrm{ days}$, $\delta^{OPT} \coloneqq \delta$) & ($\delta^{SAR} \coloneqq \delta$, $\delta^{OPT} \coloneqq \delta$)\\
\multirow{1}{*}[22ex]{\rotatebox[origin=c]{90}{\parbox[c]{3cm}{\centering $\delta = 120 \textrm{ days}$}}} & \includegraphics[width=4.5cm]{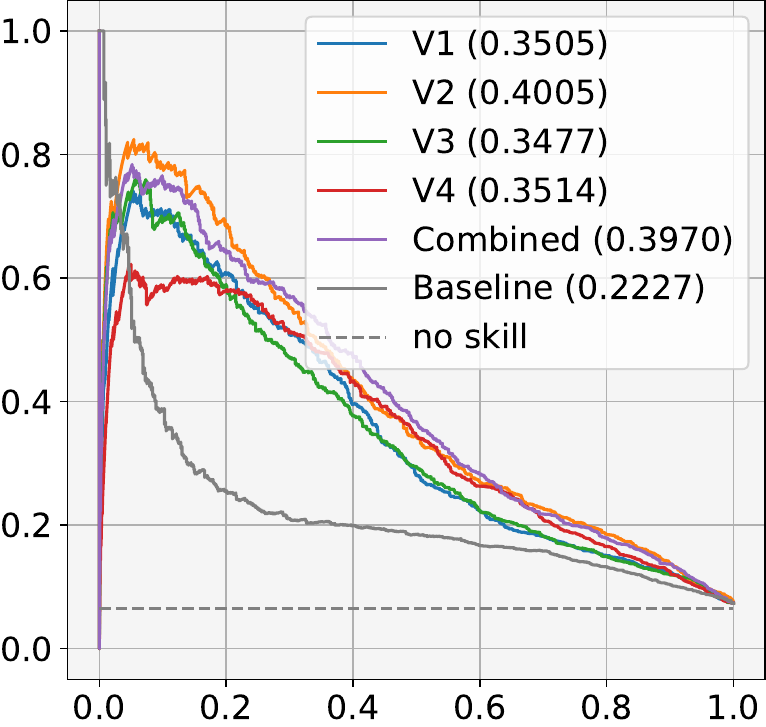} & \includegraphics[width=4.5cm]{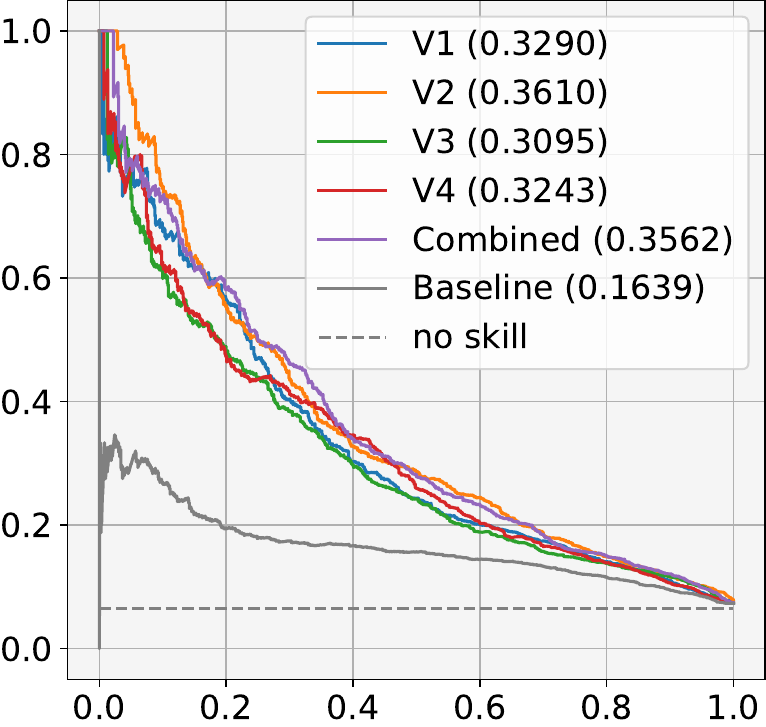} & \includegraphics[width=4.5cm]{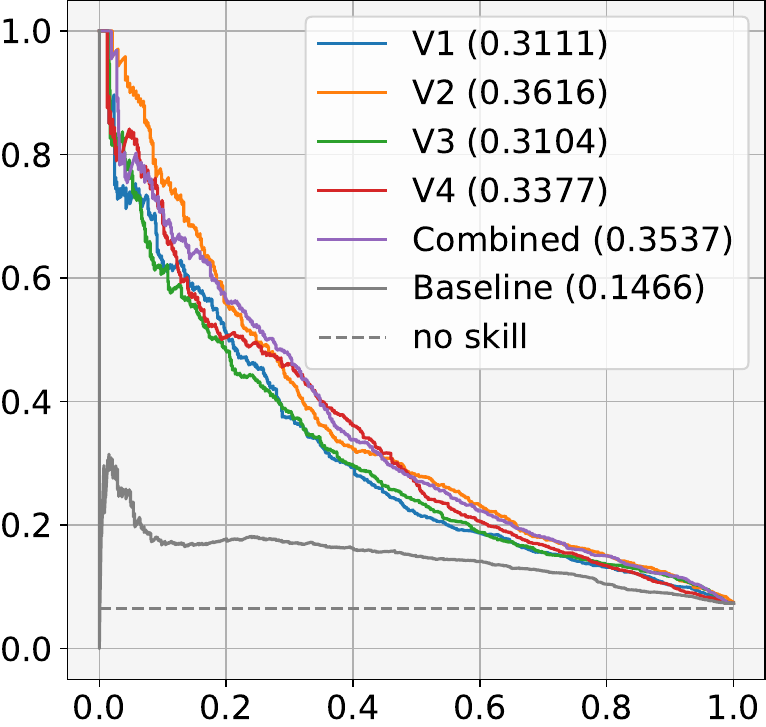}\\
\multirow{1}{*}[22ex]{\rotatebox[origin=c]{90}{\parbox[c]{3cm}{\centering $\delta = \infty \textrm{ days}$}}} & \includegraphics[width=4.5cm]{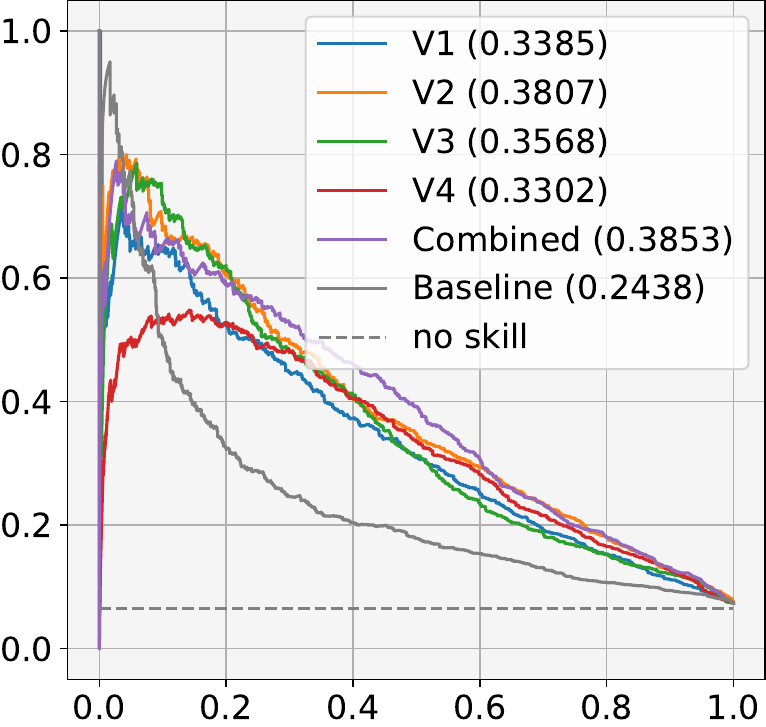} & \includegraphics[width=4.5cm]{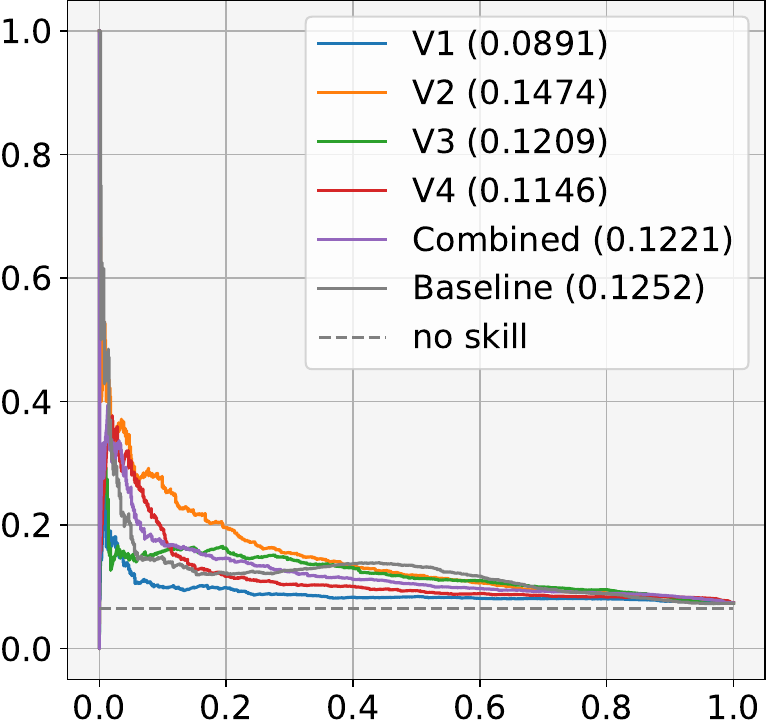} & \includegraphics[width=4.5cm]{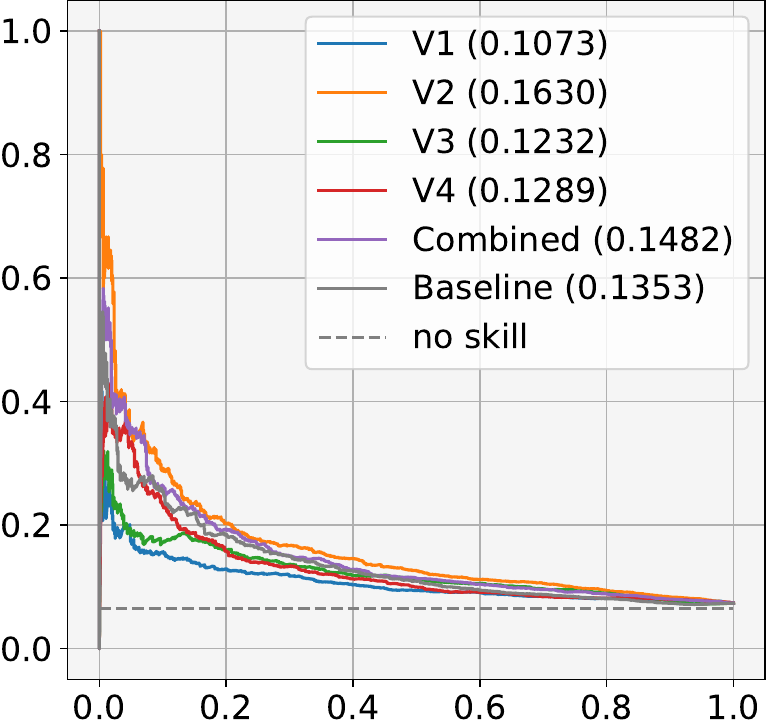}\\
\end{tabular}
\end{center}
\caption{PR curves of the different models applied on \textit{testing}$^-$ with different sampling steps.}\label{ablation_pr_testing}
\end{figure*}

\begin{figure*}[h!tb]
\begin{center}
\begin{tabular}{cccc}
& SAR & Optical & SAR \& Optical\\
& ($\delta^{SAR} \coloneqq \delta$, $\delta^{OPT} \coloneqq 2 \textrm{ days}$) & ($\delta^{SAR} \coloneqq 2 \textrm{ days}$, $\delta^{OPT} \coloneqq \delta$) & ($\delta^{SAR} \coloneqq \delta$, $\delta^{OPT} \coloneqq \delta$)\\
\multirow{1}{*}[22ex]{\rotatebox[origin=c]{90}{\parbox[c]{3cm}{\centering $\delta = 120 \textrm{ days}$}}} & \includegraphics[width=4.5cm]{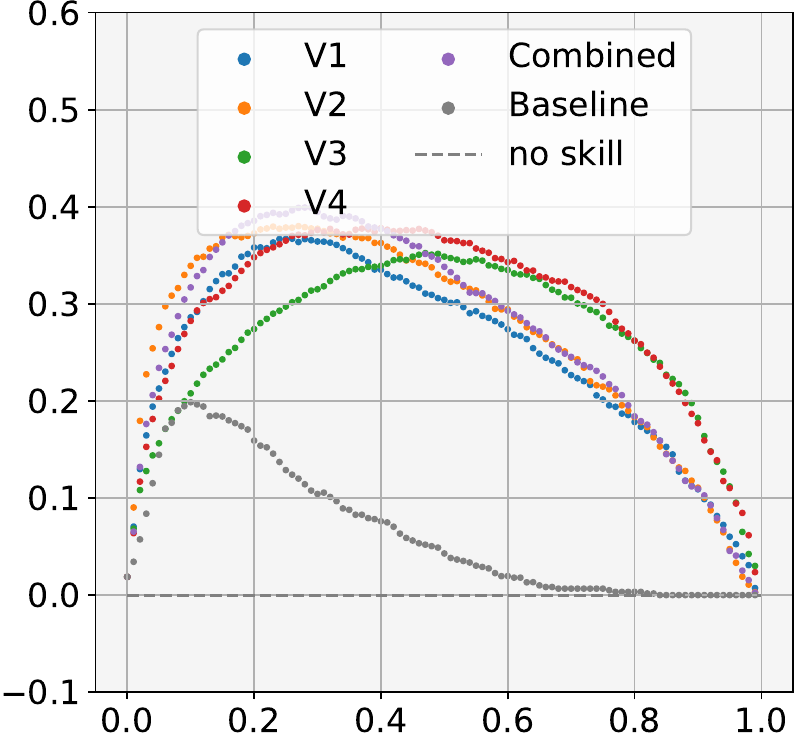} & \includegraphics[width=4.5cm]{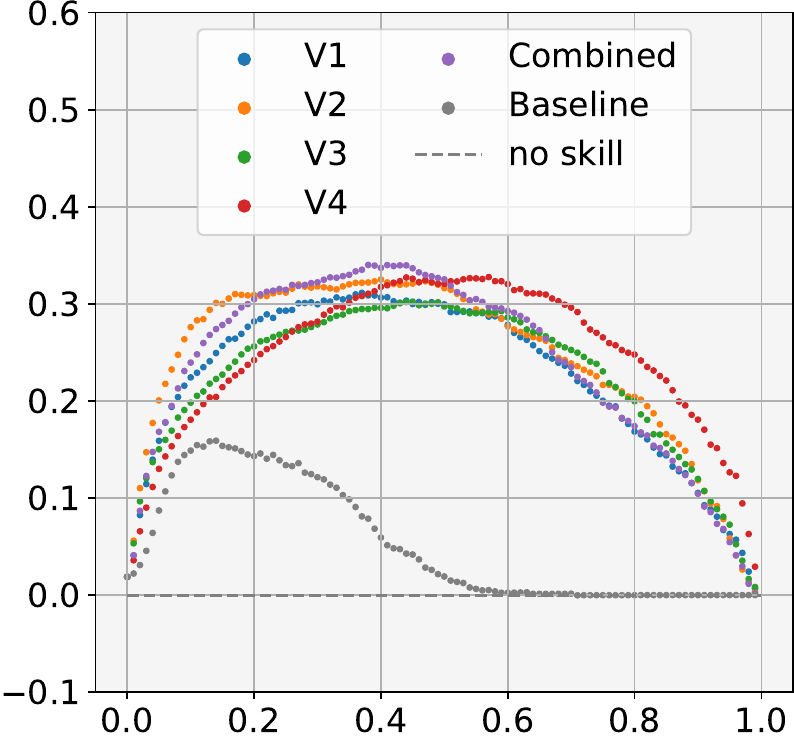} & \includegraphics[width=4.5cm]{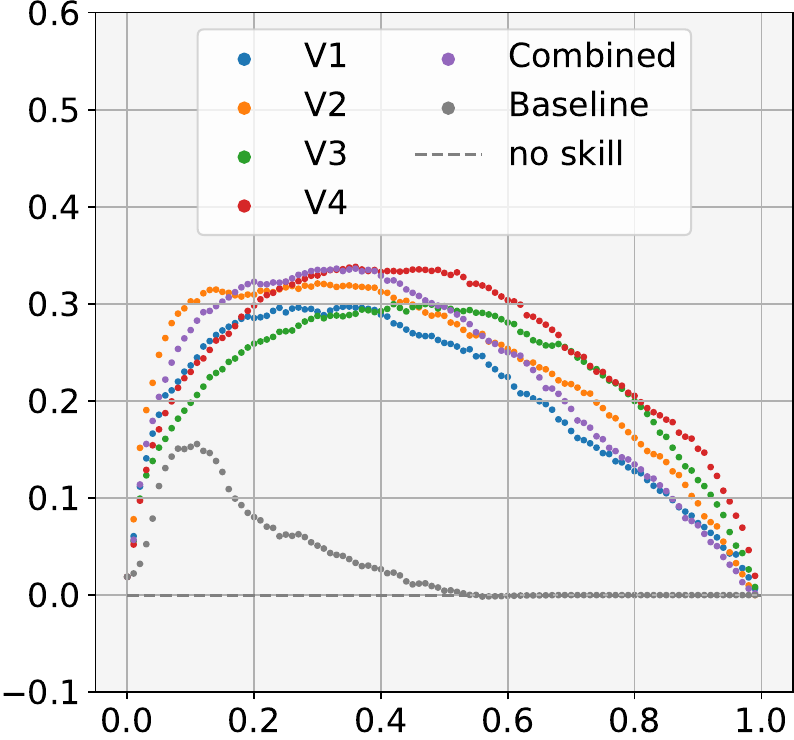}\\
\multirow{1}{*}[22ex]{\rotatebox[origin=c]{90}{\parbox[c]{3cm}{\centering $\delta = \infty \textrm{ days}$}}} & \includegraphics[width=4.5cm]{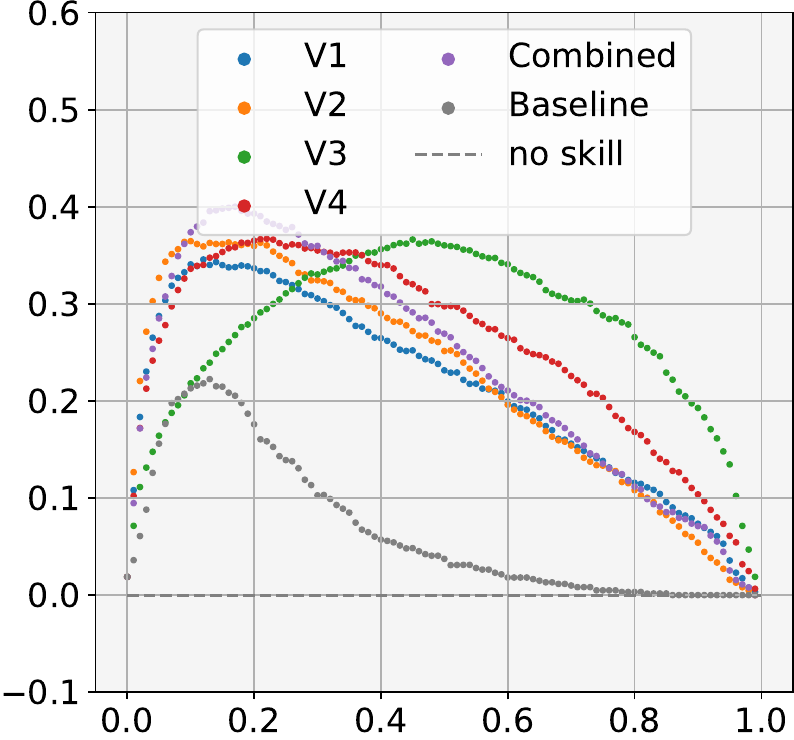} & \includegraphics[width=4.5cm]{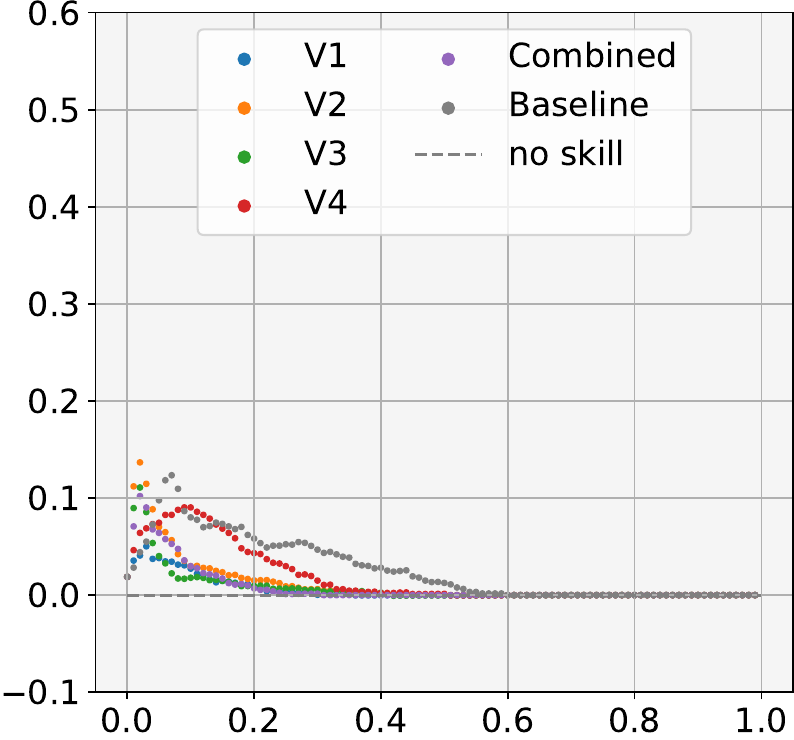} & \includegraphics[width=4.5cm]{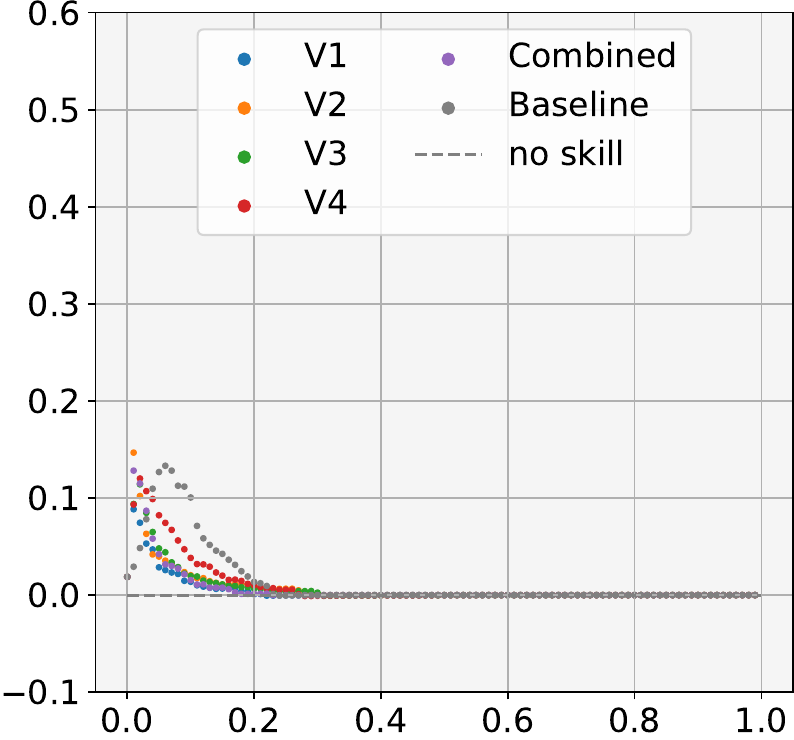}\\
\end{tabular}
\end{center}
\caption{Cohen's Kappa of the different models applied on \textit{testing}$^-$ with different sampling steps and a changing threshold.}\label{ablation_kappa_testing}
\end{figure*}

A similar ablation study was done for the \textit{testing}$^-$ dataset, covering the 2022/23 monitoring period. Due to the shorter time frame, only 2, 120 and $\infty$ days were evaluated as sampling step sizes. The ROC, PR and $\kappa$ curves are shown in Figures~\ref{ablation_roc_testing}, \ref{ablation_pr_testing}, and \ref{ablation_kappa_testing} accordingly.

The analyses showed three different effects. First, optical observations had a larger impact on the prediction performance. On the other hand, SAR observations tended to increase the threshold for which the highest $\kappa$ value was received. This was due to a larger spread of values between changes and no-changes. Hence, SAR observations added more confidence on changes, but they added less in terms of identifying the right changes.

Second, with less or no optical changes, the predictions broke down quicker, which suggested that SAR observations could not compensate. This might have been caused by the lack of differentiation of change of materials.

Third, the reduction of observations and the drop in performance was not linear. Even with few observations, predictions were possible. This was a result of the binary ground truth, which did not have or require information on the frequency or intensity of changes. It was possible for the model to still classify changes with a significantly reduced set of observations. However, fewer observations reduced the temporal resolution and had an impact on localizing a change over time.

The ablation results with the \textit{testing}$^-$ set showed differences to the ones from \textit{trainval}. With reaching $\delta = \infty$ for optical observations, the scores of the ROC and PR curves were not as close to a no-skill model as they were for \textit{trainval}. Similarly, the best threshold for the $\kappa$ values moved towards zero yet were not as low. Also, the baseline performed noticeably better for the \textit{testing}$^-$ dataset. These effects were related to the smaller dataset size and the comparably larger changes that took multiple months. The pre-trained ERCNN-DRS worked well for cases with spatio-temporal large scale changes but underestimated shorter and smaller changes (cf. \cite{doi:10.1080/01431161.2023.2243021}). Nevertheless, the transferred variants and their combination still provide the best performance for the 2022/23 monitoring period. The ablation study results for the full \textit{testing} dataset can be found in Figures~\ref{ablation_roc_full_testing}, \ref{ablation_pr_full_testing}, and \ref{ablation_kappa_full_testing} in the Appendix.

\section{Limitations}\label{limitations}
In this study, we observed the following three limitations. First, building damages were only recognized if multiple pixels were involved and there was a significant change in the surrounding area (e.g., piles of debris, cleanup efforts). With the used resolution of, at best, 10 m/pixel, this translates to approximately an area of $30 \textrm{ m} \times 30 \textrm{ m}$ and upwards.

Furthermore, changes were only detected if learned during the transfer phase. Since the changes in 2022/23 were of a different origin, we might not have been able to detect all changes like the ones shown in the example i) in Figure~\ref{example2} (tile 42:30). Similarly, the selection of samples for the transfer had a direct impact on the detected change patterns. Changes of similar types and patterns biased the trained models and reduced sensitivity to other change types and patterns. This is why large and homogeneous changes were removed from the random tile selection process when creating the \textit{trainval} dataset. Ideally, a better balance of changes and no-changes would help. This, however, is constrained by the available real changes in the transfer period. Therefore, our method works best for cases where urban changes constantly happen.

Lastly, changes were temporally localizable but not down to one month. The reason is the use of six month windows and the unclear response to changes over time. In our previous study~\cite{rs13153000}, we addressed this partially. However, more work is needed to help narrow down changes to the granularity of one month or even less.

\section{Discussion}\label{discussion}
To give more insight into the selected methods and obtained results we elaborate on five different aspects. These are the right choice of a threshold value, how changes are detectable, the effects of lowering the sampling rate, the amount of transfer samples in relation to overfitting, and the size of our network which requires a large dropout rate. With the discussion topics we also aim to help with adopting our methods for other areas and scenarios.

\subsection{Choosing a Threshold}
The quantitative analysis in section~\mbox{\ref{qanalysis}} was carried out with ROC and PR curves. These do not require to select a specific threshold when comparing the binary ground truth with the network's continuous and probabilistic output. They rather describe the network's performance irrespective of a threshold value which gives a more detailed insight. For practical use, however, a threshold needs to be chosen at which an urban change is considered as such. The earlier description of \mbox{$\kappa$} with a sweeping threshold aids in selecting the best value. The \mbox{$\kappa$} values in Figure~\mbox{\ref{trainval_curves}} for the \textit{trainval} set indicate that a threshold value in the range of \mbox{$[0.6,0.9]$} delivers the best performance, depending on the model variant. For the \mbox{\textit{testing}$^-$} dataset, the best \mbox{$\kappa$} values are suggested in the range of \mbox{$[0.3,0.6]$} in Figure~\mbox{\ref{testing_curves}}. Due to the small number of samples in that set, these values are subject to bias. A conservative joint threshold hence would be closer to the value of \mbox{$0.6$} for the combined models.

\subsection{Detectable Changes}
The use of a window in our method provides context to the neural network to identify changes resiliently, while working with error prone level 1 observations~\mbox{\cite{COLUZZI2018426}}. As a result, changes are only detected as they appear within that window. Theoretically, changes can occur that barely are detectable within this limited context. If that happens continuously, i.e., a very slow change over decades, even with a sliding window approach, such a change will likely not be clearly detectable. The window duration of six months was chosen to be a good compromise of being short for localizing changes within a year, and long enough to detect man made construction/destruction activities without a large influence of level 1 data outliers. If very slow long term changes, like decaying buildings over decades, shall be detectable, we suggest to consider a larger window duration, if observational data allows. This would, however, require a new pre-training of ERCNN-DRS to cater for a different window duration (\mbox{$\Delta$}) and also a coarser sampling rate (\mbox{$\delta$}) to reduce memory requirements.

In addition, the pre-training of ERCNN-DRS~\mbox{\cite{rs13153000}} relied on the Enhanced Normalized Difference Impervious Surface Index (ENDISI)~\mbox{\cite{10.1117/1.JRS.13.016502, doi:10.1080/22797254.2020.1820383}} by aliasing impervious surfaces as urban. The transfer learning step then enables to specify more precise patterns that are intended as urban changes and are not (only) driven by impervious surface characteristics. Due to the limitations in time and space, from when and where samples are drawn, which is approx. 536 \mbox{$\textnormal{km}^2$} from Mariupol in the period 2017-2020 for the case at hand, the selected samples impact the overall performance. If little to no urban changes are present in that time and space, the transfer will be limited in tailoring the change patterns of interest. Similarly, the diversity of changes in the transfer period impacts the type of detectable change patterns as well. Hence, our method requires sufficient urban changes to be present before the time frame of interest for the same area~\footnote{We did not yet explore the feasibility of using nearby regions as proxies.} for a useful transfer.

\subsection{Changing the Sampling Rate}
The original two day sampling rate served as a measure to reduce memory requirements. As overlapping swaths of nearby rows can, dependent on the selected region, result in more frequent updates than the expected repeat cycle, redundancies occur. In the scope of urban change monitoring, redundant and partially overlapping observations within a day are less likely to unveil urban changes. Hence, we decided to use a two day sampling step and merged overlapping observations within that step duration. This reduces significantly the memory footprint for each window (from over 150 observations down to less than 80; cf. Figure~\mbox{\ref{num_obs}}). As demonstrated in the ablation studies, a sub-linear decline in performance was observed as the sampling step has been increased. In turn, a coarser sampling can help to reduce the memory needs and still deliver acceptable performance. However, further analyses have to be done to understand if different change patterns are affected differently by a change in the sampling step.

\subsection{Size of Transfer Set and Overfitting}
The herein selected transfer data set size has been chosen to demonstrate the low labeling efforts while providing a converging transfer. We restricted the selected samples to contain a maximum of 15\% of all pixels in a tile as a subject of change. This avoided an over-representation of similar change patterns that would bias the transfer. Our mitigation of bias was to select smaller and thus diverse change patterns instead. As discussed above, the selection of samples influences the transfer and so do their immanent change patterns. For example, a large area construction of similar buildings with the same materials would be over-represented in the transfer samples and thus learned predominantly~\mbox{\cite{7780458}} with the tendency of memorization~\mbox{\cite{pmlr-v70-arpit17a}} that lowers the network's generalization.

\subsection{Size of ERCNN-DRS and Large Dropout Rates}
The ERCNN-DRS is an intentionally small network with less than 100k parameters. It aims at identifying urban changes less by shapes but more by the spectral (optical) or polarized backscatter energy (SAR) responses with a limited receptive field. This was needed due to missing details in the medium resolution observation data and to cater for the limited memory available on GPUs. Since the ratio of the amount of training data to model size under a limited receptive field is large, overfitting happens very early, resulting in a large variance. A large dropout rate was used as a countermeasure. In earlier experiments, we considered more layers (i.e., a deeper architecture) and more convolutional filters. This, however, increased the memory needs so that the resulting network did not converge due to a too small batch size and increased gradient noise~\mbox{\cite{l.2018dont}} or consumed too much memory to be trainable on state of the art GPUs (even if data parallelism was employed).

\section{Conclusion}\label{conclusion}
In this study, we demonstrated the applicability of detecting and monitoring urban changes for the AoI of Mariupol, Ukraine by using transfer learning. It was shown that transferring for the years 2017-2020 with publicly available historic VHR data enabled monitoring during the times of war in 2022/23. During that time frame availability of VHR data was limited and a transfer for that time frame would only have been possible with significant costs. We applied four different transfer variants and their bagged ensemble to both the transfer and monitoring periods, for which the ensemble provided robust results.

We further analyzed the impact of the frequency of available observations in an ablation study. It showed that our method was resilient to even a large loss of observations. However, it also indicated that our method, despite the multi-mode input, is more dependent on optical observations than SAR observations. With this understanding, we can conclude that the loss of Sentinel 1B at the end of December 2021 did not significantly impact the monitoring capabilities of our method.

\subsection*{Disclosures}
No potential conflict of interest was reported by the authors.

\subsection*{Ethical Statement}
Due to the ongoing Russo-Ukrainian War, the selection of locations of visual samples was done with care to minimize risks of influence and harm. To the best of our knowledge, we only selected locations and data that did not give direct insight to the ongoing war, but only documented the resulting (urban) changes. We also would like to underline that our monitoring methods used six-month windows and hence did not and shall not provide real-time information that could be used for military purposes. Our methods were optimized for inertial urban changes that manifest over longer periods.

\subsection* {Acknowledgments}
This research was funded by the Ministry of Education, Youth and Sports from the National Programme of Sustainability (NPS II) project “IT4Innovations excellence in science - LQ1602” and by the IT4Innovations Infrastructure, which is supported by the Ministry of Education, Youth and Sports of the Czech Republic through the e-INFRA CZ (ID:90140), and via the Open Access Grant Competition (OPEN-25-24 and OPEN-27-1). This work was also supported by ESA Network of Resources Initiative (ID:2923ca) to provide access to Sentinel Hub, and Airbus Pl\'eiades.

We also would like to thank CESNET MetaCentrum for providing us access to a DGX H100 node.

\section*{Data Availability Statement}
The labeled data used in this work and trained network models are available on Github \url{https://github.com/It4innovations/urban_change_monitoring_mariupol_ua}. We also provide collateral information like GeoTIFF files of the prediction outputs.

\section*{CRediT Authorship Contribution Statement}
\textbf{G. Zitzlsberger:} Conceptualization, Methodology, Software, Investigation, Data Curation, Writing - Original Draft, Writing - Review \& Editing, Visualization, Funding acquisition, Resources, Supervision, Project administration (90\%). \textbf{M. Podhoranyi:} Validation, Writing - Review \& Editing (10\%).

\bibliographystyle{IEEEtran}
\bibliography{ieee}

\begin{thebibliography}{10}
\providecommand{\url}[1]{#1}
\csname url@samestyle\endcsname
\providecommand{\newblock}{\relax}
\providecommand{\bibinfo}[2]{#2}
\providecommand{\BIBentrySTDinterwordspacing}{\spaceskip=0pt\relax}
\providecommand{\BIBentryALTinterwordstretchfactor}{4}
\providecommand{\BIBentryALTinterwordspacing}{\spaceskip=\fontdimen2\font plus
\BIBentryALTinterwordstretchfactor\fontdimen3\font minus
  \fontdimen4\font\relax}
\providecommand{\BIBforeignlanguage}[2]{{%
\expandafter\ifx\csname l@#1\endcsname\relax
\typeout{** WARNING: IEEEtran.bst: No hyphenation pattern has been}%
\typeout{** loaded for the language `#1'. Using the pattern for}%
\typeout{** the default language instead.}%
\else
\language=\csname l@#1\endcsname
\fi
#2}}
\providecommand{\BIBdecl}{\relax}
\BIBdecl

\bibitem{Shepard1964}
J.~R. Shepard, ``A concept of change detection,'' in \emph{In proceedings 30th
  Annual Meeting of the American Society of Photogrammetry}, vol.~30, no.~4,
  17--20 March 1964, pp. 648--651.

\bibitem{doi:10.1080/01431168908903939}
\BIBentryALTinterwordspacing
A.~SINGH, ``Review article digital change detection techniques using
  remotely-sensed data,'' \emph{International Journal of Remote Sensing},
  vol.~10, no.~6, pp. 989--1003, 1989. [Online]. Available:
  \url{https://doi.org/10.1080/01431168908903939}
\BIBentrySTDinterwordspacing

\bibitem{rs13152869}
\BIBentryALTinterwordspacing
M.~Hemati, M.~Hasanlou, M.~Mahdianpari, and F.~Mohammadimanesh, ``A systematic
  review of landsat data for change detection applications: 50 years of
  monitoring the earth,'' \emph{Remote Sensing}, vol.~13, no.~15, 2021.
  [Online]. Available: \url{https://www.mdpi.com/2072-4292/13/15/2869}
\BIBentrySTDinterwordspacing

\bibitem{rs12101688}
\BIBentryALTinterwordspacing
W.~Shi, M.~Zhang, R.~Zhang, S.~Chen, and Z.~Zhan, ``Change detection based on
  artificial intelligence: State-of-the-art and challenges,'' \emph{Remote
  Sensing}, vol.~12, no.~10, 2020. [Online]. Available:
  \url{https://www.mdpi.com/2072-4292/12/10/1688}
\BIBentrySTDinterwordspacing

\bibitem{rs12152460}
\BIBentryALTinterwordspacing
Y.~You, J.~Cao, and W.~Zhou, ``A survey of change detection methods based on
  remote sensing images for multi-source and multi-objective scenarios,''
  \emph{Remote Sensing}, vol.~12, no.~15, 2020. [Online]. Available:
  \url{https://www.mdpi.com/2072-4292/12/15/2460}
\BIBentrySTDinterwordspacing

\bibitem{9136674}
L.~Khelifi and M.~Mignotte, ``Deep learning for change detection in remote
  sensing images: Comprehensive review and meta-analysis,'' \emph{IEEE Access},
  vol.~8, pp. 126\,385--126\,400, 2020.

\bibitem{doi:10.1080/10095020.2022.2085633}
\BIBentryALTinterwordspacing
T.~Bai, L.~Wang, D.~Yin, K.~Sun, Y.~Chen, W.~Li, and D.~Li, ``Deep learning for
  change detection in remote sensing: a review,'' \emph{Geo-spatial Information
  Science}, vol.~0, no.~0, pp. 1--27, 2022. [Online]. Available:
  \url{https://doi.org/10.1080/10095020.2022.2085633}
\BIBentrySTDinterwordspacing

\bibitem{rs14040871}
\BIBentryALTinterwordspacing
A.~Shafique, G.~Cao, Z.~Khan, M.~Asad, and M.~Aslam, ``Deep learning-based
  change detection in remote sensing images: A review,'' \emph{Remote Sensing},
  vol.~14, no.~4, 2022. [Online]. Available:
  \url{https://www.mdpi.com/2072-4292/14/4/871}
\BIBentrySTDinterwordspacing

\bibitem{rs14071552}
\BIBentryALTinterwordspacing
H.~Jiang, M.~Peng, Y.~Zhong, H.~Xie, Z.~Hao, J.~Lin, X.~Ma, and X.~Hu, ``A
  survey on deep learning-based change detection from high-resolution remote
  sensing images,'' \emph{Remote Sensing}, vol.~14, no.~7, 2022. [Online].
  Available: \url{https://www.mdpi.com/2072-4292/14/7/1552}
\BIBentrySTDinterwordspacing

\bibitem{rs15082092}
\BIBentryALTinterwordspacing
E.~J. Parelius, ``A review of deep-learning methods for change detection in
  multispectral remote sensing images,'' \emph{Remote Sensing}, vol.~15, no.~8,
  2023. [Online]. Available: \url{https://www.mdpi.com/2072-4292/15/8/2092}
\BIBentrySTDinterwordspacing

\bibitem{rs11020173}
\BIBentryALTinterwordspacing
A.~Lehner and T.~Blaschke, ``A generic classification scheme for urban
  structure types,'' \emph{Remote Sensing}, vol.~11, no.~2, 2019. [Online].
  Available: \url{https://www.mdpi.com/2072-4292/11/2/173}
\BIBentrySTDinterwordspacing

\bibitem{rs13153000}
\BIBentryALTinterwordspacing
G.~Zitzlsberger, M.~Podhorányi, V.~Svatoň, M.~Lazecký, and J.~Martinovič,
  ``Neural network-based urban change monitoring with deep-temporal
  multispectral and sar remote sensing data,'' \emph{Remote Sensing}, vol.~13,
  no.~15, 2021. [Online]. Available:
  \url{https://www.mdpi.com/2072-4292/13/15/3000}
\BIBentrySTDinterwordspacing

\bibitem{doi:10.1080/01431161.2023.2243021}
G.~Zitzlsberger, M.~Podhoranyi, and J.~Martinovic, ``A practically feasible
  transfer learning method for deep-temporal urban change monitoring,''
  \emph{International Journal of Remote Sensing}, 2023.

\bibitem{ZITZLSBERGER2023101369}
\BIBentryALTinterwordspacing
G.~Zitzlsberger, M.~Podhoranyi, and J.~Martinovič, ``rsdtlib: Remote sensing
  with deep-temporal data library,'' \emph{SoftwareX}, vol.~22, p. 101369,
  2023. [Online]. Available:
  \url{https://www.sciencedirect.com/science/article/pii/S2352711023000651}
\BIBentrySTDinterwordspacing

\bibitem{10.5555/1953048.2078186}
R.~Collobert, J.~Weston, L.~Bottou, M.~Karlen, K.~Kavukcuoglu, and P.~Kuksa,
  ``Natural language processing (almost) from scratch,'' \emph{J. Mach. Learn.
  Res.}, vol.~12, no. null, p. 2493–2537, nov 2011.

\bibitem{breiman96}
L.~Breiman, ``Bagging predictors,'' \emph{Machine Learning}, vol.~24, no.~2,
  pp. 123--140, 1996.

\bibitem{sergeev2018horovod}
A.~Sergeev and M.~D. Balso, ``Horovod: fast and easy distributed deep learning
  in {TensorFlow},'' \emph{arXiv preprint arXiv:1802.05799}, 2018.

\bibitem{chen:2016:iclr}
\BIBentryALTinterwordspacing
J.~Chen, R.~Monga, S.~Bengio, and R.~Jozefowicz, ``Revisiting distributed
  synchronous {SGD},'' in \emph{Workshop Track of the International Conference
  on Learning Representations, {ICLR}}, 2016. [Online]. Available:
  \url{publications/ps/chen_2016_iclr.ps.gz}
\BIBentrySTDinterwordspacing

\bibitem{NIPS2010_abea47ba}
\BIBentryALTinterwordspacing
M.~Zinkevich, M.~Weimer, L.~Li, and A.~Smola, ``Parallelized stochastic
  gradient descent,'' in \emph{Advances in Neural Information Processing
  Systems}, J.~Lafferty, C.~Williams, J.~Shawe-Taylor, R.~Zemel, and
  A.~Culotta, Eds., vol.~23.\hskip 1em plus 0.5em minus 0.4em\relax Curran
  Associates, Inc., 2010. [Online]. Available:
  \url{https://proceedings.neurips.cc/paper/2010/file/abea47ba24142ed16b7d8fbf2c740e0d-Paper.pdf}
\BIBentrySTDinterwordspacing

\bibitem{tanimoto-compl}
F.~Diakogiannis, F.~Waldner, P.~Caccetta, and C.~Wu,
  ``\BIBforeignlanguage{English}{Resunet-a: A deep learning framework for
  semantic segmentation of remotely sensed data},''
  \emph{\BIBforeignlanguage{English}{ISPRS Journal of Photogrammetry and Remote
  Sensing}}, vol. 162, pp. 94--114, Apr. 2020.

\bibitem{10.3389/fnins.2020.00065}
\BIBentryALTinterwordspacing
G.~A. Reina, R.~Panchumarthy, S.~P. Thakur, A.~Bastidas, and S.~Bakas,
  ``Systematic evaluation of image tiling adverse effects on deep learning
  semantic segmentation,'' \emph{Frontiers in Neuroscience}, vol.~14, p.~65,
  2020. [Online]. Available:
  \url{https://www.frontiersin.org/article/10.3389/fnins.2020.00065}
\BIBentrySTDinterwordspacing

\bibitem{DBLP:journals/corr/abs-1805-12219}
\BIBentryALTinterwordspacing
B.~Huang, D.~Reichman, L.~M. Collins, K.~Bradbury, and J.~M. Malof, ``Dense
  labeling of large remote sensing imagery with convolutional neural networks:
  a simple and faster alternative to stitching output label maps,''
  \emph{CoRR}, vol. abs/1805.12219, 2018. [Online]. Available:
  \url{http://arxiv.org/abs/1805.12219}
\BIBentrySTDinterwordspacing

\bibitem{huang2019tiling}
------, ``Tiling and stitching segmentation output for remote sensing: Basic
  challenges and recommendations,'' 2019.

\bibitem{Isensee2021}
\BIBentryALTinterwordspacing
F.~Isensee, P.~F. Jaeger, S.~A.~A. Kohl, J.~Petersen, and K.~H. Maier-Hein,
  ``nnu-net: a self-configuring method for deep learning-based biomedical image
  segmentation,'' \emph{Nature Methods}, vol.~18, no.~2, pp. 203--211, Feb
  2021. [Online]. Available: \url{https://doi.org/10.1038/s41592-020-01008-z}
\BIBentrySTDinterwordspacing

\bibitem{10.5555/2969033.2969197}
J.~Yosinski, J.~Clune, Y.~Bengio, and H.~Lipson, ``How transferable are
  features in deep neural networks?'' in \emph{Proceedings of the 27th
  International Conference on Neural Information Processing Systems - Volume
  2}, ser. NIPS'14.\hskip 1em plus 0.5em minus 0.4em\relax Cambridge, MA, USA:
  MIT Press, 2014, p. 3320–3328.

\bibitem{10.5555/2595566.2595576}
M.~Majnik and Z.~Bosni\'{c}, ``Roc analysis of classifiers in machine learning:
  A survey,'' \emph{Intell. Data Anal.}, vol.~17, no.~3, p. 531–558, may
  2013.

\bibitem{10.1145/65943.65945}
\BIBentryALTinterwordspacing
V.~Raghavan, P.~Bollmann, and G.~S. Jung, ``A critical investigation of recall
  and precision as measures of retrieval system performance,'' \emph{ACM Trans.
  Inf. Syst.}, vol.~7, no.~3, p. 205–229, jul 1989. [Online]. Available:
  \url{https://doi.org/10.1145/65943.65945}
\BIBentrySTDinterwordspacing

\bibitem{10.1145/1143844.1143874}
\BIBentryALTinterwordspacing
J.~Davis and M.~Goadrich, ``The relationship between precision-recall and roc
  curves,'' in \emph{Proceedings of the 23rd International Conference on
  Machine Learning}, ser. ICML '06.\hskip 1em plus 0.5em minus 0.4em\relax New
  York, NY, USA: Association for Computing Machinery, 2006, p. 233–240.
  [Online]. Available: \url{https://doi.org/10.1145/1143844.1143874}
\BIBentrySTDinterwordspacing

\bibitem{cohen1960coefficient}
J.~Cohen, ``A coefficient of agreement for nominal scales,'' \emph{Educational
  and psychological measurement}, vol.~20, no.~1, pp. 37--46, 1960.

\bibitem{COLUZZI2018426}
\BIBentryALTinterwordspacing
R.~Coluzzi, V.~Imbrenda, M.~Lanfredi, and T.~Simoniello, ``A first assessment
  of the sentinel-2 level 1-c cloud mask product to support informed surface
  analyses,'' \emph{Remote Sensing of Environment}, vol. 217, pp. 426--443,
  2018. [Online]. Available:
  \url{https://www.sciencedirect.com/science/article/pii/S0034425718303742}
\BIBentrySTDinterwordspacing

\bibitem{10.1117/1.JRS.13.016502}
\BIBentryALTinterwordspacing
J.~Chen, K.~Yang, S.~Chen, C.~Yang, S.~Zhang, and L.~He, ``{Enhanced normalized
  difference index for impervious surface area estimation at the plateau basin
  scale},'' \emph{Journal of Applied Remote Sensing}, vol.~13, no.~1, pp. 1 --
  19, 2019. [Online]. Available: \url{https://doi.org/10.1117/1.JRS.13.016502}
\BIBentrySTDinterwordspacing

\bibitem{doi:10.1080/22797254.2020.1820383}
\BIBentryALTinterwordspacing
J.~Chen, S.~Chen, C.~Yang, L.~He, M.~Hou, and T.~Shi, ``A comparative study of
  impervious surface extraction using sentinel-2 imagery,'' \emph{European
  Journal of Remote Sensing}, vol.~53, no.~1, pp. 274--292, 2020. [Online].
  Available: \url{https://doi.org/10.1080/22797254.2020.1820383}
\BIBentrySTDinterwordspacing

\bibitem{7780458}
A.~Shrivastava, A.~Gupta, and R.~Girshick, ``Training region-based object
  detectors with online hard example mining,'' in \emph{2016 IEEE Conference on
  Computer Vision and Pattern Recognition (CVPR)}, 2016, pp. 761--769.

\bibitem{pmlr-v70-arpit17a}
\BIBentryALTinterwordspacing
D.~Arpit, S.~Jastrz{\k{e}}bski, N.~Ballas, D.~Krueger, E.~Bengio, M.~S. Kanwal,
  T.~Maharaj, A.~Fischer, A.~Courville, Y.~Bengio, and S.~Lacoste-Julien, ``A
  closer look at memorization in deep networks,'' in \emph{Proceedings of the
  34th International Conference on Machine Learning}, ser. Proceedings of
  Machine Learning Research, D.~Precup and Y.~W. Teh, Eds., vol.~70.\hskip 1em
  plus 0.5em minus 0.4em\relax PMLR, 06--11 Aug 2017, pp. 233--242. [Online].
  Available: \url{https://proceedings.mlr.press/v70/arpit17a.html}
\BIBentrySTDinterwordspacing

\bibitem{l.2018dont}
\BIBentryALTinterwordspacing
S.~L. Smith, P.-J. Kindermans, and Q.~V. Le, ``Don't decay the learning rate,
  increase the batch size,'' in \emph{International Conference on Learning
  Representations}, 2018. [Online]. Available:
  \url{https://openreview.net/forum?id=B1Yy1BxCZ}
\BIBentrySTDinterwordspacing

\end{thebibliography}

\newpage
\onecolumn
\appendix

\section{ERCNN-DRS Architecture}
\label{ercnn-drs}
\newsavebox{\tempbox}
\begin{figure*}[hbt]
\newcolumntype{R}[2]{%
    >{\adjustbox{angle=#1,lap=\width-(#2)}\bgroup}%
    l%
    <{\egroup}%
}
\newcommand*\rot{\multicolumn{1}{R{45}{1em}}}
\centering
\subfigure{
\begin{minipage}{0.42\textwidth}
\includegraphics[width=\textwidth]{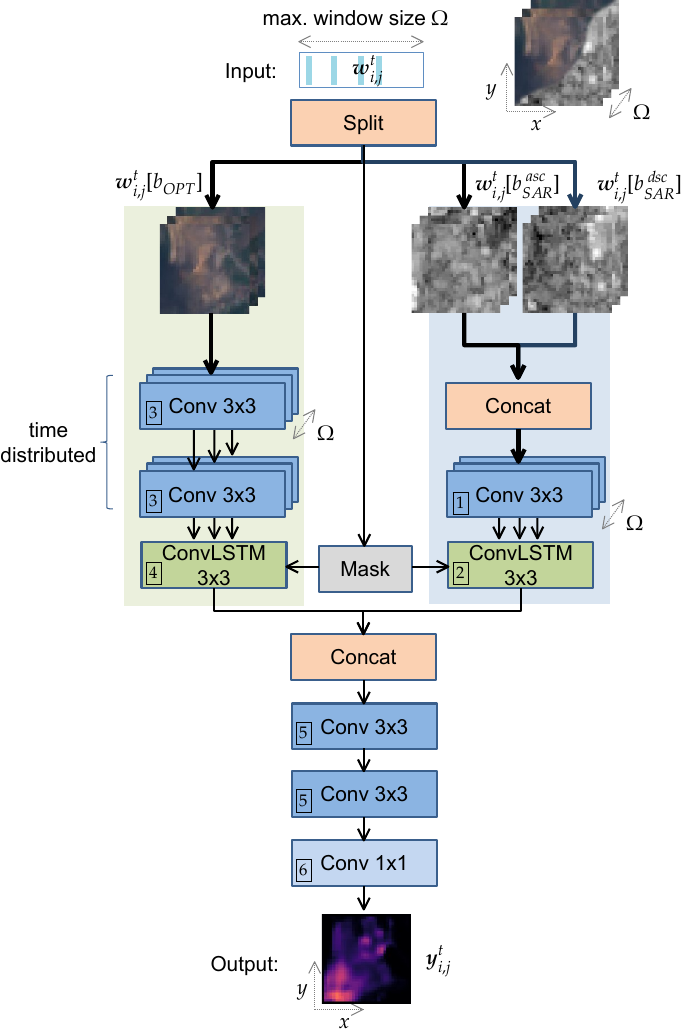}
\end{minipage}\hspace{5mm}}
\subfigure{
\begin{minipage}{0.50\textwidth}
{\small
\begin{tabular}{cl|lllll}
\toprule
&                        & \multicolumn{5}{c}{\textbf{Hyper-Parameters}}\\
&                        & \rot{Filters} & \rot{Kernel}       & \rot{Stride}       & \rot{Activation(s)} & \rot{Dropout}\\
\midrule
\multirow{10}{0mm}{\rotatebox[origin=c]{90}{\parbox[c]{4cm}{\centering \textbf{Configurations}}}}
&\textbf{\boxed{1}}      & 10      & $3 \times 3$ & $1 \times 1$ & ReLU          &\\
&\textbf{\boxed{2}}      & 10      & $3 \times 3$ & $1 \times 1$ & tanh,         & 0.4\\
&                        &         &              &              & hard          &\\
&                        &         &              &              & sigmoid       &\\
&\textbf{\boxed{3}}      & 26      & $3 \times 3$ & $1 \times 1$ & ReLU          &\\
&\textbf{\boxed{4}}      & 26      & $3 \times 3$ & $1 \times 1$ & tanh,         & 0.4\\
&                        &         &              &              & hard          &\\
&                        &         &              &              & sigmoid       &\\
&\textbf{\boxed{5}}      &  8      & $3 \times 3$ & $1 \times 1$ & ReLU          &\\
&\textbf{\boxed{6}}      &  1      & $1 \times 1$ & $1 \times 1$ & sigmoid       &\\
\bottomrule
\end{tabular}}
\end{minipage}}
\caption{The architecture as inherited from the pre-training stage. In the transfer learning stage, all layers were trained. A windowed multi-modal input was expected (green background: multispectral optical; blue background: SAR in ascending and descending orbit directions). Hyper-parameters of the respective layers are detailed in the table on the right.}
\label{ercnn-drs_arch}
\end{figure*}

\section{Metrics for the \textit{testing} dataset}
\begin{figure*}
\begin{center}
\includegraphics[height=0.3\textwidth]{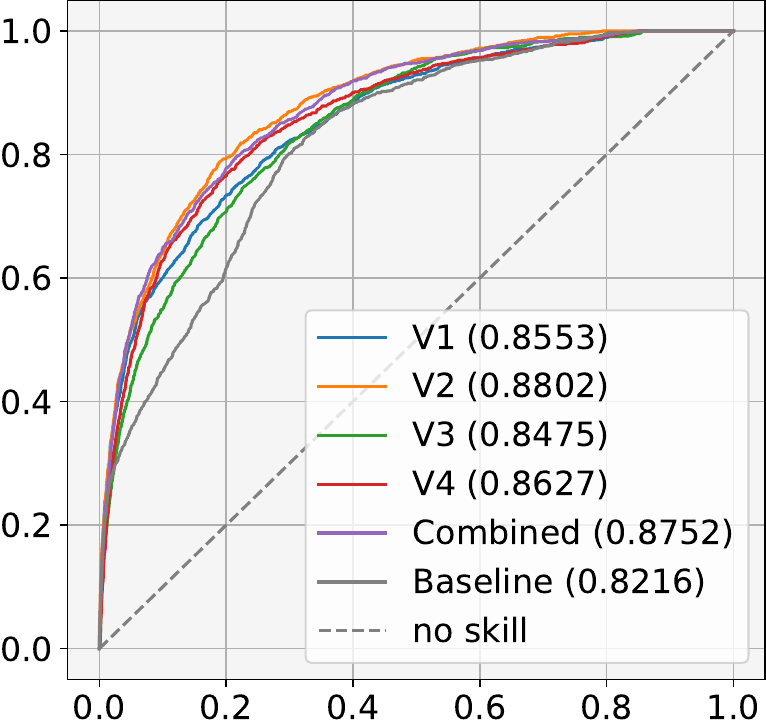}\hspace{2mm}\includegraphics[height=0.3\textwidth]{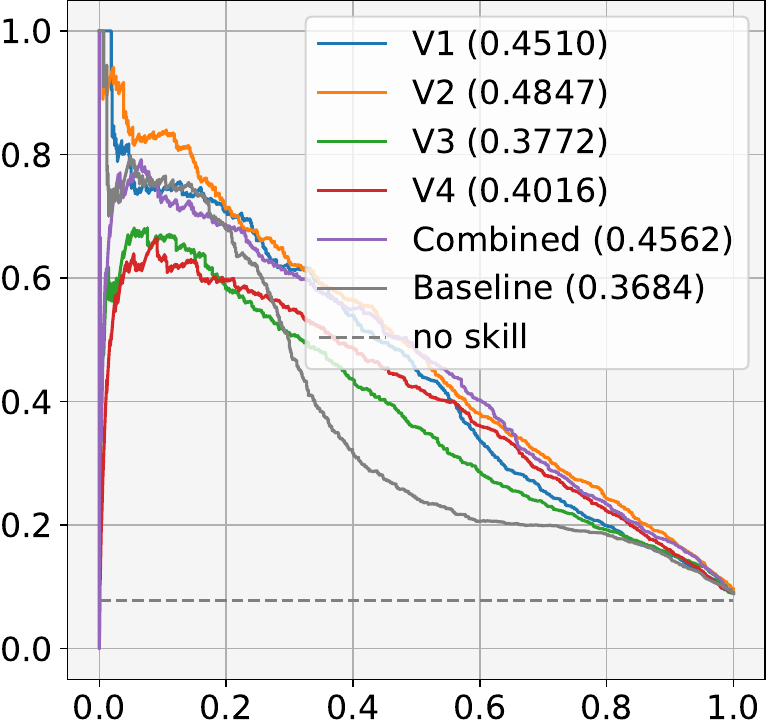}\hspace{2mm}\includegraphics[height=0.3\textwidth]{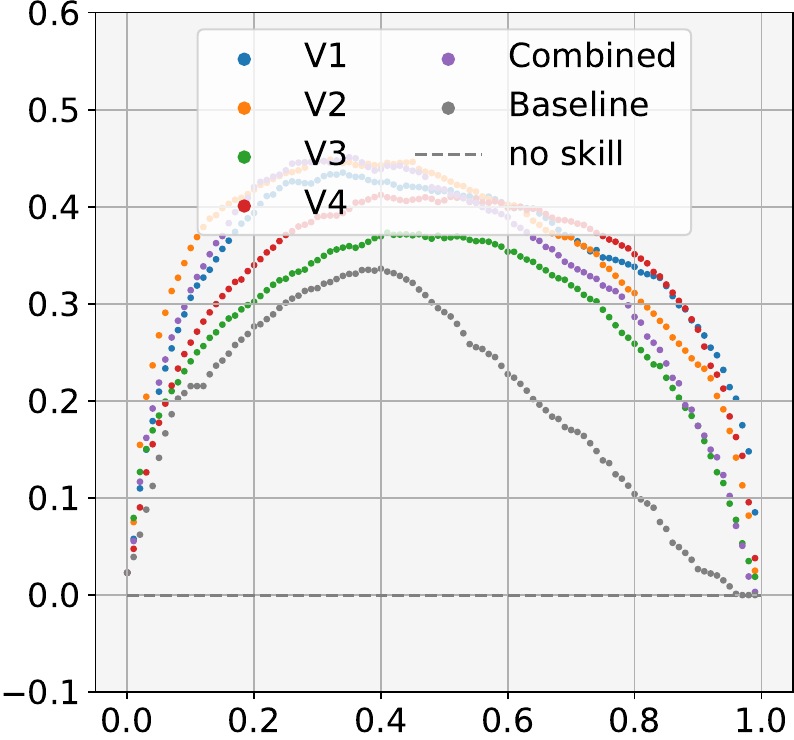}
\end{center} \caption{For the \textit{testing} dataset set: ROC (left) and PR (middle) curves; Cohen's Kappa is shown for different thresholds (right). Area under the ROC/PR curves are in parenthesis.}\label{full_testing_curves}
\end{figure*}

\section{Ablation metrics for the \textit{testing} dataset}
\begin{figure*}
\begin{center}
\begin{tabular}{cccc}
& SAR & Optical & SAR \& Optical\\
& ($\delta^{SAR} \coloneqq \delta$, $\delta^{OPT} \coloneqq 2 \textrm{ days}$) & ($\delta^{SAR} \coloneqq 2 \textrm{ days}$, $\delta^{OPT} \coloneqq \delta$) & ($\delta^{SAR} \coloneqq \delta$, $\delta^{OPT} \coloneqq \delta$)\\
\multirow{1}{*}[22ex]{\rotatebox[origin=c]{90}{\parbox[c]{3cm}{\centering $\delta = 120 \textrm{ days}$}}} & \includegraphics[width=4.5cm]{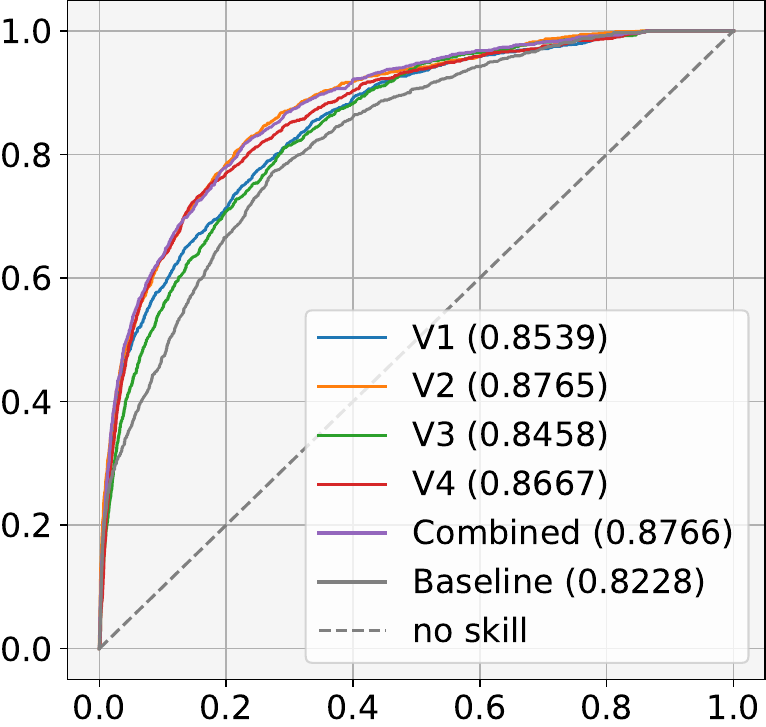} & \includegraphics[width=4.5cm]{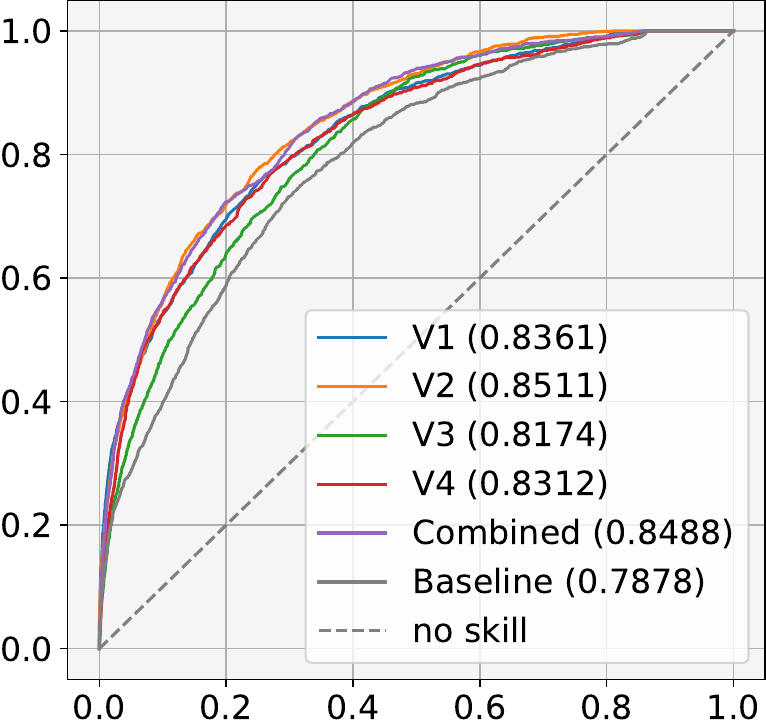} & \includegraphics[width=4.5cm]{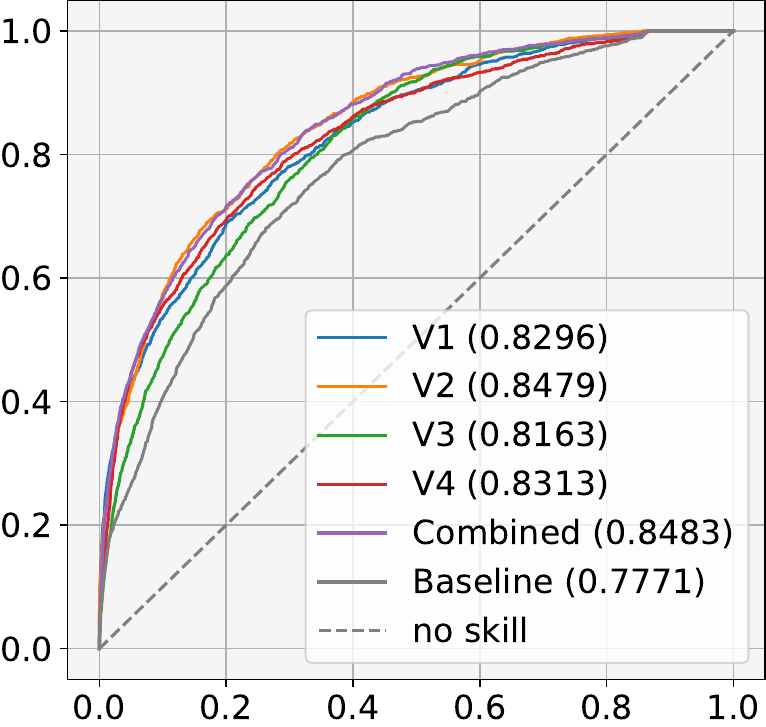}\\
\multirow{1}{*}[22ex]{\rotatebox[origin=c]{90}{\parbox[c]{3cm}{\centering $\delta = \infty \textrm{ days}$}}} & \includegraphics[width=4.5cm]{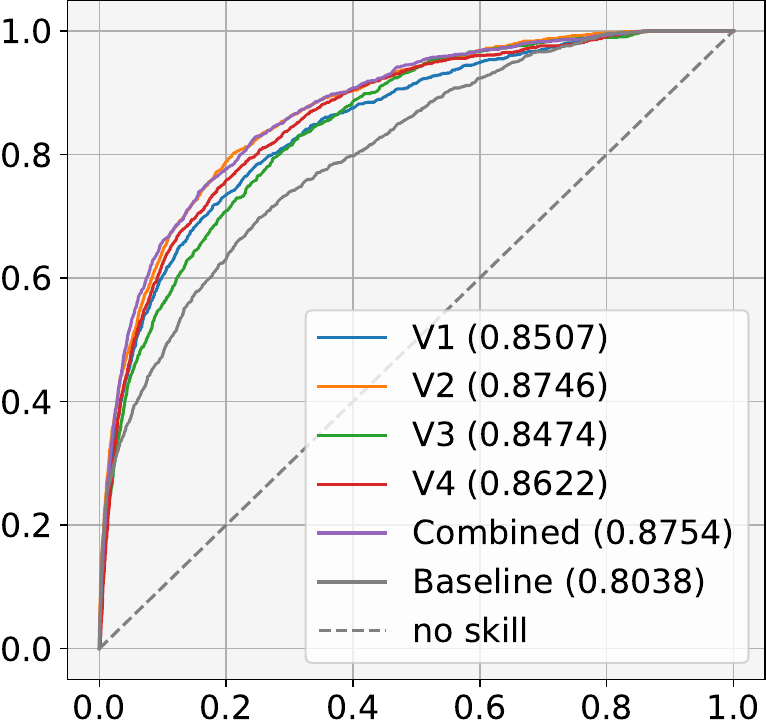} & \includegraphics[width=4.5cm]{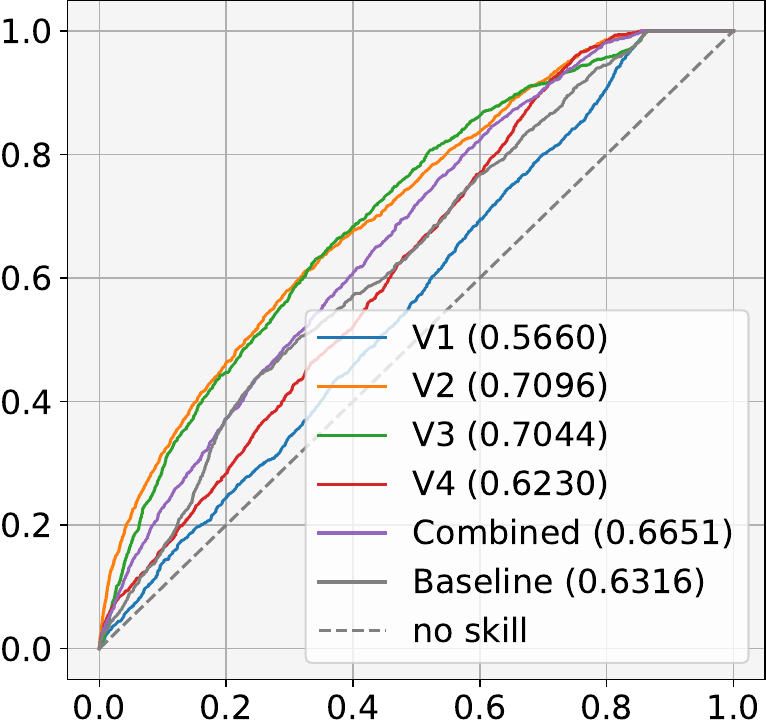} & \includegraphics[width=4.5cm]{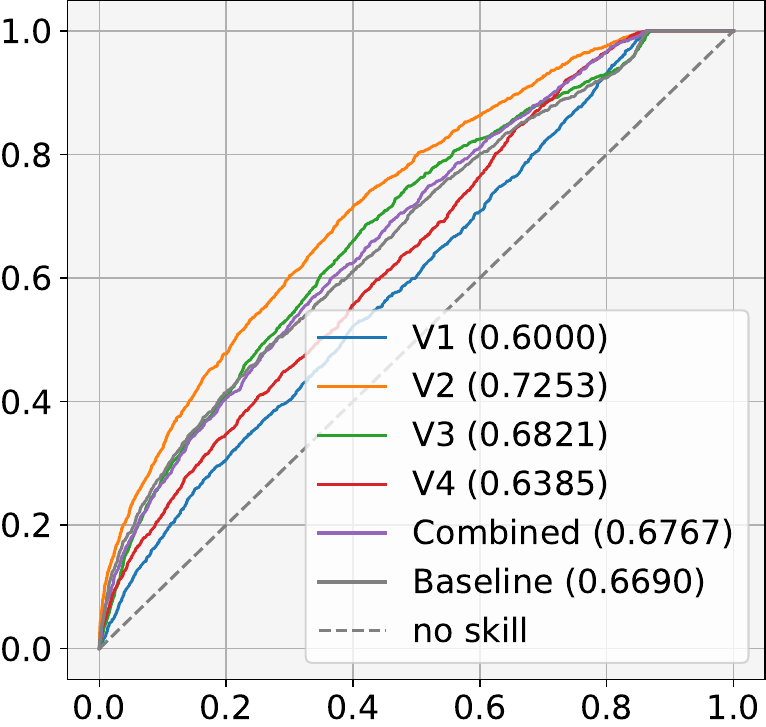}\\
\end{tabular}
\end{center}
\caption{ROC curves of the different models applied on \textit{testing} with different sampling steps.}\label{ablation_roc_full_testing}
\end{figure*}

\begin{figure*}
\begin{center}
\begin{tabular}{cccc}
& SAR & Optical & SAR \& Optical\\
& ($\delta^{SAR} \coloneqq \delta$, $\delta^{OPT} \coloneqq 2 \textrm{ days}$) & ($\delta^{SAR} \coloneqq 2 \textrm{ days}$, $\delta^{OPT} \coloneqq \delta$) & ($\delta^{SAR} \coloneqq \delta$, $\delta^{OPT} \coloneqq \delta$)\\
\multirow{1}{*}[22ex]{\rotatebox[origin=c]{90}{\parbox[c]{3cm}{\centering $\delta = 120 \textrm{ days}$}}} & \includegraphics[width=4.5cm]{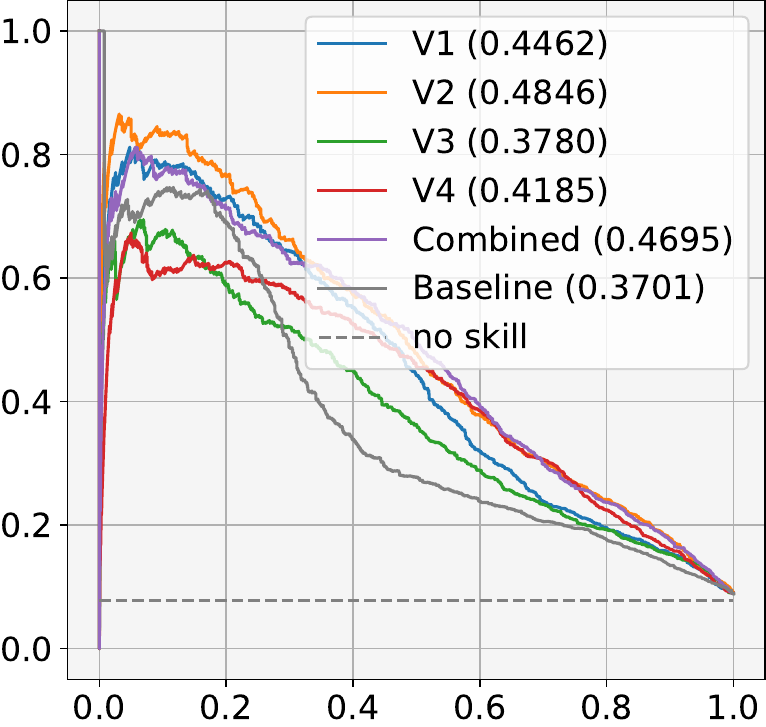} & \includegraphics[width=4.5cm]{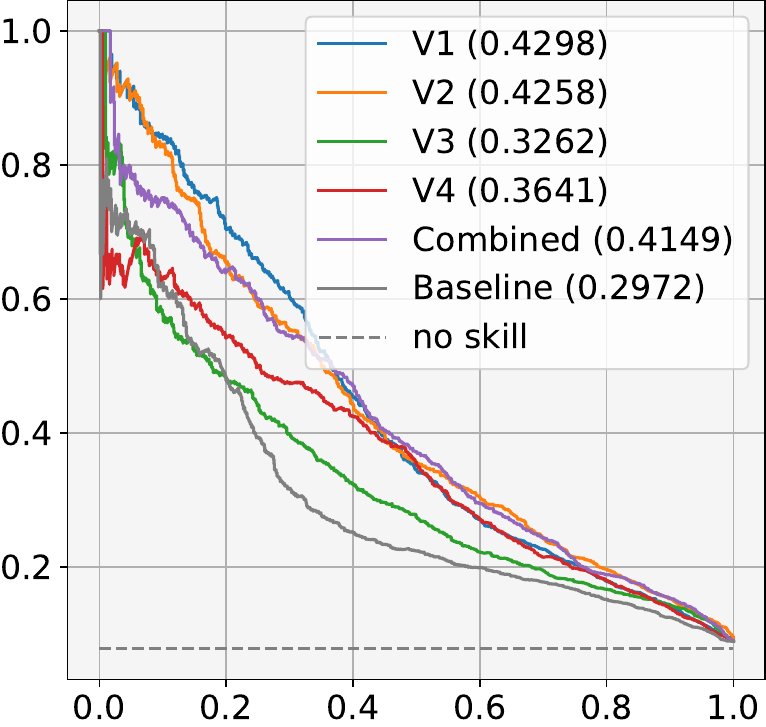} & \includegraphics[width=4.5cm]{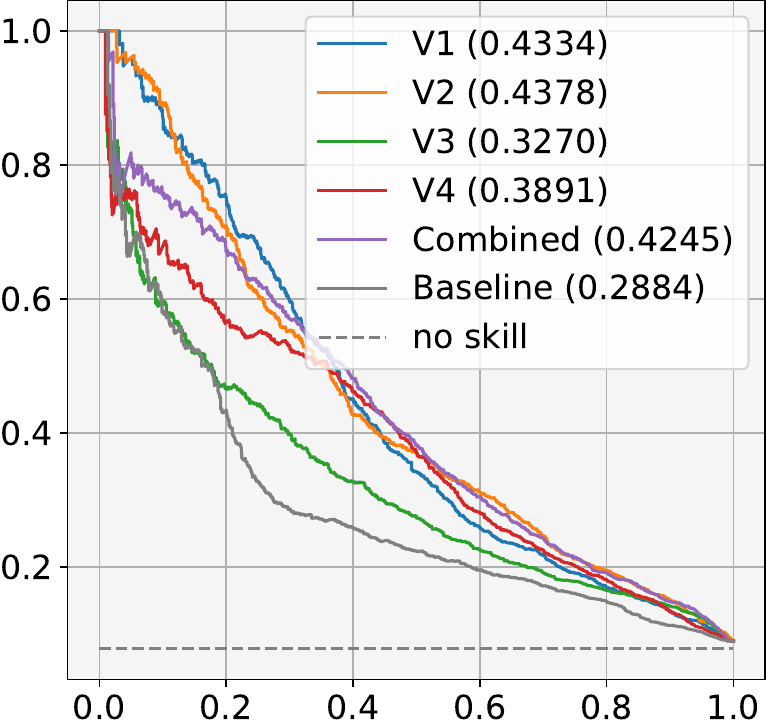}\\
\multirow{1}{*}[22ex]{\rotatebox[origin=c]{90}{\parbox[c]{3cm}{\centering $\delta = \infty \textrm{ days}$}}} & \includegraphics[width=4.5cm]{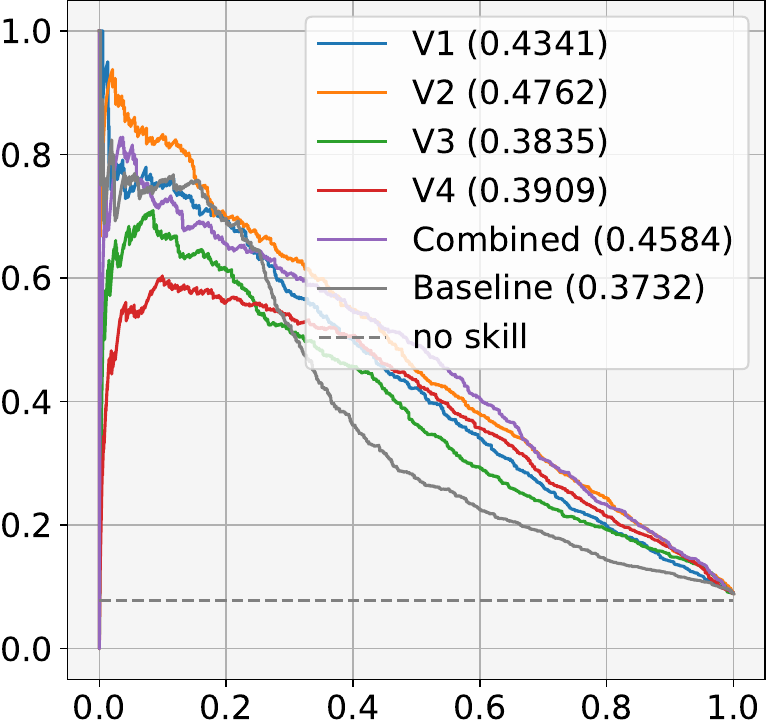} & \includegraphics[width=4.5cm]{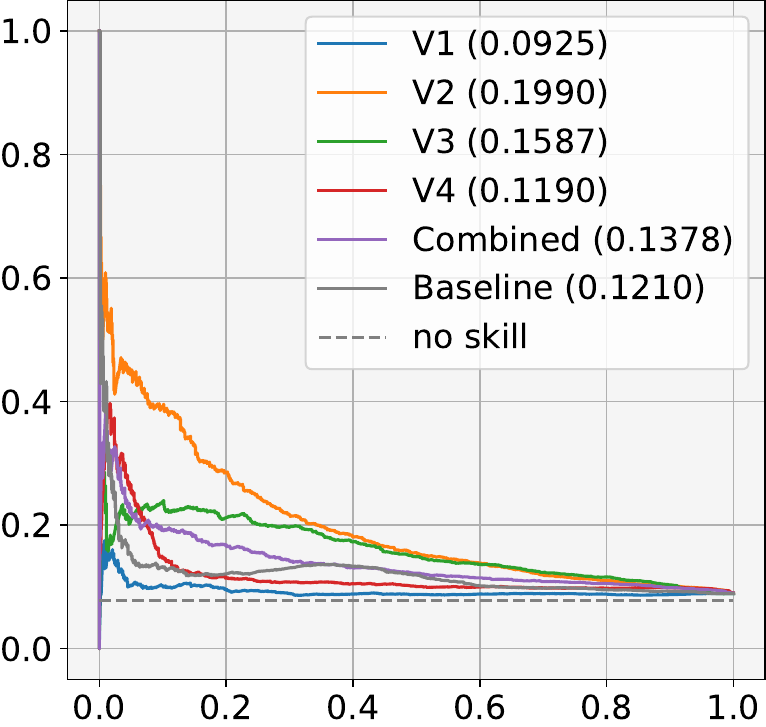} & \includegraphics[width=4.5cm]{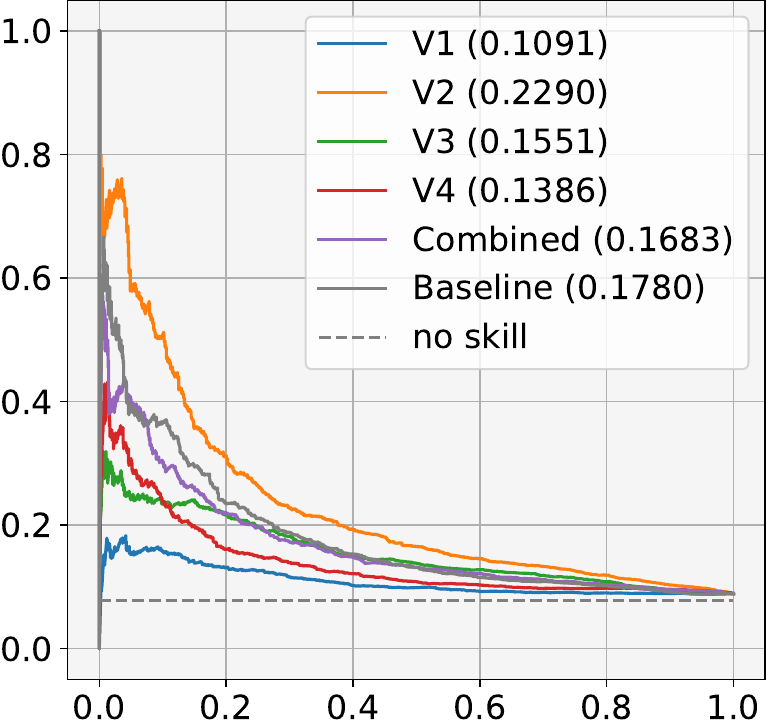}\\
\end{tabular}
\end{center}
\caption{PR curves of the different models applied on \textit{testing} with different sampling steps.}\label{ablation_pr_full_testing}
\end{figure*}

\begin{figure*}
\begin{center}
\begin{tabular}{cccc}
& SAR & Optical & SAR \& Optical\\
& ($\delta^{SAR} \coloneqq \delta$, $\delta^{OPT} \coloneqq 2 \textrm{ days}$) & ($\delta^{SAR} \coloneqq 2 \textrm{ days}$, $\delta^{OPT} \coloneqq \delta$) & ($\delta^{SAR} \coloneqq \delta$, $\delta^{OPT} \coloneqq \delta$)\\
\multirow{1}{*}[22ex]{\rotatebox[origin=c]{90}{\parbox[c]{3cm}{\centering $\delta = 120 \textrm{ days}$}}} & \includegraphics[width=4.5cm]{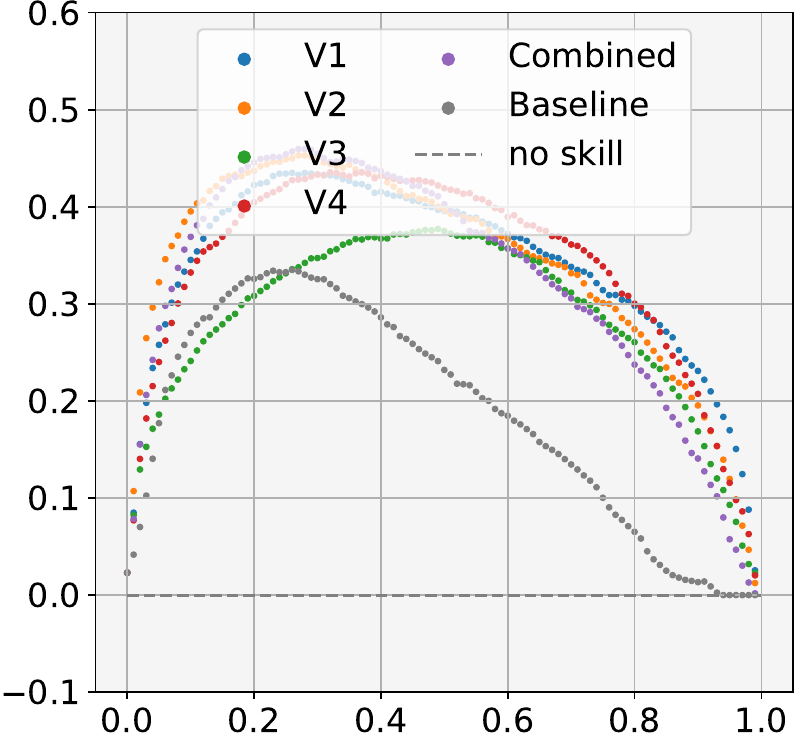} & \includegraphics[width=4.5cm]{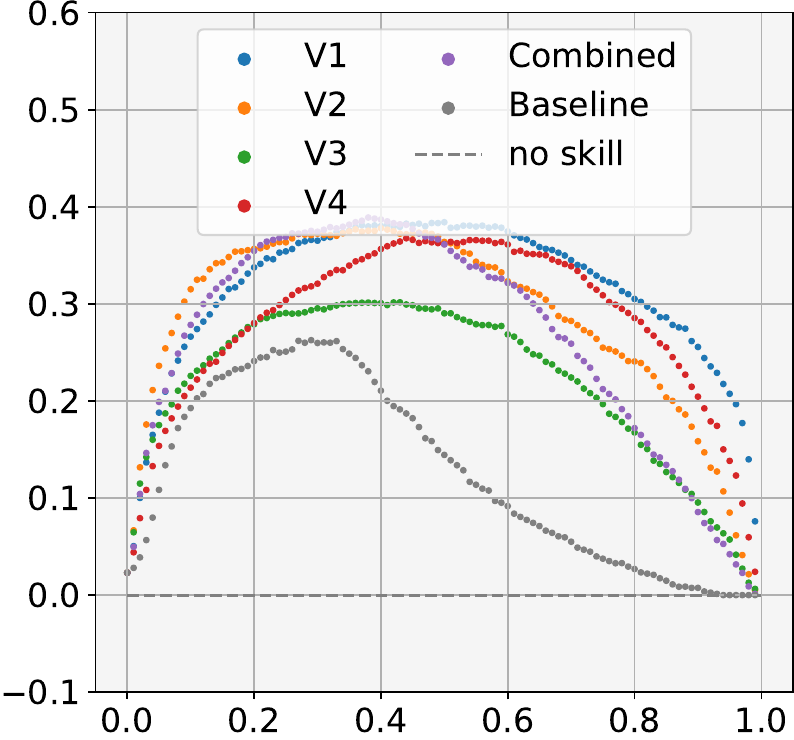} & \includegraphics[width=4.5cm]{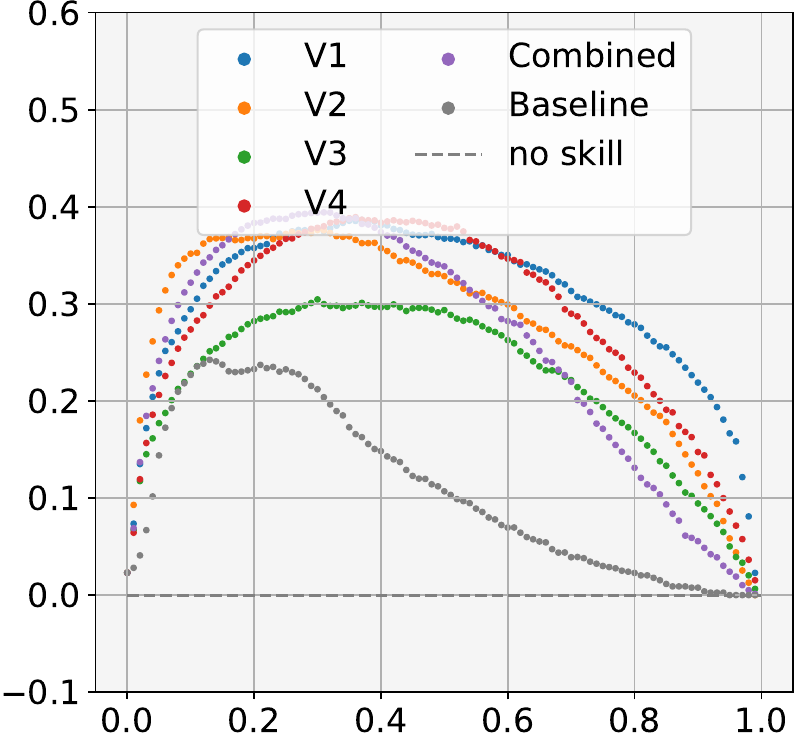}\\
\multirow{1}{*}[22ex]{\rotatebox[origin=c]{90}{\parbox[c]{3cm}{\centering $\delta = \infty \textrm{ days}$}}} & \includegraphics[width=4.5cm]{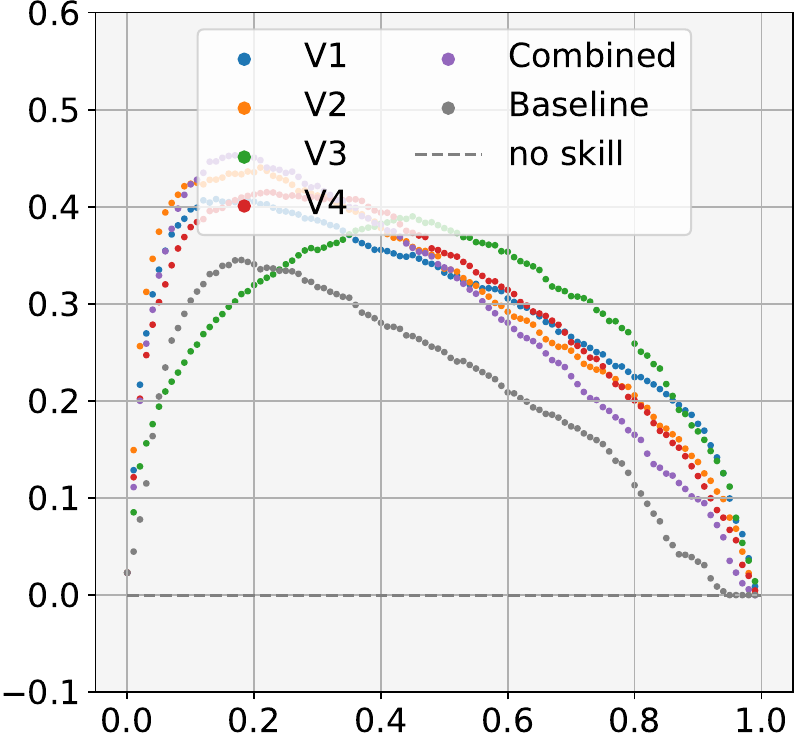} & \includegraphics[width=4.5cm]{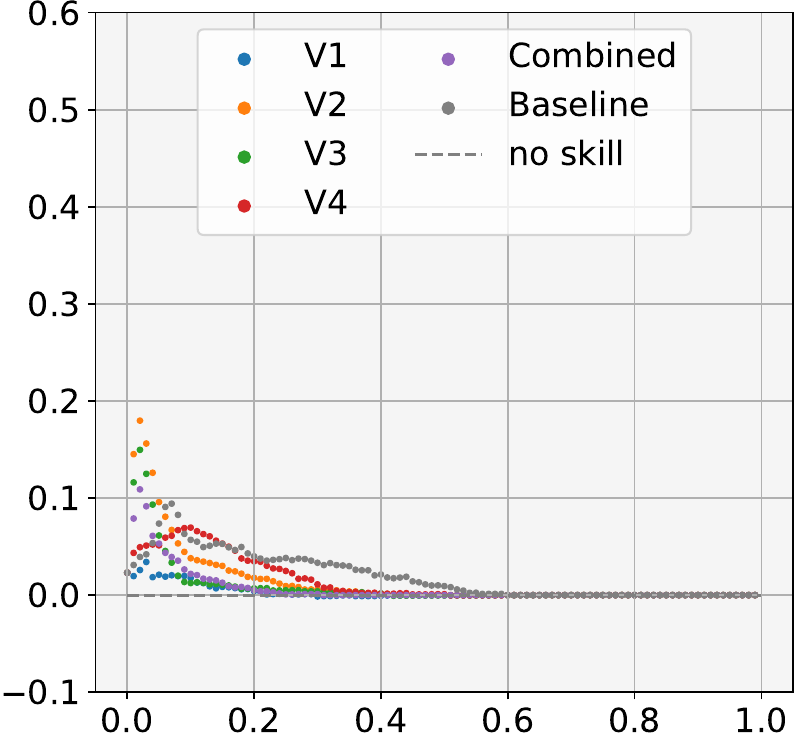} & \includegraphics[width=4.5cm]{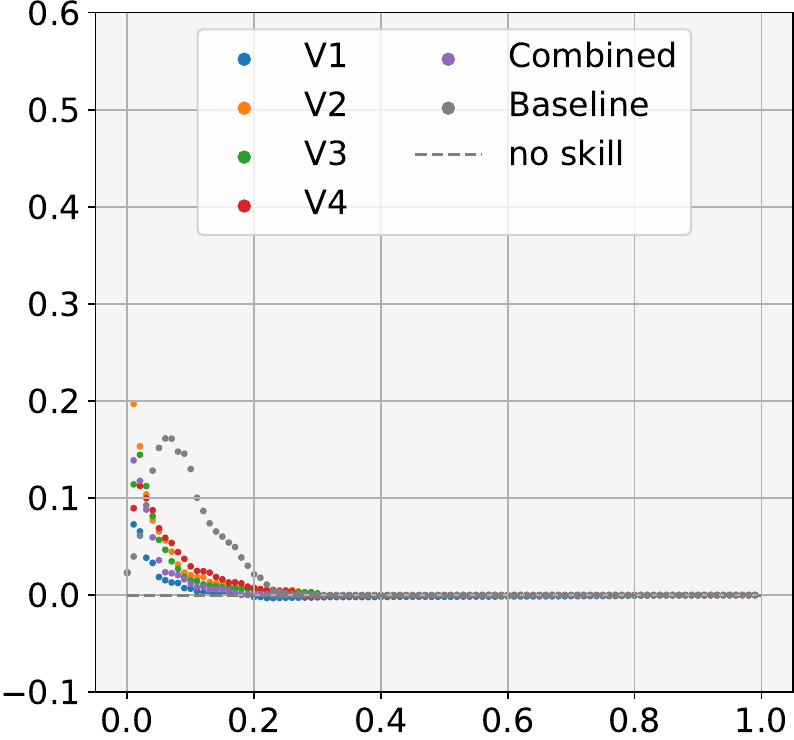}\\
\end{tabular}
\end{center}
\caption{Cohen's Kappa of the different models applied on \textit{testing} with different sampling steps and a changing threshold.}\label{ablation_kappa_full_testing}
\end{figure*}

\section{Urban Changes in Mariupol 2022/23}
\begin{figure*}[htb]
\FrameSep0pt
\centering
\begin{minipage}[c]{0.7\textwidth}
\includegraphics[width=\textwidth]{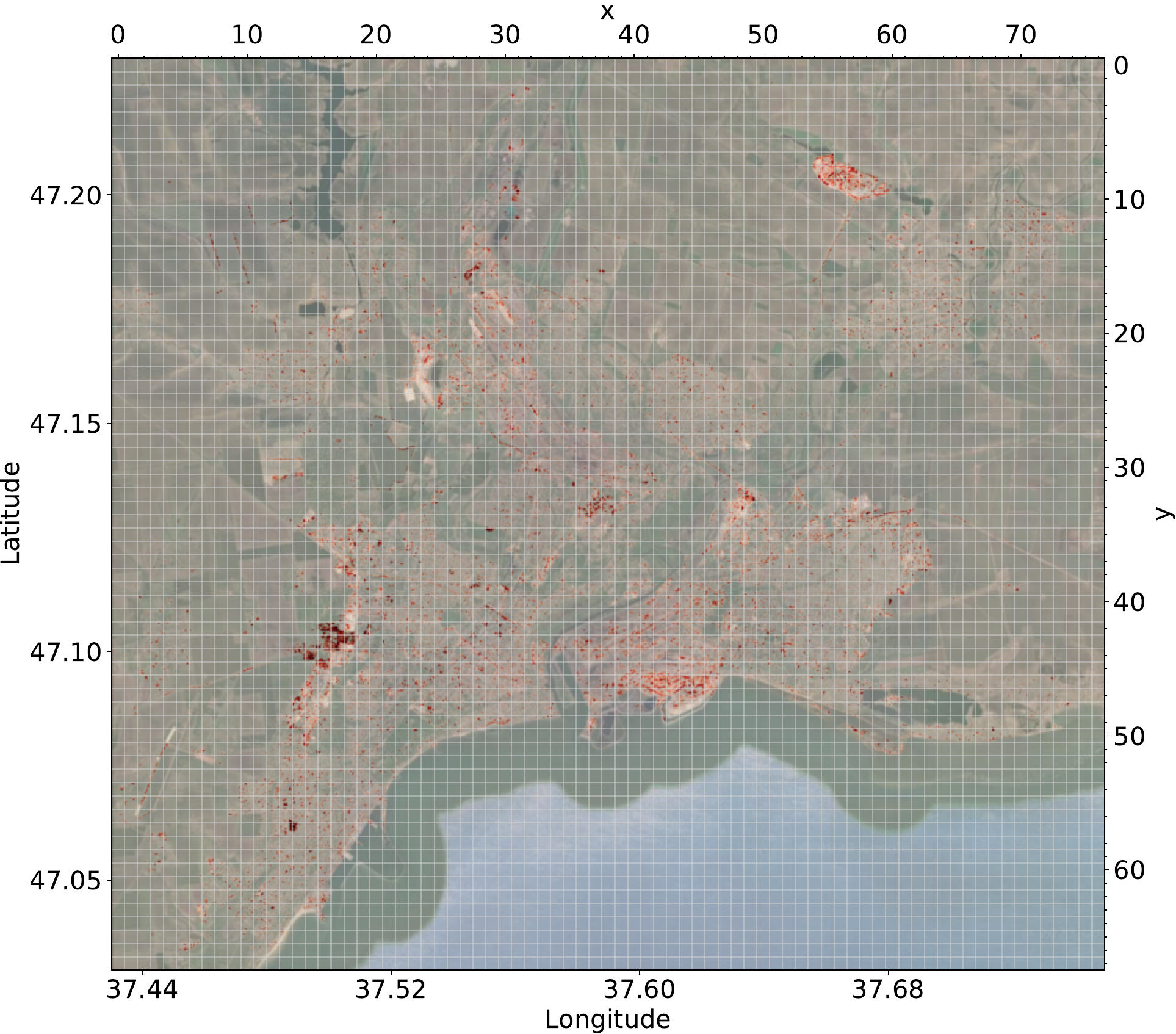}
\caption[Urban changes in Mariupol during the Russian invasion 2022/23 with combined models.]{Urban changes in Mariupol during the Russian invasion 2022/23 with combined models. Highlights in red show identified urban changes using $\bm{y}_{i,j}^{C}$ for every tile. Background image \copyright 2019/20 Google Earth, for reference only.}\label{mariupol_urban_changes}
\end{minipage}
\end{figure*}
\end{document}